\documentclass[
aps,prd,reprint,onecolumn,amsmath,amssymb
]{revtex4-2}
\usepackage[utf8]{inputenc}
\usepackage[T1]{fontenc}
\usepackage{amsmath}
\usepackage{autobreak}
\usepackage{amssymb}
\usepackage{amsthm}
\usepackage{amsfonts}
\usepackage{braket}
\usepackage{bm}
\usepackage{graphicx}
\usepackage{float}
\usepackage{dashrule}
\usepackage{subcaption}
\usepackage{color}
\usepackage{framed}
\usepackage{mathtools}
\usepackage{mathrsfs}
\usepackage{booktabs}
\usepackage{array}
\usepackage{makecell}
\usepackage{setspace}
\usepackage{multirow}
\usepackage[colorlinks, citecolor=blue,anchorcolor=red,menucolor=red, linkcolor=red,filecolor=red,runcolor=red,urlcolor=blue,frenchlinks=red]{hyperref} 
\usepackage{cleveref} 
\allowdisplaybreaks[2]

\begin{document}
	\title{P-wave fully charm and fully bottom tetraquark states}

	\author{Zhi-Zhong Chen$^1$}
\author{Xu-Liang Chen$^1$}
\author{Peng-Fei Yang$^1$}
\author{Wei Chen$^{1,\, 2}$}
\email{chenwei29@mail.sysu.edu.cn}
\affiliation{$^1$School of Physics, Sun Yat-sen University, Guangzhou 510275, China \\ 
$^2$Southern Center for Nuclear-Science Theory (SCNT), Institute of Modern Physics, 
Chinese Academy of Sciences, Huizhou 516000, Guangdong Province, China}
\begin{abstract}
We have studied the mass spectra of the P-wave fully charm and fully bottom tetraquark states in the framework of QCD sum rules. We construct the interpolating currents by inserting the covariant derivative operator $\overset{ \leftrightarrow } { \mathcal D }_{ \mu }$ between the S-wave diquark and antidiquark fields. The excitation structures show that the pure $\lambda$-mode excited P-wave fully heavy tetraquarks exist for the quantum numbers  $J^{PC}=1^{--}, 1^{-+}, 2^{--}, 2^{-+}$ and $3^{--}$, while it is difficult to separate the $\lambda$-mode and $\rho$-mode excitations in the $0^{-+}$ channel. Within three Lorentz indices, there is no pure $\lambda$-mode excited P-wave fully charm/bottom tetraquark operators with $J^{PC}=0^{--}$ and $3^{-+}$. Our results support that the recent observed $X(6900)$ and $X(7200)$ resonances could be interpreted as the P-wave fully charm $cc \bar c \bar c$ tetraquark states with $J^{PC}=1^{-+}$ and $2^{-+}$, respectively. Some P-wave fully bottom $bb\bar b\bar b$ tetraquark states are predicted to be lower than the di-$\eta_b(1S)$ and di-$\Upsilon(1S)$ mass thresholds. Hopefully our calculations will be useful for identifying the nature of new exotic tetraquark states. 
\end{abstract}
\maketitle
\section{Introduction \label{Introduction}}
The investigations of exotic hadrons outside the conventional quark model could provide insightful perspective for extending our understanding of non-perturbative behaviors of QCD. Since 2003, there have been observed lots of new hadron states including the XYZ states, hidden-charm pentaquark states, doubly charmed tetraquark $T_{cc}$ state, doubly charged $T_{c\bar s}(2900)^{++}$ state, etc. Since their masses and decay properties are different from the predictions of potential model, they have been considered as good candidates of exotic hadron states, such as tetraquark states, pentaquark states, hybrids and so on~\cite{Chen:2016qju,Guo:2017jvc,Liu:2019zoy,Brambilla:2019esw,Chen:2022asf}. 

The story is ongoing with the observations of the fully heavy tetraquark candidates. In 2017, the CMS Collaboration reported an exotic structure around $18.4$ GeV in the $\Upsilon(1S)\mu^+\mu^-$ channel~\cite{CMS:2016liw}. The ANDY Collaboration at RHIC reported an evidence of a new resonance at 
$18.12$ GeV in 2019~\cite{ANDY:2019bfn}. Although these structures were not confirmed by the later experiments~\cite{LHCb:2018uwm,CMS:2020qwa}, 
they still triggered lots of investigations on the fully heavy tetraquark states~\cite{Chen:2016jxd,Anwar:2017toa,Esposito:2018cwh,Hughes:2017xie,Karliner:2016zzc,Wu:2016vtq,Richard:2017vry,Bai:2016int,Chen:2019dvd,Debastiani:2017msn}. Actually, there were already some theoretical studies on such topics much earlier~\cite{Chao:1980dv,Ader:1981db,Silvestre-Brac:1992kaa,Lloyd:2003yc,Barnea:2006sd,Berezhnoy:2011xn}. 

In 2020, the LHCb Collaboration observed a narrow structure $X(6900)$ with a global significance larger than 5$\sigma$ and a broad structure ranging from 6.2 to 6.8 GeV in the $J/\psi J/\psi$ invariant mass spectrum~\cite{LHCb:2020bwg}. They also reported a hint for a structure around 7.2 GeV~\cite{LHCb:2020bwg}. Recently, the ATLAS and CMS Collaborations confirmed the existence of $X(6900)$ in the $J/\psi J/\psi$ invariant mass spectrum with high significance far exceeding $5\sigma$~\cite{ATLAS:2023bft,CMS:2023owd}. In the same $J/\psi J/\psi$ channel, ATLAS also observed another two structures  $X(6400)$ and $X(6600)$~\cite{ATLAS:2023bft}, while CMS observed $X(6600)$ and $X(7200)$ in two models with and without considering interference effects between the resonances~\cite{CMS:2023owd}. Meanwhile, the ATLAS Collaboration also studied the $J/\psi\psi(2S)$ decay channel and found the signals of $X(6900)$ and $X(7200)$ with a $4.7\sigma$ excess of events~\cite{ATLAS:2023bft}. For convenience, we collect the experimental information of these resonance structures in Table~\ref{X(6900)EXPERIMENT}. 
 
The observations of these exotic resonance structures have immediately attracted great theoretical interests to study the fully heavy tetraquarks for their mass spectra~\cite{Albuquerque:2020hio,Giron:2020wpx,Gordillo:2020sgc,Guo:2020pvt,Jin:2020jfc,Karliner:2020dta,Faustov:2020qfm,Ke:2021iyh,Li:2021ygk,Liang:2021fzr,Liu:2019zuc,liu:2020eha,Pal:2021gkr,Sonnenschein:2020nwn,Wan:2020fsk,Wang:2018poa,Wang:2019rdo,Wang:2020ols,Wang:2021kfv,Weng:2020jao,Yang:2020rih,Yang:2020wkh,Zhang:2020xtb,Zhao:2020zjh,Zhao:2020nwy,Zhu:2020xni,Cao:2020gul,Lesteiro-Tejeda:2021wne,Albuquerque:2021erv,Wang:2021mma,Wang:2021taf,Yang:2021zrc,An:2022qpt,Wu:2022qwd,Kuang:2022vdy,Faustov:2022mvs,Dong:2022sef,Lu:2023ccs,Feng:2023agq,Wang:2023igz,Mutuk:2021hmi,Yang:2021hrb,Galkin:2023wox,Anwar:2023fbp,Wu:2024euj,Agaev:2023tzi,Agaev:2024pej}, decay properties~\cite{Guo:2020pvt,Lu:2020cns,Li:2019uch,Chen:2020xwe,Becchi:2020uvq,Sonnenschein:2020nwn,Chen:2022sbf,Biloshytskyi:2022pdl,Sang:2023ncm,Anwar:2023fbp,Agaev:2023tzi,Agaev:2024pej,Agaev:2023rpj} and their production mechanisms~\cite{Huang:2021vtb,Feng:2020riv,Wang:2020gmd,Feng:2020qee,Maciula:2020wri,Goncalves:2021ytq,Ma:2020kwb,Wang:2020tpt,Zhao:2020nwy,Zhu:2020sn,Gong:2020bmg,Feng:2023ghc}. 
A fully heavy tetraquark is more likely to form a compact tetraquark state via the short range gluon-exchange color interaction rather than a loosely bound hadron molecule due to the absence of light quarks~\cite{Maiani:2020pur,Chao:2020dml,Richard:2020hdw}. There are also some different interpretations such as a cusp effect~\cite{Zhuang:2021pci}, a dynamically resonance pole structure from the coupled-channel effects~\cite{Gong:2020bmg} , hybrid tetraquarks $c c \bar c \bar c g$ \cite{Wan:2020fsk}, a BSM $0^{++}$ Higgs-like boson \cite{Zhu:2020snb}. To date, the quantum numbers of these exotic states have not yet been determined. In Refs.~\cite{Chen:2016jxd,Faustov:2020qfm,liu:2020eha,Zhang:2022qtp,Yan:2023lvm,Yu:2022lak,Ke:2021iyh,Li:2021ygk}, $X(6600)$ may be interpreted as the $S$-wave $0^{++}$ fully charm tetraquark state. For the $X(6900)$ state, it may be explained as the P-wave $c c \bar c \bar c$ tetraquark with $J^{PC}=0^{-+}$, $1^{-+}$ or $2^{-+}$~\cite{Chen:2016jxd,Bedolla:2019zwg,Debastiani:2017msn,Tiwari:2021tmz,Yu:2022lak,Kuchta:2023anj,Niu:2022cug}, or the $S$-wave $c c \bar c \bar c$ state with $J^{PC}=0^{++}$~\cite{Ke:2021iyh,liu:2020eha,Giron:2020wpx,Zhao:2020nwy,Li:2021ygk,Karliner:2020dta,Zhou:2022xpd,Kuang:2023vac,Niu:2022cug}, or $1^{++}$~\cite{Ke:2021iyh,Zhao:2020jvl}. There are also many different  interpretations for $X(7200)$, such as the 2$S$, 3$S$, 1$D$ state and so on~\cite{Faustov:2020qfm,Li:2021ygk,Zhao:2020jvl,Liang:2021fzr,Liu:2021rtn,Yu:2022lak,Agaev:2023rpj}. 

	\begin{table}[t]
		\setstretch{1.2}
		\small
		\caption{Experimental information for the candidates of fully charm tetraquark states.} \label{X(6900)EXPERIMENT}
		\begin{ruledtabular}
			\begin{tabular}{c c c c c}
				Collaborations & Resonances & Masses (MeV) & Widths (MeV) & Observed Channels \\
				\midrule[0.05pt] $\makecell[l]{\text{LHCb (model I)} \\ \text{LHCb (model II)}}$ & $X(6900)$ & $\begin{array}{c} 6905 \pm 11 \pm 7 \\ 6886 \pm 11 \pm 1 \end{array}$ & $\begin{array}{c} 80 \pm 19 \pm 33 \\ 168 \pm 33 \pm 69 \end{array}$ & $\text{di-}J/\psi$~\cite{LHCb:2020bwg}\\
				\midrule[0.05pt] ATLAS (model A) & $\begin{array}{c} X(6400) \\ X(6600) \\ X(6900) \end{array}$ & $\begin{array}{c} 6410 \pm 80^{+80}_{-30} \\ 6630 \pm 50^{+80}_{-10} \\ 6860 \pm 30^{+10}_{-20} \end{array}$ & $\begin{array}{c} 590 \pm 350^{+120}_{-200} \\ 350 \pm 110^{+110}_{-40} \\ 110 \pm 50^{+20}_{-10} \end{array}$ & $\text{di-}J/\psi$~\cite{ATLAS:2023bft} \\
				\midrule[0.05pt] ATLAS (model B) & $\begin{array}{c}  X(6600) \\ X(6900) \end{array}$ & $\begin{array}{c} 6650 \pm 20^{+30}_{-20} \\ 6910 \pm 10\pm 10 \end{array}$ & $\begin{array}{c} 440 \pm 50^{+60}_{-50} \\ 150 \pm 30\pm 10 \end{array}$ & $\text{di-}J/\psi$~\cite{ATLAS:2023bft} \\
				\midrule[0.05pt] $\makecell[l]{\text{ATLAS (model $\beta$)} \\ \text{ATLAS (model $\alpha$)}}$  & $\begin{array}{c} X(6900) \\ X(7200) \end{array}$ & $\begin{array}{c} 6960 \pm 50\pm 30 \\ 7220 \pm 30^{+10}_{-40} \end{array}$ & $\begin{array}{c} 510 \pm 170^{+110}_{-100} \\ 90\pm 60^{+60}_{-50} \end{array}$ & $J/\psi \psi(2S)$~\cite{ATLAS:2023bft} \\
				\midrule[0.05pt] CMS (No-interference) & $\begin{array}{c} X(6600) \\ X(6900) \\ X(7200) \end{array}$ & $\begin{array}{c} 6552\pm 10\pm 12 \\ 6927\pm 9\pm 4 \\ 7287^{+20}_{-18}\pm 5 \end{array}$ & $\begin{array}{c} 124^{+32}_{-26}\pm 33 \\ 122^{+24}_{-21}\pm 18 \\ 95^{+59}_{-40}\pm 19 \end{array}$ & $\text{di-}J/\psi$~\cite{CMS:2023owd} \\
				\midrule[0.05pt] CMS (Interference) & $\begin{array}{c} X(6600) \\ X(6900) \\ X(7200) \end{array}$ & $\begin{array}{c} 6638^{+43+16}_{-38-31} \\ 6847^{+44+48}_{-28-20} \\ 7134^{+48+41}_{-25-15} \end{array}$ & $\begin{array}{c} 440^{+230+110}_{-200-240} \\ 191^{+66+25}_{-49-17} \\ 97^{+40+29}_{-29-26} \end{array}$ & $\text{di-}J/\psi$~\cite{CMS:2023owd} \\
			\end{tabular}
		\end{ruledtabular}
	\end{table}

In this work, we shall study the P-wave fully charm and fully bottom compact tetraquark states carrying various quantum numbers within the framework of QCD sum rules. We construct the non-local interpolating tetraquark currents with negative parities by inserting the covariant derivative operator between the S-wave diquark and antidiquark fields. These currents are different from our previous studies in which the P-wave fully heavy tetraquarks were in pure $\rho$-mode excitation~\cite{Chen:2016jxd}. 

This paper is organized as follows. In Sec.\;\ref{ICPFCTS}, we construct the compact P-wave fully heavy interpolating currents in the diquark-antidiquark configuration. In Sec.\;\ref{QCDSR}, we introduce the formalism of QCD sum rules and calculate the two-point correlation functions and spectral functions for all these currents. We perform numerical analyses for the P-wave $cc\bar c\bar c$ and $bb\bar b\bar b$ tetraquark systems and predict their mass spectra in Sec.\;\ref{NA}. In the last section we give a brief summary. There is an additional supplementary file ``SpectralFunction.pdf'' containing the spectral functions for all interpolating currents.

\begin{table}
	\setstretch{1.3}
	\small
	\caption{Properties of the diquark operators ($i, j=1, 2, 3$) without and with one covariant derivative. The braces $\{...\}$ and square brackets $[...]$ in the nonlocal diquark operators mean that the inside Lorentz indices are symmetrized and antisymmetrized, respectively. ``TP'' represents the trace part of corresponding operator.} \label{DIQUARK_CURRENTS}
\begin{ruledtabular}
\begin{tabular}{lccc}
\text{Operators} & \makecell[c]{$J^P$} & $\text { States }$ & $(\text {Flavor},\text{Color})$ \\
\midrule[0.01pt] 
$q_a^T C \gamma_5 q_b$ &$\ \;\;0^{+}$ & ${ }^1 S_0$ & $(6_\textbf f,6_\textbf c)$ or $(\bar 3_\textbf f,\bar 3_\textbf c)$ \\
$q_a^T C q_b$ & $\ \;\;0^{-}$ & ${ }^3 P_0$ & $(6_\textbf f,6_\textbf c)$ or $(\bar 3_\textbf f,\bar 3_\textbf c)$ \\
$q_a^T C \gamma_\mu \gamma_5 q_b$ &
			$
\begin{dcases}
				1^{-}, & \text {for } \mu=i \\
				0^{+}, & \text {for } \mu=0
			\end{dcases}
			$ & $
			\begin{array}{c}
				{ }^3 P_1 \\[0.68ex] { }^1 S_0
			\end{array}
			$ & $(6_\textbf f,6_\textbf c)$ or $(\bar 3_\textbf f,\bar 3_\textbf c)$ \\ [4ex]
			$q_a^T C \gamma_\mu q_b$ & 
			$
			\begin{dcases}
				1^{+}, & \text {for } \mu=i \\
				0^{-}\,\footnote{These channels exist only for the unequal mass diquarks containing two quarks with different flavors. They vanish for the equal mass diquarks due to the Fermi-Dirac statistics.}, & \text {for } \mu=0\
			\end{dcases}
			$ & $
			\begin{array}{c}
				{ }^3 S_1 \\[0.68ex] { }^3 P_0
			\end{array}
			$ & $(6_\textbf f,\bar 3_\textbf c)$ or $(\bar 3_\textbf f,6_\textbf c)$ \\ [4ex]
			$q_a^T C \sigma_{\mu \nu} q_b
			$ & $
			\begin{dcases} \setstretch{2}
				1^{-}, & \text {for } \mu=i, \nu=j \\
				1^{+}, & \text {for } \mu=0, \nu=i\text{ or }\mu=i, \nu=0
			\end{dcases}
			$ & $
			\begin{array}{c}
				{ }^1 P_1 \\[0.68ex]
				{ }^3 S_1
			\end{array}
			$ & $(6_\textbf f,\bar 3_\textbf c)$ or $(\bar 3_\textbf f,6_\textbf c)$ \\ [4ex]
			$q_a^T C \sigma_{\mu \nu} \gamma_5 q_b
			$ &$
			\begin{dcases} \setstretch{2}
				1^{+}, & \text {for } \mu=i, \nu=j \\
				1^{-}, & \text {for } \mu=0, \nu=i\text{ or }\mu=i, \nu=0
			\end{dcases}
			$ & $
			\begin{array}{c}
				{ }^3 S_1 \\[0.68ex]{ }^1 P_1
			\end{array}
			$ & $(6_\textbf f,\bar 3_\textbf c)$ or $(\bar 3_\textbf f,6_\textbf c)$\\ [2.8ex]
			\midrule[0.01pt]  
			$q_a^T C \gamma_5 \overset{ \leftrightarrow } { \mathcal D }_{ \mu } q_b$ & $
			\begin{dcases}
				1^{-}, & \text {for } \mu=i \\
				0^{+}\,\footnotemark[1], & \text {for } \mu=0
			\end{dcases}
			$ & $
			\begin{array}{c}
				{ }^1 P_1 \\[0.68ex] { }^1 S_0
			\end{array}
			$ & $(6_\textbf f,\bar 3_\textbf c)$ or $(\bar 3_\textbf f,6_\textbf c)$\\ [4ex]
			$q_a^T C \overset{ \leftrightarrow } { \mathcal D }_{ \mu } q_b$ & $
			\begin{dcases}
				1^{+}, & \text {for } \mu=i \\
				0^{-}\,\footnotemark[1], & \text {for } \mu=0\
			\end{dcases}
			$ & $
			\begin{array}{c}
				{ }^3 D_1 \\[0.68ex] { }^3 P_0
			\end{array}
			$ & $(6_\textbf f,\bar 3_\textbf c)$ or $(\bar 3_\textbf f,6_\textbf c)$ \\ [4ex]
			$q_a^T C \gamma_{ \{ \mu } {}_{,} \overset{ \leftrightarrow } { \mathcal D }_{ \nu \} } q_b
			$ & $
			\begin{dcases} \setstretch{2}
				2^{-}, & \text {for } \mu=i, \nu=j \\
				1^{+}\,\footnotemark[1], & \text{for } \mu=0, \nu=i\text{ or }\mu=i, \nu=0 \\
				0^{-}, & \text {for } \mu=i, \nu=j\text{ or } \text{TP}
			\end{dcases}
			$ & $
			\begin{array}{c}
				{ }^3 P_2 \\[0.68ex]
				{ }^3 S_1 \\[0.68ex]
				{ }^3 P_0
			\end{array}
			$ & $(6_\textbf f,6_\textbf c)$ or $(\bar 3_\textbf f,\bar 3_\textbf c)$ \\ [4ex]
			$q_a^T C \gamma_{ [ \mu } {}_{,} \overset{ \leftrightarrow } { \mathcal D }_{ \nu ] } q_b$ & $
			\begin{dcases} \setstretch{2}
				1^{-}, & \text {for } \mu=i, \nu=j \\
				1^{+}\,\footnotemark[1], & \text {for } \mu=0, \nu=i\text{ or }\mu=i, \nu=0
			\end{dcases}
			$ & $
			\begin{array}{c}
				{ }^3 P_1 \\[0.68ex]
				{ }^3 S_1
			\end{array}
			$ & $(6_\textbf f,6_\textbf c)$ or $(\bar 3_\textbf f,\bar 3_\textbf c)$ \\ [4ex]
			$q_a^T C \gamma_{ \{ \mu } {}_{,} \overset{ \leftrightarrow } { \mathcal D }_{ \nu \} } \gamma_5 q_b
			$ & $
			\begin{dcases} \small
				2^{+}, & \text {for } \mu=i, \nu=j \\
				1^{-}, & \text {for } \mu=0, \nu=i\text{ or }\mu=i, \nu=0 \\
				0^{+}\,\footnotemark[1], & \text {for } \mu=i, \nu=j\text{ or }\text{TP}
			\end{dcases}
			$ & $
			\begin{array}{c}
				{ }^3 D_2 \\ [0.68ex]
				{ }^1 P_1 \\ [0.68ex]
				{ }^1 S_0
			\end{array}
			$ & $(6_\textbf f,\bar 3_\textbf c)$ or $(\bar 3_\textbf f,6_\textbf c)$ \\ [5.8ex]
			$q_a^T C \gamma_{ [ \mu } {}_{,} \overset{ \leftrightarrow } { \mathcal D }_{ \nu ] } \gamma_5 q_b
			$ & $
			\begin{dcases} \setstretch{2}
				1^{+}, & \text {for } \mu=i, \nu=j \\
				1^{-}, & \text {for } \mu=0, \nu=i\text{ or }\mu=i, \nu=0
			\end{dcases}
			$ & $
			\begin{array}{c}
				{ }^3 D_1 \\[0.68ex]
				{ }^1 P_1
			\end{array}
			$ & $(6_\textbf f,\bar 3_\textbf c)$ or $(\bar 3_\textbf f,6_\textbf c)$
		\end{tabular}
	\end{ruledtabular}
\end{table}
\section{Interpolating Currents for P-wave fully heavy Tetraquark States \label{ICPFCTS}}
\begin{figure}[b]
\centering
\includegraphics[width=5.5cm]{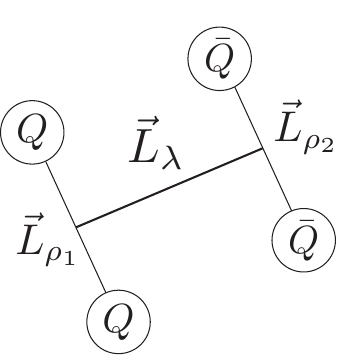}
\caption{Excitation configuration of the compact diquark-antidiquark tetraquark system, in which $\vec{L}_{\rho1}$($ \vec L_{\rho2}$) describes the intrinsic orbital excitation inside of the diquark (antidiquark), and $ \vec L_{\lambda}$ represents the relative orbital excitation between diquark and antidiquark fields.} \label{ccccL}
\end{figure}

We construct the interpolating currents for the P-wave fully heavy tetraquark states in the compact $[QQ] [\bar Q \bar Q]$ diquark-antidiquark configuration. As shown in Fig.~\ref{ccccL}, the orbital momentum in a comp act tetraquark system can be decomposed as 
\begin{equation}
\vec L = \vec{L}_{\rho} + \vec L_{\lambda}=\vec{L}_{\rho1} +\vec{L}_{\rho2} + \vec L_{\lambda}\, ,
\end{equation}
where $\vec{L}_{\rho1}$($ \vec L_{\rho2}$) is the intrinsic orbital angular momentum in the diquark $[QQ]$ (antidiquark $[\bar Q \bar Q]$) field, while $ \vec L_{\lambda}$ is the relative orbital angular momentum between diquark and antidiquark fields. The P-wave tetraquark operator can be in either the $\rho$-mode excitation ($L_{\rho1}+L_{\rho2}=L_{\rho}=1$, $L_{\lambda}=0$) or $\lambda$-mode excitation ($L_{\lambda}=1$, $L_\rho=0$). 

A diquark operator can be composed of two quark fields and insertions of the covariant derivative. We collect the properties of the diquark operators without (local) and with one covariant derivative (nonlocal) in Table~\ref{DIQUARK_CURRENTS}, where the local ones can also be found in Refs.~\cite{Jaffe:2004ph,Du:2012wp}. In the SU(3) theory, the color structure of a diquark field could be symmetric sextet $\bf 6_\textbf c$ or antisymmetric triplet $\bf \bar 3_\textbf c$
\begin{equation}
\bf 3_\textbf{c} \otimes 3_\textbf c = 6_\textbf c \oplus \bar 3_\textbf c\, ,
\end{equation}
while sextet $\bf\bar 6_\textbf c$ or triplet $\bf 3_\textbf c$ for an antidiquark field 
\begin{equation}
\bf \bar 3_\textbf c \otimes \bar 3_\textbf c = \bar 6_\textbf c \oplus 3_\textbf c\, .
\end{equation}

The flavor structure for the light diquark system is similar to the above color decomposition according to the SU(3) flavor symmetry. As indicated in Ref.~\cite{Du:2012wp}, the color and flavor structures are always entangled for these diquark operators due to the Fermi-Dirac statistics. For the $[cc] [\bar c \bar c]$ fully charm tetraquark systems, the flavor structures of both diquark and antidiquark fields are symmetric $\bf 6_\textbf f$ so that their color structures are also determined. As shown in Table~\ref{DIQUARK_CURRENTS}, the vector and tensor diquark operators can couple to various channels with different spin-parity quantum numbers. However, some of these channels shall vanish for the diquarks consisting of two identical quark fields, since the restriction of the Fermi-Dirac statistics. They exist only for the unequal mass diquarks, which contain two quarks with different flavors. It is easy to find that all the nonlocal diquarks in the equal mass systems couple to P-wave or D-wave states. Interested in potentially low energy configurations, we shall use only the S-wave local diquarks $Q_a^T C \gamma_5 Q_b$, $Q_a^T C \gamma_\mu Q_b$ and $Q_a^T C \sigma_{\mu\nu}\gamma_5 Q_b$ to construct the P-wave tetraquark currents by introducing the covariant derivative operator
\begin{equation}
\begin{aligned}
\overset{ \leftrightarrow } { \mathcal D }_{ \mu } &= \overset{ \rightarrow } { \mathcal D }_{ \mu } - \overset{ \leftarrow } { \mathcal D }_{ \mu }\, , \\
\overset{ \rightarrow } { \mathcal D }_{ \mu } &= \overset{ \rightarrow } { \partial }_{ \mu } - i g_s A_{ \mu }\, ,~\overset{ \leftarrow } { \mathcal D }_{ \mu } = \overset{ \leftarrow } { \partial }_{ \mu } + i g_s A_{ \mu }\, ,
\end{aligned} \label{covariantderivative}
\end{equation}
in which $g_s$ is the strong coupling constant, and $A_\mu$ is the gluon field. Inserting $\overset{ \leftrightarrow } { \mathcal D }_{ \mu }$ between the diquark and antidiquark fields, we can obtain the interpolating currents for the P-wave fully heavy tetraquark states with $J^{PC}=0^{- \pm},\,1^{- \pm},\, 2^{- \pm},\, 3^{--}$ as the following 
\begin{equation}
\begin{aligned}
		j^{ 1,A } &= ( Q_a^T C \sigma^{ \mu \nu } \gamma_5 Q_b ) \overset{ \leftrightarrow } { \mathcal D }{}_{ \mu } ( \bar Q_a \gamma_\nu C \bar Q_b^T ) + ( Q_a^T C \gamma_\nu Q_b ) \overset{ \leftrightarrow } { \mathcal D }{}_{ \mu } ( \bar Q_a \sigma^{ \mu \nu } \gamma_5 C\bar Q_b^T ), &J^{ PC } = 0^{ -- };
		\\
		j^{ 2,A } &= ( Q_a^T C \sigma^{ \mu \nu } \gamma_5 Q_b ) \overset{ \leftrightarrow } { \mathcal D }{}_{ \mu } ( \bar Q_a \gamma_\nu C \bar Q_b^T ) - ( Q_a^T C \gamma_\nu Q_b ) \overset{ \leftrightarrow } { \mathcal D }{}_{ \mu } ( \bar Q_a \sigma^{ \mu \nu } \gamma_5 C\bar Q_b^T ), &J^{ PC } = 0^{ -+ };
		\\
		j^{ 3,S }_{ \mu } &= ( Q_a^T C \gamma_5 Q_b ) \overset{ \leftrightarrow } { \mathcal D }_{ \mu } ( \bar Q_a \gamma_5 C \bar Q_b^T ), &J^{ PC } = 1^{ -- };
		\\
		j^{ 4,A }_{ \mu } &= ( Q_a^T C \gamma_\alpha Q_b ) \overset{ \leftrightarrow } { \mathcal D }_{ \mu } ( \bar Q_a \gamma^\alpha C \bar Q_b^T ),& J^{ PC } = 1^{ -- };
		\\
		j^{ 5,A }_{ \mu } &= ( Q_a^T C \gamma_{ \mu } Q_b ) \overset{ \leftrightarrow } { \mathcal D }_{ \alpha } ( \bar Q_a \gamma^{ \alpha } C \bar Q_b^T ) + ( Q_a^T C \gamma^{ \alpha } Q_b ) \overset{ \leftrightarrow } { \mathcal D }_{ \alpha } ( \bar Q_a \gamma_{ \mu } C \bar Q_b^T ), &J^{ PC } = 1^{ -- };
		\\
		j^{ 6,A }_{ \mu } &= ( Q_a^T C \gamma_\mu Q_b ) \overset{ \leftrightarrow } { \mathcal D }_{ \alpha } ( \bar Q_a \gamma^\alpha C \bar Q_b^T ) - ( Q_a^T C \gamma^\alpha Q_b ) \overset{ \leftrightarrow } { \mathcal D }_{ \alpha } ( \bar Q_a \gamma_\mu C \bar Q_b^T ), &J^{ PC } = 1^{ -+ };
		\\
		j^{ 7,A }_{ \mu \nu } &= ( Q_a^T C \sigma_{ \mu \alpha } \gamma_5 Q_b ) \overset{ \leftrightarrow } { \mathcal D }{}^{ \alpha } ( \bar Q_a \gamma_\nu C \bar Q_b^T ) + ( Q_a^T C \gamma_\nu Q_b ) \overset{ \leftrightarrow } { \mathcal D}{}^{ \alpha } ( \bar Q_a \sigma_{ \mu \alpha } \gamma_5 C \bar Q_b^T ), &J^{ PC } = 2^{ -- };
		\\
	      j^{ 8,A }_{ \mu \nu } &= ( Q_a^T C \sigma_{ \alpha \mu } \gamma_5 Q_b ) \overset{ \leftrightarrow } { \mathcal D }_{ \nu } ( \bar Q_a \gamma^\alpha C \bar Q_b^T ) + ( Q_a^T C \gamma^\alpha Q_b ) \overset{ \leftrightarrow } { \mathcal D }_{ \nu } ( \bar Q_a \sigma_{ \alpha \mu } \gamma_5 C \bar Q_b^T ), &J^{ PC } = 2^{ -- };
		\\
		j^{ 9,A }_{ \mu \nu } &= ( Q_a^T C \sigma_{ \mu \alpha } \gamma_5 Q_b ) \overset{ \leftrightarrow } { \mathcal D }{}^{ \alpha } ( \bar Q_a \gamma_\nu C \bar Q_b^T ) - ( Q_a^T C \gamma_\nu Q_b ) \overset{ \leftrightarrow } { \mathcal D }{}^{ \alpha } ( \bar Q_a \sigma_{ \mu \alpha } \gamma_5 C \bar Q_b^T ), &J^{ PC } = 2^{ -+ };
		\\
		j^{ 10,A }_{ \mu \nu } &= ( Q_a^T C \sigma_{ \alpha \mu } \gamma_5 Q_b ) \overset{ \leftrightarrow } { \mathcal D }_{ \nu } ( \bar Q_a \gamma^\alpha C \bar Q_b^T ) - ( Q_a^T C \gamma^\alpha Q_b ) \overset{ \leftrightarrow } { \mathcal D }_{ \nu } ( \bar Q_a \sigma_{ \alpha \mu } \gamma_5 C \bar Q_b^T ), &J^{ PC } = 2^{ -+ };
		\\
		j^{ 11,A }_{ \mu \nu \rho } &= ( Q_a^T C \gamma_\mu Q_b ) \overset{ \leftrightarrow } { \mathcal D }_{ \rho } ( \bar Q_a \gamma_\nu C \bar Q_b^T ) + ( Q_a^T C \gamma_\nu Q_b ) \overset{ \leftrightarrow } { \mathcal D }_{ \rho } ( \bar Q_a \gamma_\mu C \bar Q_b^T ), &J^{ PC } = 3^{ -- },
\end{aligned} \label{tetraquarkcurrents}
\end{equation}
in which $Q=c/b$ is a charm or bottom quark field, the subscripts $a, b$ are color indices, and the superscripts $S$ and $A$ denote that the color structures of these diquark-antidiquark currents are symmetric $\bf 6_\textbf c \otimes \bar 6_\textbf c $ and antisymmetric $\bf \bar 3_\textbf c \otimes 3_\textbf c $ respectively. As shown in Eq.~\eqref{tetraquarkcurrents}, all interpolating currents except for $j^{ 3,S }_{ \mu }$ and $j^{ 4,A }_{ \mu }$ contain two parts, which are actually charge-conjugation for each other to make definite C-parity for the corresponding tetraquark currents. The vector currents $j^{ 3,S }_{ \mu }$ and $j^{ 4,A }_{ \mu }$ carry the negative $C$-parity with only one term. One can also find the $c \bar c u \bar d$ tetraquark currents with similar Lorentz structures in Ref.~\cite{Wang:2018ejf}. In our calculations, we don't consider the $A_\mu$ term in the covariant derivative, which actually leads to the hybrid tetraquark operator with an excited gluon field. 
\begin{table}[t]
		\setstretch{1.4}
		\small
		\caption{The $J^{PC}$ quantum numbers, color structures and excitation structures for the P-wave fully heavy tetraquark currents in Eq.~\eqref{tetraquarkcurrents}.} \label{CURRENT_STRUCTURES}
		\begin{ruledtabular}
			\begin{tabular}{cccc}
				$\text{Currents}$ & $J^{PC}$ & $\text {Color Structure}$ & Excitation Structure: $[L_{\lambda}, L_{\rho}]$\ \\
				\colrule
				$J^{1,A}$ & $0^{ -- }$ & $\bar 3_\textbf c \otimes  3_\textbf c$ &  $[0,1]$\\
				$J^{2,A}$ & $0^{ -+ }$ & $\bar 3_\textbf c \otimes  3_\textbf c$ & $[1,0] \oplus [0,1]$\\
				$J^{3,S}_{ \mu }$ & $1^{ -- }$ & $ 6_\textbf c \otimes \bar 6_\textbf c$ &  [1,0]\\
				$J^{4,A}_{ \mu }$ & $1^{ -- }$ & $\bar 3_\textbf c \otimes  3_\textbf c$ &  [1,0]\\
				$J^{5,A}_{ \mu }$ & $1^{ -- }$ & $\bar 3_\textbf c \otimes  3_\textbf c$ &  [1,0]\\
				$J^{6,A}_{ \mu }$ & $1^{ -+ }$ & $\bar 3_\textbf c \otimes  3_\textbf c$ &  [1,0]\\
				$J^{7,A}_{ \mu \nu }$ & $2^{ -- }$ & $\bar 3_\textbf c \otimes  3_\textbf c$ & $[1,0] \oplus [0,1]$\\
				$J^{8,A}_{ \mu \nu }$ & $2^{ -- }$ & $\bar 3_\textbf c \otimes  3_\textbf c$ & $[1,0]$\\
				$J^{9,A}_{ \mu \nu }$ & $2^{ -+ }$ & $\bar 3_\textbf c \otimes  3_\textbf c$ & $[1,0] \oplus [0,1]$\\
				$J^{10,A}_{ \mu \nu }$ & $2^{ -+ }$ & $\bar 3_\textbf c \otimes  3_\textbf c$ & $[1,0]$\\
				$J^{11,A}_{ \mu \nu \rho }$ & $3^{ -- }$ & $\bar 3_\textbf c \otimes  3_\textbf c$ &  [1,0]
			\end{tabular}
		\end{ruledtabular}
	\end{table}

One may wonder if the excitation structures of all interpolating currents in Eq.~\eqref{tetraquarkcurrents} be $\lambda$-mode with $[L_{\lambda},\, L_{\rho}]=[1,\, 0]$ since the existence of $\overset{ \leftrightarrow } { \mathcal D }_{\mu}$ between the diquark and antidiquark fields. However, the situation is more complicated due to the Lorentz contractions in these currents. We consider the following fully charm tetraquark operator 
\begin{equation}
\mathcal{O}_\mu=\mathcal{F}\overset{ \leftrightarrow } {\mathcal D }_{ \mu }\mathcal{\bar F}=( c_a^T C \Gamma_1 c_b ) \overset{ \leftrightarrow } { \mathcal D }_{ \mu } ( \bar c_a \Gamma_2 C \bar c_b^T )\, ,
\end{equation}
where $\mathcal{F}$ and $\mathcal{\bar F}$ are the diquark and antidiquark fields, respectively. The Lorentz structures of $\mathcal{F}$ and $\mathcal{\bar F}$ are contained in the Dirac matrices $\Gamma_1, \Gamma_2$, which may be contracted with $\overset{ \leftrightarrow } { \mathcal D }_{\mu}$ as the currents in Eq.~\eqref{tetraquarkcurrents}. Considering the definition of $\overset{ \leftrightarrow } { \mathcal D }_{\mu}$ in Eq.~\eqref{covariantderivative},  the $\mathcal{O}_i (i=1, 2, 3)$ component of $\mathcal{O}_\mu$ shall provide one relative orbital excitation $L_{\lambda}=1$ while $\mathcal{O}_0$ component gives no orbital excitation. This is very important for being able to separate the pure $\lambda$-mode excited P-wave contributions from the correlations functions induced by the interpolating currents $j^{ 3,S }_{ \mu }$, $j^{ 4,A }_{ \mu }$, $j^{ 5,A }_{ \mu }$, $j^{ 6,A }_{ \mu }$, $j^{8,A }_{ \mu\nu }$, $j^{10, A }_{ \mu\nu }$ and $j^{11,A }_{ \mu\nu\rho }$. However, it is not able to separate the pure ${\lambda}$-mode P-wave contributions for the currents $j^{ 2,A }$, $j^{7,A }_{ \mu\nu }$ and $j^{9, A }_{ \mu\nu }$ since the special way of Lorentz contractions in these operators. The P-wave tetraquarks extracted from these currents are mixed ones containing both ${\lambda}$-mode and ${\rho}$-mode excitations. For the current $j^{1,A }$, it contains only the ${\rho}$-mode excited P-wave component $\mathcal{O}_0$ while the $\lambda$-mode excited component $\mathcal{O}_i$ vanishes due to the Fermi-Dirac statistics. This is consistent with the results in Ref.~\cite{PhysRevD.104.036016} that only the ${\rho}$-mode P-wave $cc\bar c\bar c$ tetraquark exist in the $0^{--}$ channel. Within three Lorentz indices, we are not able to construct the $\lambda$-mode excited P-wave fully charm/bottom tetraquark current with $J^{PC}=3^{-+}$. We show the properties of these interpolating currents in Table~\ref{CURRENT_STRUCTURES}, including their $J^{PC}$ quantum numbers, color structures and excitation structures. 

\section{Formalism of QCD Sum Rules \label{QCDSR}}
In this section, we give a brief introduction to the method of QCD sum rules, which have been widely used to study the hadron properties, such as hadron masses, coupling constants, decay widths, magnetic moments and so on~\cite{Shifman:1978bx,Reinders:1984sr}. We start with the two-point correlation functions induced by the interpolating currents $J(x)$, $J_\mu(x)$, $J_{\mu\nu}(x)$ and $J_{\mu\nu\rho}(x)$ with various Lorentz indices
\begin{align}
\Pi(q^2)& =i \int d^4 x e^{i q \cdot x}\bra0 T[J(x) J^{\dagger}(0)]\ket0,\\
\nonumber \Pi_{\mu \nu}(q^2) & =i \int d^4 x e^{i q \cdot x}\bra0T [J_\mu(x) J_\nu^{\dagger}(0)]\ket0\\ 
&=\left(\frac{q_\mu q_\nu}{q^2}-g_{ \mu \nu }\right)\Pi_1(q^2)+\frac{q_\mu q_\nu}{q^2} \Pi_0(q^2), \label{Pivector} \\
\nonumber\Pi_{\mu \nu, \rho \sigma}(q^2) & =i \int d^4 x e^{i q \cdot x}\bra0 T[J_{\mu \nu}(x) J_{\rho \sigma}^{\dagger}(0)]\ket0\\ 
&= \left(\eta_{\mu \rho} \eta_{\nu \sigma}+\eta_{\mu \sigma} \eta_{\nu \rho}-\frac{2}{3} \eta_{\mu \nu} \eta_{\rho \sigma}\right) \Pi_2(q^2) +\cdots, \label{Pitensor2} \\ 
\nonumber\Pi_{\mu \nu \rho, \alpha \beta \sigma}(q^2) &=i \int d^4 x e^{i q \cdot x}\bra0 T[J_{\mu \nu \rho}(x) J_{\alpha \beta \sigma}^{\dagger}(0)]\ket0 \\ 
&= \theta_{\mu \nu \rho, \alpha \beta \sigma} \Pi_3(q^2)+\cdots , \label{Pitensor3}
\end{align}
where
\begin{equation}
\eta_{ \mu \nu } = \frac{q_\mu q_\nu}{q^2}-g_{ \mu \nu },
\end{equation}
\begin{equation}
\begin{aligned}
\theta_{\mu \nu \rho, \alpha \beta \sigma} = -\frac{1}{6} &\left[\eta_{\mu \alpha}\left(\eta_{\nu \sigma} \eta_{\rho \beta}+\eta_{\nu \beta} \eta_{\rho \sigma}\right)+\eta_{\mu \beta}\left(\eta_{\nu \sigma} \eta_{\rho \alpha}+\eta_{\nu \alpha} \eta_{\rho \sigma}\right) +\eta_{\mu \sigma}\left(\eta_{\nu \beta} \eta_{\rho \alpha}+\eta_{\nu \alpha} \eta_{\rho \beta}\right)\right] \\
			+\frac{1}{15} & \left[\eta_{\mu \nu}\left(\eta_{\beta \sigma} \eta_{\rho \alpha}+\eta_{\alpha \sigma} \eta_{\rho \beta}+\eta_{\alpha \beta} \eta_{\rho \sigma}\right)+\eta_{\mu \rho}\left(\eta_{\beta \sigma} \eta_{\nu \alpha}+\eta_{\alpha \sigma} \eta_{\nu \beta}+\eta_{\alpha \beta} \eta_{\nu \sigma}\right) \right.\\
			+ &\left.\eta_{\nu \rho} \left(\eta_{\beta \sigma} \eta_{\mu \alpha}+\eta_{\alpha \sigma} \eta_{\mu \beta}+\eta_{\alpha \beta} \eta_{\mu \sigma}\right) \right].
\end{aligned}
\end{equation}
The interpolating currents $J(x)$, $J_{\mu}(x)$, $J_{\mu \nu}(x)$ and $J_{\mu\nu\rho}(x)$ can couple to the potentially interested hadron states carrying the same quantum numbers with the operators. The invariant function $\Pi_i(q^2) (i=0, 1, 2, 3)$ in Eqs.~\eqref{Pivector}-\eqref{Pitensor3} denotes the pure spin-$i$ contribution to the correlation function. The ``$\cdots$'' in Eqs.~\eqref{Pitensor2}-\eqref{Pitensor3} contain contributions from different spins, which are omitted because we don't extract these invariant functions in this work. 

	The two-point correlation function can be calculated at the quark-gluonic level via the method of operator product expansion (OPE)~\cite{Wilson:1969zs,Ioffe:1981kw,Reinders:1984sr} by Wick's theorem:
\begin{equation}
\begin{aligned}
T_{\mu\nu...}\Pi(q^2)=i \int d^4 x e^{i q \cdot x}\bra 0 T[J_\Gamma(x) J_\Gamma^{\dagger}(0)]\ket 0 = \sum_n C^\Gamma_n(Q^2) \langle O_n \rangle, \, Q^2=-q^2\, ,
\end{aligned}
\end{equation}
in which $\langle O_n \rangle$ represents the vacuum expectation value of the local gauge invariant operator $O_n$ constructed from the quark and gluon fields, $C_n(Q^2)$ is the Wilson coefficient~\cite{Wilson:1969zs}. In QCD sum rules, the local gauge invariant operator $O_n$ contains only long distance nonperturbative effects. The Wilson coefficient $C_n(Q^2)$ describes short distance behaviors and can be calculated via the perturbative theory. In our calculations, we adopt the following propagator for heavy quarks up to dimension-4 gluon condensate
	\begin{equation}
		\begin{aligned}
			i S_Q^{a b}(p)=&\frac{i \delta^{a b}}{\hat{p}-m_Q}-\frac{i}{4} g_s \frac{\lambda_{a b}^n}{2} G_{\mu \nu}^n \frac{\sigma^{\mu \nu}\left(\hat{p}+m_Q\right)+\left(\hat{p}+m_Q\right) \sigma^{\mu \nu}}{(p^2-m_Q^2)^2}+\frac{i \delta^{a b}}{12}\left\langle g_s^2 G G\right\rangle m_Q \frac{p^2+m_Q \hat{p}}{(p^2-m_Q^2)^4}\, ,
			\label{HQP}
		\end{aligned}
	\end{equation}
where $\hat{p}=p^{\mu} \gamma_{\mu}$ and $\lambda^n$ is the Gell-Mann matrix.

At the hadronic level, one can describe the correlation function by using the dispersion relation 
\begin{align}
\Pi(q^2)= \frac{\left(q^2\right)^N}{\pi} \int_{s<}^{\infty} \frac{\operatorname{Im} \Pi(s)}{s^N\left(s-q^2-i \epsilon\right)} d s+\sum_{n=0}^{N-1} b_n\left(q^2\right)^n\, ,\label{OPE_DR_SF}
\end{align}
where $b_n$ is the subtraction constant which can be eliminated later by taking Borel transform. 
The imaginary part of $\Pi(q^2)$ can be defined as the spectral function
\begin{align}
\nonumber \rho(s) \equiv \frac{\operatorname{Im} \Pi(s)}{\pi}&=\sum_n \delta\left(s-m_n^2\right)\langle 0|J| n\rangle \langle n |J^{\dagger} | 0\rangle \\
&=f_X^2 \delta\left(s-m_X^2\right)+\cdots,
\end{align}
in which $m_X$ and $f_X$ denote the hadron mass and decay constant for the lowest-lying state. In the last step, the narrow resonance approximation is adopted, and ``$\cdots$'' contains contributions from the continuum and higher excited states. The Borel transform is usually applied to the correlation function to suppress the contributions from higher excited states and eliminate the unknown subtraction terms in Eq.~\eqref{OPE_DR_SF}
	\begin{equation}
		\hat B\left[\Pi\left(q^2\right)\right]=\lim _{\substack{-q^2, n \rightarrow \infty \\-q^2 / n \equiv M_B^2}} \frac{(-q^2)^{n+1}}{n !}\left(\frac{d}{d q^2}\right)^n \Pi\left(q^2\right),
	\end{equation}
in which the Borel parameter $M_B^2$ is introduced. Then one can establish the QCD sum rules by equating the correlation functions at both quark-gluonic and hadronic levels
\begin{equation}
\begin{aligned}
f_X^2 m_X^{2 N} e^{-m_X^2 / M_B^2}=&\int_{16m_Q^2}^{s_0} d s e^{-s / M_B^2} \rho(s) s^N \equiv \mathcal{L}_N\left(s_0, M_B^2\right), \label{sumrules}
\end{aligned}
\end{equation}
where $s_0$ is the continuum threshold above which the contributions from the higher excited states and continuum can be approximated well by the QCD spectral function. Then the hadron mass of the lowest-lying state can be obtained as the function of $s_0$ and $M_B^2$
	\begin{equation}
		m_X\left(s_0, M_B^2\right)=\sqrt{\frac{\mathcal{L}_1\left(s_0, M_B^2\right)}{\mathcal{L}_0\left(s_0, M_B^2\right)}}.
		\label{mass_of_X}
	\end{equation}
For the interpolating currents in Eq.~\eqref{tetraquarkcurrents}, we have calculated their two-point correlation functions and spectral functions up to dimension-4 gluon condensate at the leading order of $\alpha_s$. In general, the spectral functions can be expressed as 
	\begin{equation}
		\begin{aligned}
			\rho(s)=\int_{x_1}^{x_2}dx \int_{y_1}^{y_2}dy \int_{z_1}^{z_2}dz\,\left[\zeta_\text{pert}(s, x, y, z)+\zeta_{GG,1}(s, x, y, z)\right] +\int_{x_1}^{x_2}dx \int_{y_1}^{y_2}dy \,\zeta_{GG,2}(s, x, y), \label{sepctraldensityG}
		\end{aligned}
	\end{equation}
where $x, y, z$ are Feynman parameters and the integration limits are 
\begin{equation}
	\begin{aligned}
		x_1&=\frac{s-8 m^2_Q}{2 s}-\frac{1}{2} \sqrt{\frac{64 m^4_Q-20 m^2_Q s+s^2}{s^2}}, \\
		x_2&=\frac{s-8 m^2_Q}{2 s}+\frac{1}{2} \sqrt{\frac{64 m^4_Q-20 m^2_Q s+s^2}{s^2}}, \\
		y_1&=\frac{2 x m^2_Q+m^2_Q+x^2 s-x s}{2\left(m^2_Q-x s\right)}-\frac{1}{2} \sqrt{\frac{8 x m^4_Q+m^4_Q+8 x^3 s m^2_Q-6 x^2 s m^2_Q-2 x s m^2_Q+x^4 s^2-2 x^3 s^2+x^2 s^2}{\left(m^2_Q-x s\right)^2}}, \\
		y_2&=\frac{2 x m^2_Q+m^2_Q+x^2 s-x s}{2\left(m^2_Q-x s\right)}+\frac{1}{2} \sqrt{\frac{8 x m^4_Q+m^4_Q+8 x^3 s m^2_Q-6 x^2 s m^2_Q-2 x s m^2_Q+x^4 s^2-2 x^3 s^2+x^2 s^2}{\left(m^2_Q-x s\right)^2}}, \\
		z_1&=\frac{1-x-y}{2}-\sqrt{\frac{(x+y-1)}{4}\left(\frac{4 x y m^2_Q}{x y s-(x+y) m^2_Q}+x+y-1\right)}, \\
		z_2&=\frac{1-x-y}{2}+\sqrt{\frac{(x+y-1)}{4}\left(\frac{4 x y m^2_Q}{x y s-(x+y) m^2_Q}+x+y-1\right)}. 
	\end{aligned}
\end{equation}
In Eq.~\eqref{sepctraldensityG}, the function $\zeta_\text{pert}(s, x, y, z)$ provides the perturbative term while $\zeta_{GG,1}(s, x, y, z)$, $\zeta_{GG,2}(s, x, y)$ give the gluon condensate contributions to the spectral function. Then functions $\zeta_\text{pert}(s, x, y, z)$ and $\zeta_{GG,1}(s, x, y, z)$ are fractional polynomials coming from the first term and last two terms of the heavy quark propagator in Eq.\,\eqref{HQP}, while the function $\zeta_{GG,2}(s, x, y)$ comes from the second term of Eq.\,\eqref{HQP}. Since these expressions are very complicated and lengthy, we  collect them in a supplementary file (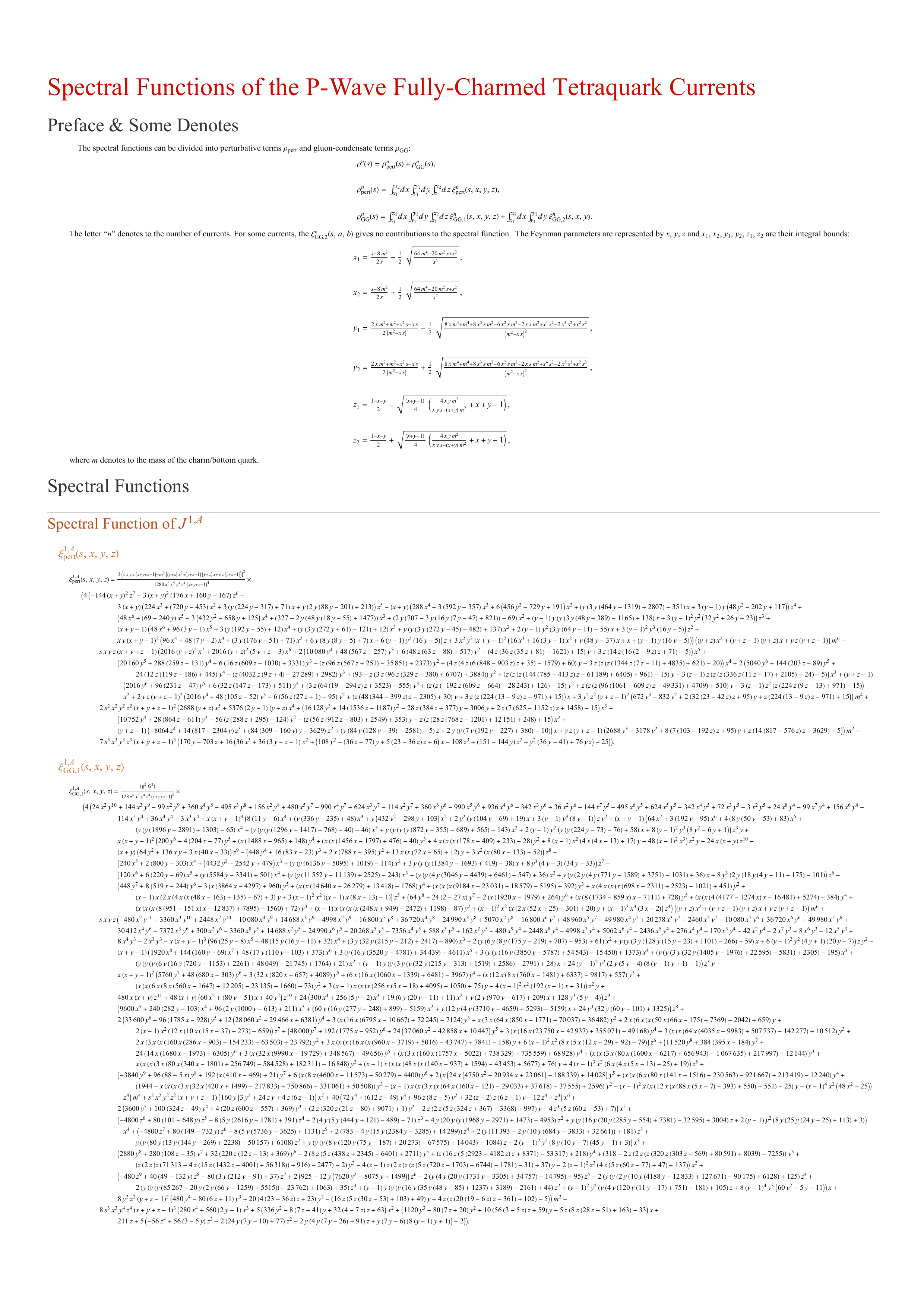).

\section{Numerical Analysis \label{NA}}
We use the following values of the heavy quark masses and the gluon condensate to perform Laplace sum rule numerical analysis~\cite{Ovchinnikov:1988gk,Narison:2018dcr,10.1093/ptep/ptac097}:
	\begin{equation}
		\begin{aligned}
			& m_c=1.27 \pm 0.02\,\mathrm{GeV}, \\
			& m_b=4.18^{+0.03}_{-0.02}\,\mathrm{GeV}, \\
			& \left\langle g_s^2 G G\right\rangle=(0.48 \pm 0.14)\ \mathrm{GeV}^4,
		\end{aligned}
	\end{equation}
where $m_c$ and $m_b$ are the running masses in the $\overline{\mathrm{MS}}$ scheme. The gluon condensate $\left\langle g_s^2 G G\right\rangle$ contributes nonperturbative effects to the correlation functions. 

It is clearly that the hadron mass is expressed as the function of the continuum threshold $s_0$ and Borel parameter $M_B^2$, as shown in Eq.~\eqref{mass_of_X}. To obtain reasonable sum rule analysis, one needs to find out the appropriate working regions for these two parameters. Since the weight function $e^{-s / M_B^2}$ in Eq.~\eqref{sumrules}, the Borel transform will improve the OPE convergence of the power series, giving the lower bound on the Borel parameter $M_B^2$. Meanwhile, its upper bound is also constrained to guarantee the pole dominance requirement. In this work, we assign that the contribution of perturbative term be three times larger than that of gluon condensate to ensure good OPE convergence and the pole contribution (PC) be larger than $50\%$
\begin{align}
\frac{\int_{16 m_Q^2}^{\infty}  e^{-s / M_B^2} \rho_\text{pert}(s) ds}{\int_{16 m_Q^2}^{\infty} e^{-s / M_B^2} \rho_{GG}(s) ds} &\geq 3\, ,
		\label{BMin}
\\
\operatorname{PC}\left(s_0, M_B^2\right)=\frac{\mathcal{L}_0\left(s_0, M_B^2\right)}{\mathcal{L}_0\left(\infty, M_B^2\right)}& \geq 50\%\, , \label{PC}
\end{align}
which will give the suitable working region of the Borel parameter $[M_{B,\text{min}}^2,\, M_{B,\text{max}}^2]$. However, the pole contribution in Eq.~\eqref{PC} depends also on the continuum threshold $s_0$, so that one needs to figure out its optimal value at the same time. Following Refs.~\cite{Chen:2017rhl,Duan:2024uuf}, we define the quantity $\chi^2(s_0)$ as
\begin{equation}
	\begin{aligned}
		\chi^2(s_0) & =\sum_{i=1}^N\left[\frac{m_X\left(s_0, M_{B,i}^2\right)}{\bar{m}_X\left(s_0\right)}-1\right]^2,
	\end{aligned}
		\label{ChiSquareMass}
\end{equation}
in which $\bar{m}_X(s_0)$ is defined as the averaged hadron mass
\begin{equation}
	\begin{aligned}
	\bar{m}_X(s_0) & =\sum_{i=1}^N \frac{m_X\left(s_0, M_{B,i}^2\right)}{N}, 
	\end{aligned}
		\label{ChiSquareMass}
\end{equation}
where $M_{B,i}^2 (i=1,2,\dots,N)$ are arbitrary $N$ points throughout the Borel window. 
\begin{figure}[H] 
\setstretch{1.5}
\centering
\subcaptionbox{$\chi^2$ Curve}{\includegraphics[width = .44\linewidth]{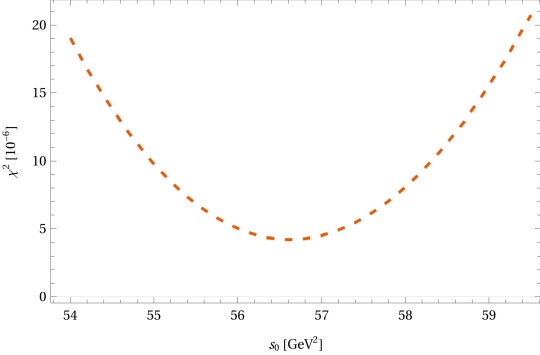}}\hspace{.5cm}
\subcaptionbox{Pole Contribution}{\includegraphics[width = .45\linewidth]{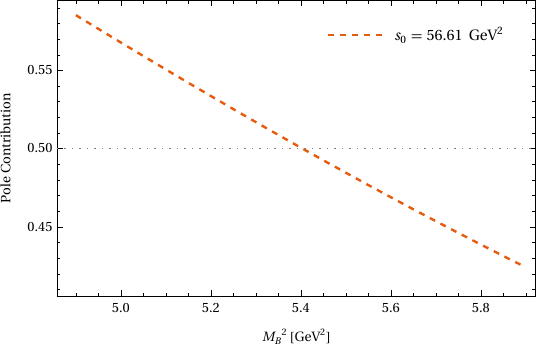}}
\caption{The $\chi^2(s_0)$ and pole contribution curves for the fully charm $cc\bar c\bar c$ tetraquark system by using the interpolating current $J^{6,A}$ with $J^{PC}=1^{-+}$.} \label{GRJ6}
\end{figure}
\begin{figure}[H] 
\setstretch{1.5}
\centering
\includegraphics[width = .44\linewidth]{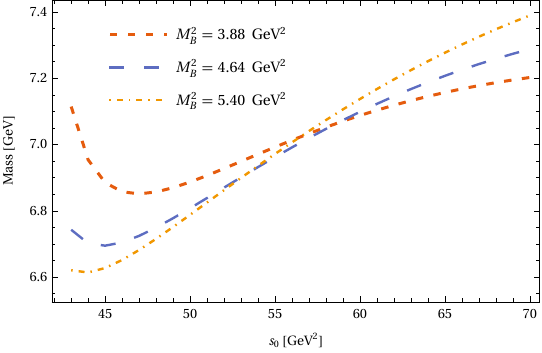}\hspace{.5cm}
\includegraphics[width = .44\linewidth]{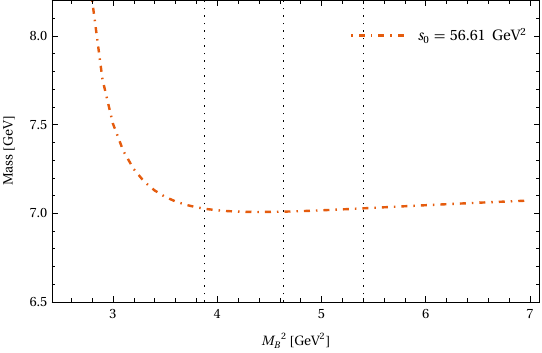}
\caption{The extracted tetraquark mass depends on the continuum threshold $s_0$ and Borel parameter $M_B^2$ for the fully charm $cc\bar c\bar c$ tetraquark state by using the interpolating current $J^{6,A}$ with $J^{PC}=1^{-+}$.} \label{GRJ6mass}
\end{figure}
We use the interpolating current $j^{ 6,A }_{ \mu }$ with $J^{PC}=1^{-+}$ as an example to illustrate the above discussions for the fully charm tetraquark system. By studying the OPE convergence and pole contribution as Eqs.~\eqref{BMin}-\eqref{PC}, we obtain the Borel window for $j^{ 6,A }_{ \mu }$ as $[M_{B,\text{min}}^2,\, M_{B,\text{max}}^2]=[3.88,\, 5.40]~\text{GeV}^2$. The optimal value of the continuum threshold can be found as $s_0=56.61^{+4.57}_{-4.96}~\text{GeV}^2$ by requiring the quantity $\chi^2(s_0)$ be minimum. We plot the $\chi^2(s_0)$ and pole contribution curves in Fig.~\ref{GRJ6} to show their behaviors. Within these parameter working regions, we show the extracted fully charm tetraquark mass depends on $s_0$ and $M_B^2$ in Fig.~\ref{GRJ6mass}. It is shown that the extracted tetraquark mass is very stable in the Borel window, so that one can obtain the hadron mass as 	\begin{gather}
m_X=7.02^{+0.26}_{-0.29} \text{ GeV}\, .
\end{gather}

Similar analyses can also be performed to the fully bottom $bb\bar b\bar b$ tetraquark systems. For all the interpolating currents in Eq.~\eqref{tetraquarkcurrents}, we collect the numerical results for the $cc\bar c\bar c$ and $bb\bar b\bar b$ tetraquark states in Table~\ref{MASSTABLEcccc} and Table~\ref{MASSTABLEbbbb}, respectively. The errors come from the uncertainties of the heavy quark masses, the gluon condensates, the continuum threshold $s_0$ and Borel parameter $M_B^2$. In appendix~\ref{MCAC}, we show the mass curves and Borel stabilities of fully charm $cc\bar c\bar c$ tetraquark systems for other interpolating currents in Eq.~\eqref{tetraquarkcurrents}. For the currents $J^{4,A}_{ \mu }$ and $J^{8,A}_{ \mu \nu }$, the pole contributions are too small to give suitable Borel windows so that the mass predictions from these two currents are unreliable. 

\begin{table}[H] 
\setstretch{1.4}
\caption{Numerical results for the P-wave fully charm $c c \bar c \bar c$ tetraquark states.} \label{MASSTABLEcccc}
	\begin{ruledtabular}
		\begin{tabular}{ccccccc}
			Currents & $J^{PC}$& Mass\,[GeV] & $s_0$\,[GeV${}^2$] & $M_{B,\text{min}}^2$\,[GeV${}^2$] & $M_{B,\text{max}}^2$\,[GeV${}^2$] & PC\,[\%]  \\
			\colrule $J^{1,A}$ & $0^{ -- }$ & $6.65^{+0.22}_{-0.25}$ & $50.12^{+3.63}_{-3.93}$ & $3.07^{+0.38}_{-0.43}$ & $4.39^{+0.49}_{-0.55}$ & $63.9^{+0.6}_{-0.3}$  \\
			$J^{2,A}$ & $0^{ -+ }$ & $9.38^{+0.37}_{-0.44}$ & $105.94^{+8.59}_{-9.99}$ & $8.53^{+0.98}_{-1.10}$ & $12.17^{+1.28}_{-2.70}$ & $64.8^{+0.5}_{-1.1}$  \\
			$J^{3,S}_{ \mu }$ & $1^{ -- }$ & $6.81^{+0.25}_{-0.37}$ & $52.33^{+4.19}_{-4.62}$ & $3.70^{+0.47}_{-0.55}$ & $4.60^{+0.57}_{-0.61}$ & $58.7 \pm 0.1$  \\
			$J^{4,A}_{ \mu }$ & $1^{ -- }$ & $-$ & $-$ & $-$ & $-$ & $-$ \\
			$J^{5,A}_{ \mu }$ & $1^{ -- }$ & $6.96^{+0.25}_{-0.27}$ & $55.24^{+4.41}_{-4.86}$ & $3.80^{+0.47}_{-0.54}$ & $5.05^{+0.60}_{-0.68}$ & $61.5\pm 0.3$ \\
			$J^{6,A}_{ \mu }$ & $1^{ -+ }$ & $7.02^{+0.26}_{-0.29}$ & $56.61^{+4.57}_{-4.96}$ & $3.88^{+0.48}_{-0.55}$ & $5.40^{+0.64}_{-0.70}$ & $63.2^{+0.3}_{-0.2}$ \\
			$J^{7,A}_{ \mu \nu }$ & $2^{ -- }$ & $7.55^{+0.32}_{-0.36}$ & $66.11^{+5.93}_{-6.51}$ & $5.29^{+0.71}_{-0.70}$ & $6.69^{+0.83}_{-0.90}$ & $59.8^{+0.6}_{-0.7}$ \\
		$J^{8,A}_{ \mu \nu }$ & $2^{ -- }$ & $-$ & $-$ & $-$ & $-$ & $-$ \\
			$J^{9,A}_{ \mu \nu }$ & $2^{ -+ }$ & $7.24^{+0.29}_{-0.34}$ & $60.05^{+5.21}_{-5.81}$ & $4.77^{+0.65}_{-0.74}$ & $5.92^{+0.75}_{-0.84}$ & $58.8^{+0.5}_{-0.6}$ \\
			$J^{10,A}_{ \mu \nu }$ & $2^{ -+ }$ & $7.19^{+0.24}_{-0.29}$ & $58.36^{+4.30}_{-4.87}$ & $3.61^{+0.41}_{-0.48}$ & $4.65^{+0.51}_{-0.60}$ & $60.9^{+0.2}_{-0.3}$ \\
			$J^{11,A}_{ \mu \nu \rho }$ & $3^{ -- }$ & $6.65^{+0.23}_{-0.16}$ & $49.61^{+3.71}_{-4.16}$ & $3.29^{+0.42}_{-0.49}$ & $4.15^{+0.49}_{-0.55}$ & $59.3^{+0.7}_{-0.4}$\\
		\end{tabular}
	\end{ruledtabular}
\end{table}
\begin{table}[H]
	\setstretch{1.4}
	\caption{Numerical results for the P-wave fully bottom $bb\bar b \bar b$ tetraquark states.} \label{MASSTABLEbbbb}
	\begin{ruledtabular}
		\begin{tabular}{ccccccc}
		Currents & $J^{PC}$& Mass\,[GeV] & $s_0$\,[GeV${}^2$] & $M_{B,\text{min}}^2$\,[GeV${}^2$] & $M_{B,\text{max}}^2$\,[GeV${}^2$] & PC\,[\%] \\
		\colrule 
		$J^{1,A}$ & $0^{ -- }$ & $18.87^{+0.31}_{-0.32}$ & $381.04^{+14.49}_{-14.6}$ & $12.35^{+1.33}_{-1.51}$ & $18.59^{+1.88}_{-2.17}$ & $65.6\pm0.2$ \\
		$J^{3,S}_{ \mu }$ & $1^{ -- }$ & $17.65^{+0.21}_{-0.19}$ & $321.75^{+8.43}_{-7.72}$ & $5.33^{+0.57}_{-0.68}$ & $7.67^{+0.78}_{-0.91}$ & $64.0^{+0.4}_{-0.1}$ \\
		$J^{4,A}_{ \mu }$ & $1^{ -- }$ & $-$ & $-$ & $-$ & $-$ & $-$ \\
		$J^{5,A}_{ \mu }$ & $1^{ -- }$ & $18.88^{+0.31}_{-0.32}$ & $381.57^{+14.52}_{-14.74}$ & $12.26^{+1.29}_{-1.48}$ & $18.29^{+1.87}_{-2.09}$ & $65.7^{+0.1}_{-0.3}$ \\
		$J^{6,A}_{ \mu }$ & $1^{ -+ }$ & $19.39\pm0.37$ & $407.64^{+17.46}_{-17.90}$ & $15.59^{+1.69}_{-1.92}$ & $23.08^{+2.32}_{-2.66}$ & $65.3\pm0.2$ \\
		$J^{7,A}_{ \mu \nu }$ & $2^{ -- }$ & $20.06^{+0.43}_{-0.44}$ & $442.34^{+21.15}_{-22.34}$ & $19.72^{+2.15}_{-2.44}$ & $28.66^{+2.90}_{-3.33}$ & $64.8^{+0.3}_{-0.2}$ \\
		$J^{8,A}_{ \mu \nu }$ & $2^{ -- }$ & $-$ & $-$ & $-$ & $-$ & $-$ \\
		$J^{9,A}_{ \mu \nu }$ & $2^{ -+ }$ & $19.11^{+0.33}_{-0.35}$ & $392.87^{+15.81}_{-16.31}$ & $13.92^{+1.53}_{-1.73}$ & $20.55^{+2.10}_{-2.42}$ & $65.0\pm0.2$ \\
		$J^{10,A}_{ \mu \nu }$ & $2^{ -+ }$ & $18.23^{+0.25}_{-0.24}$ & $346.22^{+10.34}_{-10.17}$ & $6.77^{+0.67}_{-0.79}$ & $9.04^{+0.87}_{-1.04}$ & $62.8\pm0.2$ \\
		$J^{11,A}_{ \mu \nu \rho }$ & $3^{ -- }$ & $17.62^{+0.20}_{-0.18}$ & $320.48^{+8.05}_{-7.46}$ & $5.05^{+0.53}_{-0.62}$ & $7.47^{+0.73}_{-0.86}$ & $65.0\pm0.3$ \\
		\end{tabular}
	\end{ruledtabular}
\end{table}

As shown in Table~\ref{MASSTABLEcccc} and Table~\ref{MASSTABLEbbbb}, there are two $QQ\bar Q\bar Q$ states with $J^{PC}=1^{--}$ from $J^{3,S}_\mu$, $J^{5,A}_\mu$ and two states with $J^{PC}=2^{-+}$ from $J^{9,A}_{\mu \nu}$, $J^{10,A}_{\mu \nu}$. To study the mixing effects between the currents with the same quantum numbers, we have calculated the off-diagonal correlators $\Pi^{3,5}_{\mu \nu}(q^2)$ and $\Pi^{9,10}_{\mu\nu,\rho\sigma}(q^2)$ defined as 
\begin{align}
\Pi^{3,5}_{\mu \nu}(q^2)& =i \int d^4 x e^{i q \cdot x}\bra 0 T[J^{3}_\mu(x) \left(J^{5}_\nu(0)\right)^{\dagger}]\ket 0, \\
\Pi^{9,10}_{\mu\nu,\rho\sigma}(q^2)& =i \int d^4 x e^{i q \cdot x}\bra 0 T[J^{9}_{\mu \nu}(x) \left(J^{10}_{\rho\sigma}(0)\right)^{\dagger}]\ket 0. 
\end{align}
The corresponding spectral functions are collected in the supplementary file. It is shown that the perturbative term in $\Pi^{3,5}_{\mu \nu}(q^2)$ vanishes since the special Lorentz structures of the currents. Comparing to the diagonal correlators $\Pi^{3,3}_{\mu \nu}(q^2)$ and $\Pi^{5,5}_{\mu \nu}(q^2)$, the contribution from this off-diagonal correlator is very small, implying that the currents $J^{3,S}_\mu$ and $J^{5,A}_\mu$ would couple to different hadron states. Actually, $J^{3,S}_\mu$ will couple to the pure ${}^1P_1$ state while $J^{5,A}_\mu$ couples to a mixture of ${}^1P_1$ and ${}^5P_1$ states, according to their different Lorentz structures in Eq.~\eqref{tetraquarkcurrents}. 

For $J^{9,A}_{\mu \nu}$ and $J^{10,A}_{\mu \nu}$ with $J^{PC}=2^{-+}$, their off-diagonal correlator is comparable to the diagonal ones, which will lead to sizable mixing effect between these two currents. Moreover, both two currents couple to the pure ${}^3P_2$ state due to similar Lorentz structures. However, the excitation structures of these two currents are different, in which $J^{10,A}_{\mu \nu}$ contains pure $\lambda$-mode excited component while $J^{9,A}_{\mu \nu}$ contains both $\lambda$-mode and $\rho$-mode excitation structures, as shown in Table~\ref{CURRENT_STRUCTURES}. 
To investigate the pure $\lambda$-mode ${}^3P_2$ fully-heavy tetraquark states, we shall not study the influence of such mixing effects to the hadron masses in our analyses.

In Table~\ref{comparecccc} and Table~\ref{comparebbbb}, we compare our calculations of the P-wave $cc \bar c\bar c$ and $bb \bar b\bar b$ tetraquarks to the results obtained in different theoretical methods. In our work, we only obtain the pure $\lambda$-mode excited tetraquarks in the $1^{--}, 1^{-+}, 2^{-+}$ and $3^{--}$ channels. In these channels, the extracted masses for the $cc \bar c\bar c$ tetraquarks are in agreement with those in Refs.~\cite{liu:2020eha,Faustov:2022mvs,Dong:2022sef} within errors, while the masses for the $bb \bar b\bar b$ tetraquarks are much lighter than the results in  Refs.~\cite{liu:2020eha,Faustov:2022mvs}. For the $0^{-+}$ and $2^{--}$ channels, we obtain the mixed P-wave tetraquark states containing both the $\lambda$-mode and $\rho$-mode excitation structures. Only pure $\rho$-mode excited P-wave tetraquarks are obtained for the $0^{--}$ channel. In Ref.~\cite{Wang:2021kfv}, the $cc \bar c\bar c$ tetraquarks are the color mixtures containing both symmetric and antisymmetric color structures. The states collected in Table~\ref{comparecccc} from Ref.~\cite{Wang:2021kfv} are  dominated by the antisymmetric color structure. 
\begin{table}[H]
	\begin{ruledtabular}
		\setstretch{1.3}
		\caption{Mass spectra (MeV) of the P-wave $cc \bar c\bar c$ tetraquarks in various theoretical works. The notation $(\lambda+\rho)$ represents the mixed P-wave tetraquark state containing both the $\lambda$-mode and $\rho$-mode excitation structures, while $(\lambda)$ and $(\rho)$ indicate the pure $\lambda$-mode and $\rho$-mode excited states respectively. } \label{comparecccc}
		\begin{tabular}{c c c c c c c c c}
			$J^{PC}$ & State  & Color &  This work  & Ref.~\cite{liu:2020eha}  & Ref.~\cite{Wang:2021kfv} & Ref.~\cite{Faustov:2022mvs}  & Ref.~\cite{Dong:2022sef} (GI) \\
			\hline  
			$0^{--} $ & ${}^3P_0$ & $\bar 3_\textbf c \otimes  3_\textbf c$ & $6650^{+220}_{-250}$ ($\rho$) 	& 	6926 ($\rho$) &	 6913($\rho$) & &  \\
			$0^{-+}$  & ${}^3P_0$ & $\bar 3_\textbf c \otimes  3_\textbf c$ &	$9380^{+370}_{-440}$ ($\lambda+\rho$)   &	6891 ($\lambda$)/ 6749 ($\rho$)& 6746($\lambda$) &	6628($\lambda$) & 6633($\lambda$) \\
			$1^{--}$ & ${}^1P_1$ & $\bar 3_\textbf c \otimes  3_\textbf c$ & $6810^{+250}_{-370}$ ($\lambda$) & 6904 ($\lambda$) & 6740($\lambda$) & 6631($\lambda$) & 6698($\lambda$) \\
			$1^{-+}$ & ${}^3P_1$ & $\bar 3_\textbf c \otimes  3_\textbf c$ & $7020^{+260}_{-290}$ ($\lambda$) &6908 ($\lambda$) &6746($\lambda$)   & 6634($\lambda$) & 6697($\lambda$) \\
			$2^{--} $ & ${}^5P_2$ & $\bar 3_\textbf c \otimes  3_\textbf c$ &	$7550^{+320}_{-360}$ ($\lambda+\rho$)	  & 6955 ($\lambda$)  &  6741($\lambda$) &6648($\lambda$) &	6712($\lambda$) \\
			$2^{-+}$ & ${}^3P_2$ & $\bar 3_\textbf c \otimes  3_\textbf c$ &	$7190^{+240}_{-290}$ ($\lambda$)  &6928 ($\lambda$)& 6746($\lambda$)   & 6644($\lambda$) &6718($\lambda$) \\
			$3^{--} $ & ${}^5P_3$ & $\bar 3_\textbf c \otimes  3_\textbf c$ &	$6650^{+230}_{-160}$ ($\lambda$) &  6801 ($\lambda$) &6741($\lambda$)    & 6664($\lambda$) &6739($\lambda$)
		\end{tabular}
	\end{ruledtabular}
\end{table}
\begin{table}[H]
	\begin{ruledtabular}
		\setstretch{1.3}
		\caption{Mass spectra (MeV)  of the P-wave $bb \bar b\bar b$ tetraquarks in various theoretical works.} \label{comparebbbb}
		\begin{tabular}{c c c c c c}
			$J^{PC}$ & State  & Color &  This work & Ref.~\cite{liu:2020eha}  & Ref.~\cite{Faustov:2022mvs}   \\
			\hline  
			$0^{--} $ & ${}^3P_0$ & $\bar 3_\textbf c \otimes  3_\textbf c$ & $18870^{+310}_{-320}$ ($\rho$) & 	19756 ($\rho$) &	  \\
			$0^{-+}$  & ${}^3P_0$ & $\bar 3_\textbf c \otimes  3_\textbf c$  &&	19739 ($\lambda$)/ 19595 ($\rho$)&	19533($\lambda$)   \\
			$1^{--}$ & ${}^1P_1$ & $\bar 3_\textbf c \otimes  3_\textbf c$ & $17650^{+210}_{-190}$ ($\lambda$)  & 19749 ($\lambda$) & 19536($\lambda$)  \\
			$1^{-+}$ & ${}^3P_1$ & $\bar 3_\textbf c \otimes  3_\textbf c$ & $19390\pm370$ ($\lambda$) &19748 ($\lambda$) & 19535($\lambda$) \\
			$2^{--} $ & ${}^5P_2$ & $\bar 3_\textbf c \otimes  3_\textbf c$ &	$20060^{+430}_{-440}$ ($\lambda+\rho$) & 19767 ($\lambda$)  &19538($\lambda$)  \\
			$2^{-+}$ & ${}^3P_2$ & $\bar 3_\textbf c \otimes  3_\textbf c$ &	$18230^{+250}_{-240}$ ($\lambda$) &19756 ($\lambda$)&19539($\lambda$)  \\
			$3^{--} $ & ${}^5P_3$ & $\bar 3_\textbf c \otimes  3_\textbf c$ &	$17620^{+200}_{-180}$ ($\lambda$) & 19617 ($\lambda$) & 19545($\lambda$) 
		\end{tabular}
	\end{ruledtabular}
\end{table}

\section{Conclusion and Discussion \label{DC}}
We have studied the mass spectra of P-wave fully heavy $c c \bar c \bar c$ and $bb\bar b\bar b$ tetraquark states with various quantum numbers in the framework of QCD sum rules. We construct the non-local interpolating tetraquark currents by inserting the covariant derivative operator $\overset{ \leftrightarrow } { \mathcal D }_{ \mu }$ between the S-wave diquark and antidiquark fields. After studying the excitation structures of these tetraquark operators, we find that it is difficult to separate the $\lambda$-mode and $\rho$-mode excitations for the P-wave tetraquaks with $J^{PC}=0^{-+}$. The pure $\lambda$-mode excited P-wave fully heavy tetraquarks exist in the $1^{--}, 1^{-+}, 2^{--}, 2^{-+}$ and $3^{--}$ channels. Within three Lorentz indices, we are not able to construct the pure $\lambda$-mode excited P-wave $c c \bar c \bar c$ and $bb\bar b\bar b$ tetraquark operators with $J^{PC}=0^{--}$ and $3^{-+}$. 

We calculate the two-point correlation functions and spectral functions up to dimension-4 gluon condensate at the leading order of $\alpha_s$. We perform Laplace sum rule analyses to the fully charm and fully bottom tetraquark systems and extract their masses in various channels, as shown in Table~\ref{MASSTABLEcccc} and Table~\ref{MASSTABLEbbbb}. The masses of the fully charm $cc\bar c\bar c$ tetraquark states with $J^{PC}=1^{-+}$ and $2^{-+}$ are about $7.02$ GeV and $7.20$ GeV respectively, which are in good agreement with the masses of $X(6900)$ and $X(7200)$ within errors. For the fully bottom systems, the predicted masses for the $bb\bar b\bar b$ tetraquark states with $J^{PC}=0^{--}, 1^{--}, 2^{-+}$ and $3^{--}$ are below or very close to the mass thresholds of di-$\eta_b(1S)$ and di-$\Upsilon(1S)$. This is different from the fully charm tetraquark systems, which are predicted to be above the di-$\eta_c(1S)$ and di-$J/\psi$ mass thresholds. These results are consistent with our previous predictions for the $\rho$-mode excited P-wave $bb\bar b\bar b$ tetraquarks in Refs.~\cite{Chen:2016jxd,Wang:2021mma}. 

The fully charm $c c \bar c \bar c$ tetraquarks can decay into the two charmonium final states via the fall apart mechanism, as shown in Table~\ref{DECAY_MODES}. We emphasize the di-$J/\psi$ and $J/\psi\psi(2S)$ modes in the bold font where the $X(6400), X(6600), X(6900)$ and $X(7200)$ were observed. It is clearly that the final states of the S-wave decays contain a P-wave charmonium, while the two S-wave charmonium final states can only appear in the P-wave decay modes. Such decay properties may lead to narrower decay widths for these P-wave $c c \bar c \bar c$ tetraquarks than the S-wave ones. More efforts in both theoretical and experimental aspects are expected to understand the fully heavy tetraquark systems in the future. Hopefully our calculations will be useful for identifying the nature of new exotic tetraquark states.

\begin{table}[t]
		\setstretch{1.4}
		\caption{Possible S-wave and P-wave decay modes of the $c c \bar c \bar c$ tetraquark states.} \label{DECAY_MODES}
	\begin{ruledtabular}
		\begin{tabular}{l c c}
				$J^{PC}$ & S-wave & P-wave\\
				\colrule  $0^{--}$ & $J/\psi \chi_{c1}$ & $\eta_c J/\psi$,\,$\eta_c \psi(2S)$\\
				$0^{-+}$ & $\eta_c \chi_{c0}$,\,$ J/\psi h_c$ & \textbf{di-}$\boldsymbol{J/\psi}$,\,$\boldsymbol{J/\psi \psi(2S)}$\\
				$1^{--}$ & \makecell[c]{$\eta_c h_c$,\,$ J/\psi \chi_{c0}$,\,$ J/\psi \chi_{c1}$,\,$ J/\psi \chi_{c2}$ } & \makecell[c]{$\eta_c J/\psi$,\,$\eta_c \psi(2S)$,\,$\eta_c (2S) J/\psi$ }\\
				$1^{-+}$ & \makecell[c]{$ J/\psi h_c$,\,$\eta_c \chi_{c1}$} & \makecell[c]{\textbf{di-}$\boldsymbol{J/\psi}$,\,di-$\eta_c$,\,di-$\chi_{c1}$,\,di-$\chi_{c2}$,\,di-$h_c$,\,$\chi_{c0} \chi_{c1}$,\,$\chi_{c0} \chi_{c2}$ }\\
				$2^{--}$ & \makecell[c]{\;$ J/\psi \chi_{c1}$,\,$ J/\psi \chi_{c2}$,\,$ \psi(2S) \chi_{c1}$,\,$ \psi(2S) \chi_{c2}$ } & \makecell[c]{\; $\eta_c J/\psi$,\,$\eta_c (2S) J/\psi$,\,$\eta_c \psi(2S)$,\,$\eta_c (2S) \psi(2S)$,\,$h_c \chi_{c0}$,\,$h_c \chi_{c1}$,\,$h_c \chi_{c2}$ }\\
				$2^{-+}$ & \makecell[c]{ $\eta_c \chi_{c2}$,\,$\eta_c (2S) \chi_{c2}$,\,$J/\psi h_c$,\,$\psi h_c$ } & \makecell[c]{\textbf{di-}$\boldsymbol{J/\psi}$,\,di-$\chi_{c1}$,\,di-$\chi_{c2}$,\,di-$h_c$,\,$\boldsymbol{J/\psi \psi(2S)}$ }\\
				$3^{--}$ & $ J/\psi \chi_{c2}$ & $-$\\
		\end{tabular}
	\end{ruledtabular}
\end{table}
	\par

\begin{acknowledgments}
Zhi-Zhong Chen thanks Qi-Nan Wang for useful discussions. This work is supported by the National Natural Science Foundation of China under Grant No. 12175318, the Natural Science Foundation of Guangdong Province of China under Grant No. 2022A1515011922, the National Key R$\&$D Program of China under Contracts No. 2020YFA0406400.
\end{acknowledgments}

\appendix
\setstretch{1.0}
\section{Mass curves and Borel stabilities for all channels \label{MCAC}}
In this appendix, we show the mass curves and Borel stabilities of fully charm $cc\bar c\bar c$ tetraquark systems for other interpolating currents in Eq.~\eqref{tetraquarkcurrents}.
\begin{figure}[H] 
	\setstretch{1.5}
	\centering
	\includegraphics[width = .44\linewidth]{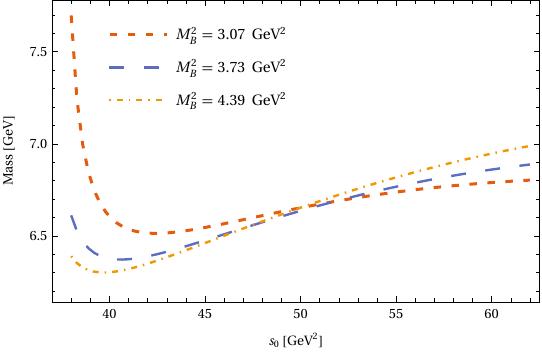}\hspace{.5cm}
	\includegraphics[width = .44\linewidth]{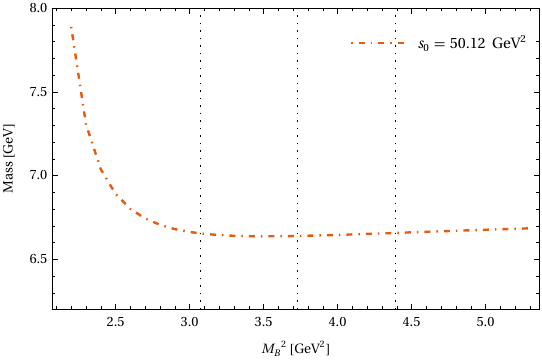}
	\caption{For the interpolating current $J^{1,A}$ with $J^{PC}=0^{--}$.}
\end{figure}
\begin{figure}[H] 
	\setstretch{1.5}
	\centering
	\includegraphics[width = .44\linewidth]{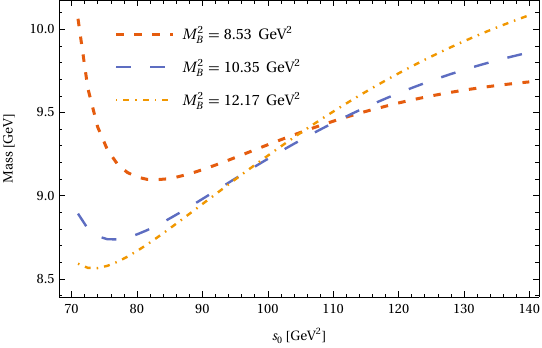}\hspace{.5cm}
	\includegraphics[width = .44\linewidth]{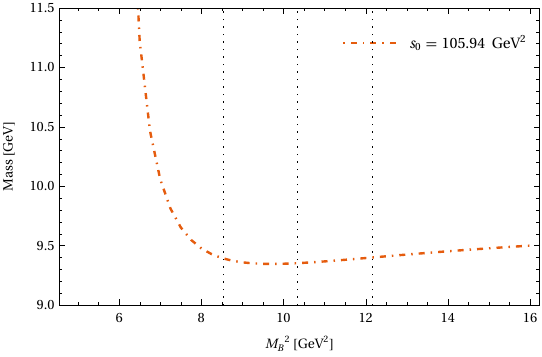}
	\caption{For the interpolating current $J^{2,A}$ with $J^{PC}=0^{-+}$.}
\end{figure}
\begin{figure}[H] 
	\setstretch{1.5}
	\centering
	\includegraphics[width = .44\linewidth]{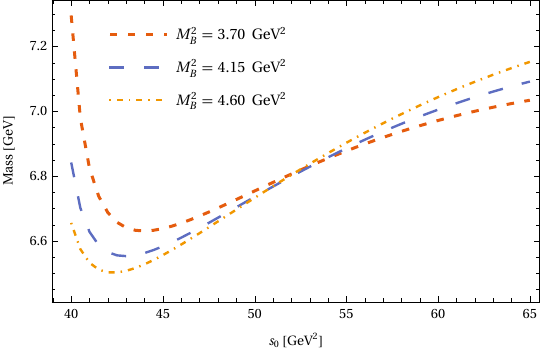}\hspace{.5cm}
	\includegraphics[width = .44\linewidth]{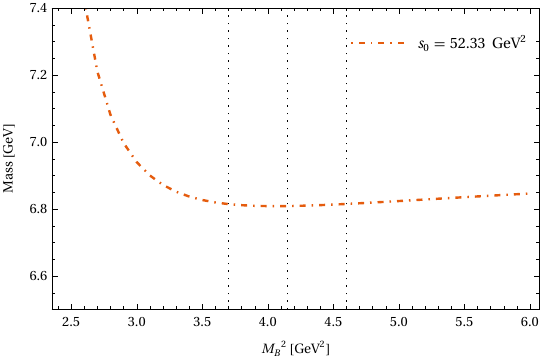}
	\caption{For the interpolating current $J_{\mu}^{3,S}$ with $J^{PC}=1^{--}$.}
\end{figure}
\begin{figure}[H] 
	\setstretch{1.5}
	\centering
	\includegraphics[width = .44\linewidth]{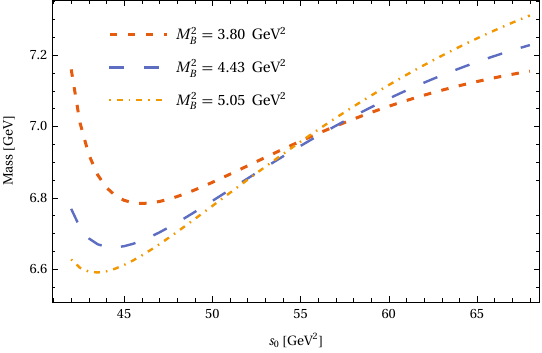}\hspace{.5cm}
	\includegraphics[width = .44\linewidth]{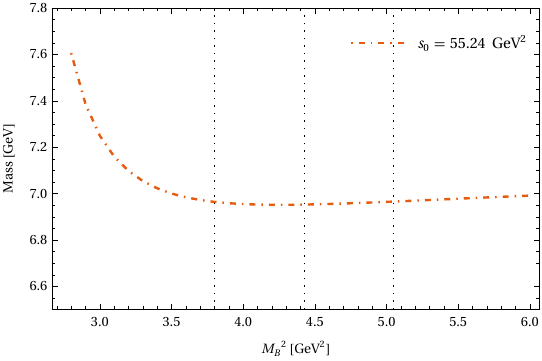}
	\caption{For the interpolating current $J_{\mu}^{5,A}$ with $J^{PC}=1^{--}$.}
\end{figure}
\begin{figure}[H] 
	\setstretch{1.5}
	\centering
	\includegraphics[width = .44\linewidth]{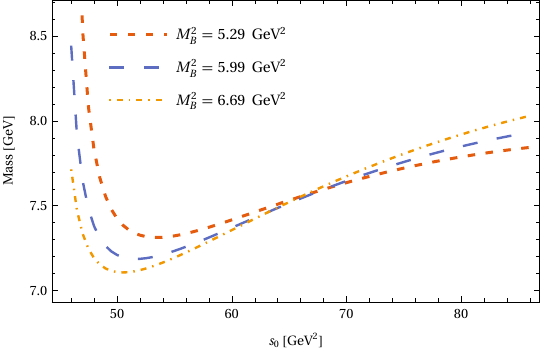}\hspace{.5cm}
	\includegraphics[width = .44\linewidth]{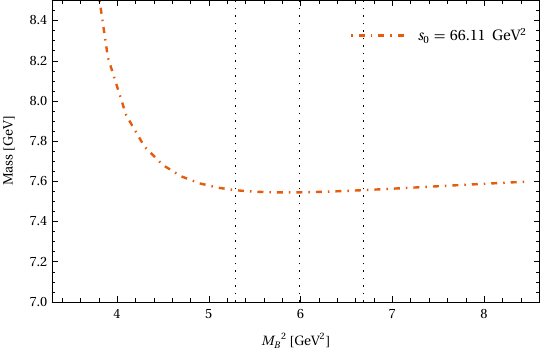}
	\caption{For the interpolating current $J_{\mu\nu}^{7,A}$ with $J^{PC}=2^{--}$.}
\end{figure}
\begin{figure}[H] 
	\setstretch{1.5}
	\centering
	\includegraphics[width = .44\linewidth]{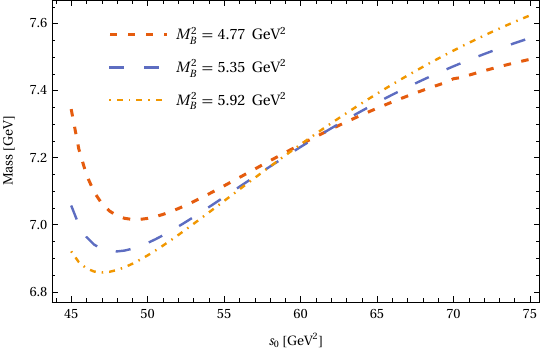}\hspace{.5cm}
	\includegraphics[width = .44\linewidth]{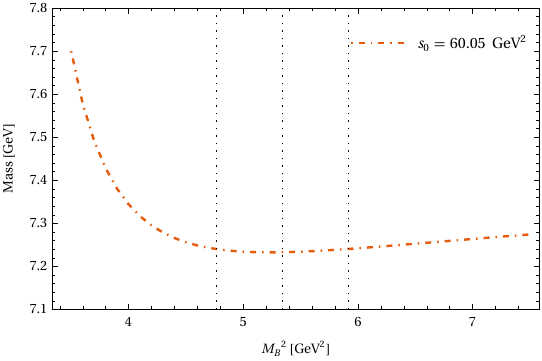}
	\caption{For the interpolating current $J_{\mu\nu}^{9,A}$ with $J^{PC}=2^{-+}$.}
\end{figure}
\begin{figure}[H] 
	\setstretch{1.5}
	\centering
	\includegraphics[width = .44\linewidth]{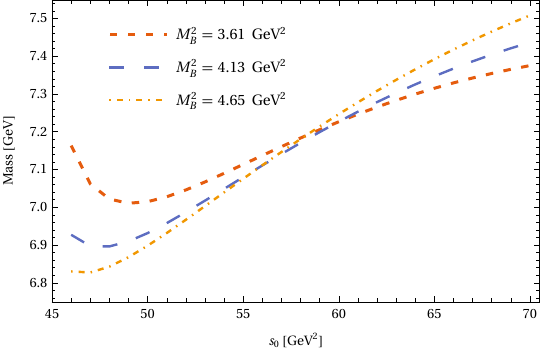}\hspace{.5cm}
	\includegraphics[width = .44\linewidth]{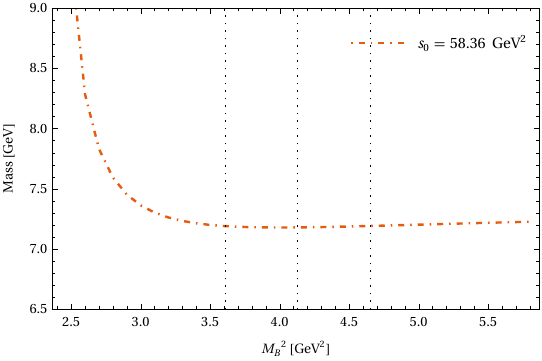}
	\caption{For the interpolating current $J_{\mu\nu}^{10,A}$ with $J^{PC}=2^{-+}$.}
\end{figure}
\begin{figure}[H] 
	\setstretch{1.5}
	\centering
	\includegraphics[width = .44\linewidth]{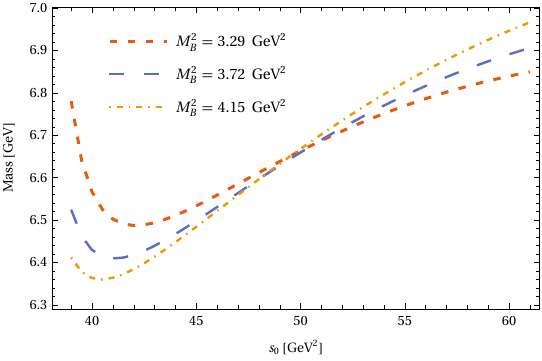}\hspace{.5cm}
	\includegraphics[width = .44\linewidth]{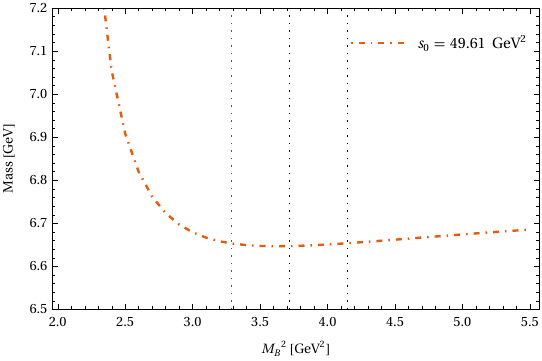}
	\caption{For the interpolating current $J_{\mu\nu\rho}^{11,A}$ with $J^{PC}=3^{--}$.}
\end{figure}

\begin{thebibliography}{150}%
\makeatletter
\providecommand \@ifxundefined [1]{%
 \@ifx{#1\undefined}
}%
\providecommand \@ifnum [1]{%
 \ifnum #1\expandafter \@firstoftwo
 \else \expandafter \@secondoftwo
 \fi
}%
\providecommand \@ifx [1]{%
 \ifx #1\expandafter \@firstoftwo
 \else \expandafter \@secondoftwo
 \fi
}%
\providecommand \natexlab [1]{#1}%
\providecommand \enquote  [1]{``#1''}%
\providecommand \bibnamefont  [1]{#1}%
\providecommand \bibfnamefont [1]{#1}%
\providecommand \citenamefont [1]{#1}%
\providecommand \href@noop [0]{\@secondoftwo}%
\providecommand \href [0]{\begingroup \@sanitize@url \@href}%
\providecommand \@href[1]{\@@startlink{#1}\@@href}%
\providecommand \@@href[1]{\endgroup#1\@@endlink}%
\providecommand \@sanitize@url [0]{\catcode `\\12\catcode `\$12\catcode
  `\&12\catcode `\#12\catcode `\^12\catcode `\_12\catcode `\%12\relax}%
\providecommand \@@startlink[1]{}%
\providecommand \@@endlink[0]{}%
\providecommand \url  [0]{\begingroup\@sanitize@url \@url }%
\providecommand \@url [1]{\endgroup\@href {#1}{\urlprefix }}%
\providecommand \urlprefix  [0]{URL }%
\providecommand \Eprint [0]{\href }%
\providecommand \doibase [0]{https://doi.org/}%
\providecommand \selectlanguage [0]{\@gobble}%
\providecommand \bibinfo  [0]{\@secondoftwo}%
\providecommand \bibfield  [0]{\@secondoftwo}%
\providecommand \translation [1]{[#1]}%
\providecommand \BibitemOpen [0]{}%
\providecommand \bibitemStop [0]{}%
\providecommand \bibitemNoStop [0]{.\EOS\space}%
\providecommand \EOS [0]{\spacefactor3000\relax}%
\providecommand \BibitemShut  [1]{\csname bibitem#1\endcsname}%
\let\auto@bib@innerbib\@empty
\bibitem [{\citenamefont {Chen}\ \emph {et~al.}(2016)\citenamefont {Chen},
  \citenamefont {Chen}, \citenamefont {Liu},\ and\ \citenamefont
  {Zhu}}]{Chen:2016qju}%
  \BibitemOpen
  \bibfield  {author} {\bibinfo {author} {\bibfnamefont {H.-X.}\ \bibnamefont
  {Chen}}, \bibinfo {author} {\bibfnamefont {W.}~\bibnamefont {Chen}}, \bibinfo
  {author} {\bibfnamefont {X.}~\bibnamefont {Liu}},\ and\ \bibinfo {author}
  {\bibfnamefont {S.-L.}\ \bibnamefont {Zhu}},\ }\href
  {https://doi.org/10.1016/j.physrep.2016.05.004} {\bibfield  {journal}
  {\bibinfo  {journal} {Phys. Rep.}\ }\textbf {\bibinfo {volume} {639}},\
  \bibinfo {pages} {1} (\bibinfo {year} {2016})}\BibitemShut {NoStop}%
\bibitem [{\citenamefont {Guo}\ \emph {et~al.}(2018)\citenamefont {Guo},
  \citenamefont {Hanhart}, \citenamefont {Mei{\ss}ner}, \citenamefont {Wang},
  \citenamefont {Zhao},\ and\ \citenamefont {Zou}}]{Guo:2017jvc}%
  \BibitemOpen
  \bibfield  {author} {\bibinfo {author} {\bibfnamefont {F.-K.}\ \bibnamefont
  {Guo}}, \bibinfo {author} {\bibfnamefont {C.}~\bibnamefont {Hanhart}},
  \bibinfo {author} {\bibfnamefont {U.-G.}\ \bibnamefont {Mei{\ss}ner}},
  \bibinfo {author} {\bibfnamefont {Q.}~\bibnamefont {Wang}}, \bibinfo {author}
  {\bibfnamefont {Q.}~\bibnamefont {Zhao}},\ and\ \bibinfo {author}
  {\bibfnamefont {B.-S.}\ \bibnamefont {Zou}},\ }\href
  {https://doi.org/10.1103/RevModPhys.90.015004} {\bibfield  {journal}
  {\bibinfo  {journal} {Rev. Mod. Phys.}\ }\textbf {\bibinfo {volume} {90}},\
  \bibinfo {pages} {015004} (\bibinfo {year} {2018})}  \BibitemShut {NoStop}%
\bibitem [{\citenamefont {Liu}\ \emph {et~al.}(2019{\natexlab{a}})\citenamefont
  {Liu}, \citenamefont {Chen}, \citenamefont {Chen}, \citenamefont {Liu},\ and\
  \citenamefont {Zhu}}]{Liu:2019zoy}%
  \BibitemOpen
  \bibfield  {author} {\bibinfo {author} {\bibfnamefont {Y.-R.}\ \bibnamefont
  {Liu}}, \bibinfo {author} {\bibfnamefont {H.-X.}\ \bibnamefont {Chen}},
  \bibinfo {author} {\bibfnamefont {W.}~\bibnamefont {Chen}}, \bibinfo {author}
  {\bibfnamefont {X.}~\bibnamefont {Liu}},\ and\ \bibinfo {author}
  {\bibfnamefont {S.-L.}\ \bibnamefont {Zhu}},\ }\href
  {https://doi.org/10.1016/j.ppnp.2019.04.003} {\bibfield  {journal} {\bibinfo
  {journal} {Prog. Part. Nucl. Phys.}\ }\textbf {\bibinfo {volume} {107}},\
  \bibinfo {pages} {237} (\bibinfo {year} {2019}{\natexlab{a}})} \BibitemShut
  {NoStop}%
\bibitem [{\citenamefont {Brambilla}\ \emph {et~al.}(2020)\citenamefont
  {Brambilla}, \citenamefont {Eidelman}, \citenamefont {Hanhart}, \citenamefont
  {Nefediev}, \citenamefont {Shen}, \citenamefont {Thomas}, \citenamefont
  {Vairo},\ and\ \citenamefont {Yuan}}]{Brambilla:2019esw}%
  \BibitemOpen
  \bibfield  {author} {\bibinfo {author} {\bibfnamefont {N.}~\bibnamefont
  {Brambilla}}, \bibinfo {author} {\bibfnamefont {S.}~\bibnamefont {Eidelman}},
  \bibinfo {author} {\bibfnamefont {C.}~\bibnamefont {Hanhart}}, \bibinfo
  {author} {\bibfnamefont {A.}~\bibnamefont {Nefediev}}, \bibinfo {author}
  {\bibfnamefont {C.-P.}\ \bibnamefont {Shen}}, \bibinfo {author}
  {\bibfnamefont {C.~E.}\ \bibnamefont {Thomas}}, \bibinfo {author}
  {\bibfnamefont {A.}~\bibnamefont {Vairo}},\ and\ \bibinfo {author}
  {\bibfnamefont {C.-Z.}\ \bibnamefont {Yuan}},\ }\href
  {https://doi.org/10.1016/j.physrep.2020.05.001} {\bibfield  {journal}
  {\bibinfo  {journal} {Phys. Rept.}\ }\textbf {\bibinfo {volume} {873}},\
  \bibinfo {pages} {1} (\bibinfo {year} {2020})},\ \Eprint
  {https://arxiv.org/abs/1907.07583} {arXiv:1907.07583 [hep-ex]} \BibitemShut
  {NoStop}%
\bibitem [{\citenamefont {Chen}\ \emph {et~al.}(2023)\citenamefont {Chen},
  \citenamefont {Chen}, \citenamefont {Liu}, \citenamefont {Liu},\ and\
  \citenamefont {Zhu}}]{Chen:2022asf}%
  \BibitemOpen
  \bibfield  {author} {\bibinfo {author} {\bibfnamefont {H.-X.}\ \bibnamefont
  {Chen}}, \bibinfo {author} {\bibfnamefont {W.}~\bibnamefont {Chen}}, \bibinfo
  {author} {\bibfnamefont {X.}~\bibnamefont {Liu}}, \bibinfo {author}
  {\bibfnamefont {Y.-R.}\ \bibnamefont {Liu}},\ and\ \bibinfo {author}
  {\bibfnamefont {S.-L.}\ \bibnamefont {Zhu}},\ }\href
  {https://doi.org/10.1088/1361-6633/aca3b6} {\bibfield  {journal} {\bibinfo
  {journal} {Rept. Prog. Phys.}\ }\textbf {\bibinfo {volume} {86}},\ \bibinfo
  {pages} {026201} (\bibinfo {year} {2023})},\ \Eprint
  {https://arxiv.org/abs/2204.02649} {arXiv:2204.02649 [hep-ph]} \BibitemShut
  {NoStop}%
\bibitem [{\citenamefont {Khachatryan}\ \emph {et~al.}(2017)\citenamefont
  {Khachatryan} \emph {et~al.}}]{CMS:2016liw}%
  \BibitemOpen
  \bibfield  {author} {\bibinfo {author} {\bibfnamefont {V.}~\bibnamefont
  {Khachatryan}} \emph {et~al.} (\bibinfo {collaboration} {CMS}),\ }\href
  {https://doi.org/10.1007/JHEP05(2017)013} {\bibfield  {journal} {\bibinfo
  {journal} {JHEP}\ }\textbf {\bibinfo {volume} {05}},\ \bibinfo {pages}
  {013}},\ \Eprint {https://arxiv.org/abs/1610.07095} {arXiv:1610.07095
  [hep-ex]} \BibitemShut {NoStop}%
\bibitem [{\citenamefont {Bland}\ \emph {et~al.}(2019)\citenamefont {Bland}
  \emph {et~al.}}]{ANDY:2019bfn}%
  \BibitemOpen
  \bibfield  {author} {\bibinfo {author} {\bibfnamefont {L.~C.}\ \bibnamefont
  {Bland}} \emph {et~al.} (\bibinfo {collaboration} {ANDY}),\ }\href@noop {} \ \Eprint {https://arxiv.org/abs/1909.03124}
  {arXiv:1909.03124 [nucl-ex]} \BibitemShut {NoStop}%
\bibitem [{\citenamefont {Aaij}\ \emph {et~al.}(2018)\citenamefont {Aaij} \emph
  {et~al.}}]{LHCb:2018uwm}%
  \BibitemOpen
  \bibfield  {author} {\bibinfo {author} {\bibfnamefont {R.}~\bibnamefont
  {Aaij}} \emph {et~al.} (\bibinfo {collaboration} {LHCb}),\ }\href
  {https://doi.org/10.1007/JHEP10(2018)086} {\bibfield  {journal} {\bibinfo
  {journal} {JHEP}\ }\textbf {\bibinfo {volume} {10}},\ \bibinfo {pages}
  {086}},\ \Eprint {https://arxiv.org/abs/1806.09707} {arXiv:1806.09707
  [hep-ex]} \BibitemShut {NoStop}%
\bibitem [{\citenamefont {Sirunyan}\ \emph {et~al.}(2020)\citenamefont
  {Sirunyan} \emph {et~al.}}]{CMS:2020qwa}%
  \BibitemOpen
  \bibfield  {author} {\bibinfo {author} {\bibfnamefont {A.~M.}\ \bibnamefont
  {Sirunyan}} \emph {et~al.} (\bibinfo {collaboration} {CMS}),\ }\href
  {https://doi.org/10.1016/j.physletb.2020.135578} {\bibfield  {journal}
  {\bibinfo  {journal} {Phys. Lett. B}\ }\textbf {\bibinfo {volume} {808}},\
  \bibinfo {pages} {135578} (\bibinfo {year} {2020})},\ \Eprint
  {https://arxiv.org/abs/2002.06393} {arXiv:2002.06393 [hep-ex]} \BibitemShut
  {NoStop}%
\bibitem [{\citenamefont {Chen}\ \emph
  {et~al.}(2017{\natexlab{a}})\citenamefont {Chen}, \citenamefont {Chen},
  \citenamefont {Liu}, \citenamefont {Steele},\ and\ \citenamefont
  {Zhu}}]{Chen:2016jxd}%
  \BibitemOpen
  \bibfield  {author} {\bibinfo {author} {\bibfnamefont {W.}~\bibnamefont
  {Chen}}, \bibinfo {author} {\bibfnamefont {H.-X.}\ \bibnamefont {Chen}},
  \bibinfo {author} {\bibfnamefont {X.}~\bibnamefont {Liu}}, \bibinfo {author}
  {\bibfnamefont {T.~G.}\ \bibnamefont {Steele}},\ and\ \bibinfo {author}
  {\bibfnamefont {S.-L.}\ \bibnamefont {Zhu}},\ }\href
  {https://doi.org/10.1016/j.physletb.2017.08.034} {\bibfield  {journal}
  {\bibinfo  {journal} {Phys. Lett. B}\ }\textbf {\bibinfo {volume} {773}},\
  \bibinfo {pages} {247} (\bibinfo {year} {2017}{\natexlab{a}})},\ \Eprint
  {https://arxiv.org/abs/1605.01647} {arXiv:1605.01647} \BibitemShut {NoStop}%
\bibitem [{\citenamefont {Anwar}\ \emph {et~al.}(2018)\citenamefont {Anwar},
  \citenamefont {Ferretti}, \citenamefont {Guo}, \citenamefont {Santopinto},\
  and\ \citenamefont {Zou}}]{Anwar:2017toa}%
  \BibitemOpen
  \bibfield  {author} {\bibinfo {author} {\bibfnamefont {M.~N.}\ \bibnamefont
  {Anwar}}, \bibinfo {author} {\bibfnamefont {J.}~\bibnamefont {Ferretti}},
  \bibinfo {author} {\bibfnamefont {F.-K.}\ \bibnamefont {Guo}}, \bibinfo
  {author} {\bibfnamefont {E.}~\bibnamefont {Santopinto}},\ and\ \bibinfo
  {author} {\bibfnamefont {B.-S.}\ \bibnamefont {Zou}},\ }\href
  {https://doi.org/10.1140/epjc/s10052-018-6073-9} {\bibfield  {journal}
  {\bibinfo  {journal} {Eur. Phys. J. C}\ }\textbf {\bibinfo {volume} {78}},\
  \bibinfo {pages} {647} (\bibinfo {year} {2018})}\BibitemShut {NoStop}%
\bibitem [{\citenamefont {Esposito}\ and\ \citenamefont
  {Polosa}(2018)}]{Esposito:2018cwh}%
  \BibitemOpen
  \bibfield  {author} {\bibinfo {author} {\bibfnamefont {A.}~\bibnamefont
  {Esposito}}\ and\ \bibinfo {author} {\bibfnamefont {A.~D.}\ \bibnamefont
  {Polosa}},\ }\href {https://doi.org/10.1140/epjc/s10052-018-6269-z}
  {\bibfield  {journal} {\bibinfo  {journal} {Eur. Phys. J. C}\ }\textbf
  {\bibinfo {volume} {78}},\ \bibinfo {pages} {782} (\bibinfo {year}
  {2018})}\BibitemShut {NoStop}%
\bibitem [{\citenamefont {Hughes}\ \emph {et~al.}(2018)\citenamefont {Hughes},
  \citenamefont {Eichten},\ and\ \citenamefont {Davies}}]{Hughes:2017xie}%
  \BibitemOpen
  \bibfield  {author} {\bibinfo {author} {\bibfnamefont {C.}~\bibnamefont
  {Hughes}}, \bibinfo {author} {\bibfnamefont {E.}~\bibnamefont {Eichten}},\
  and\ \bibinfo {author} {\bibfnamefont {C.~T.~H.}\ \bibnamefont {Davies}},\
  }\href {https://doi.org/10.1103/PhysRevD.97.054505} {\bibfield  {journal}
  {\bibinfo  {journal} {Phys. Rev. D}\ }\textbf {\bibinfo {volume} {97}},\
  \bibinfo {pages} {054505} (\bibinfo {year} {2018})}\BibitemShut {NoStop}%
\bibitem [{\citenamefont {Karliner}\ \emph {et~al.}(2017)\citenamefont
  {Karliner}, \citenamefont {Nussinov},\ and\ \citenamefont
  {Rosner}}]{Karliner:2016zzc}%
  \BibitemOpen
  \bibfield  {author} {\bibinfo {author} {\bibfnamefont {M.}~\bibnamefont
  {Karliner}}, \bibinfo {author} {\bibfnamefont {S.}~\bibnamefont {Nussinov}},\
  and\ \bibinfo {author} {\bibfnamefont {J.~L.}\ \bibnamefont {Rosner}},\
  }\href {https://doi.org/10.1103/PhysRevD.95.034011} {\bibfield  {journal}
  {\bibinfo  {journal} {Phys. Rev. D}\ }\textbf {\bibinfo {volume} {95}},\
  \bibinfo {pages} {034011} (\bibinfo {year} {2017})}\BibitemShut {NoStop}%
\bibitem [{\citenamefont {Wu}\ \emph {et~al.}(2018)\citenamefont {Wu},
  \citenamefont {Liu}, \citenamefont {Chen}, \citenamefont {Liu},\ and\
  \citenamefont {Zhu}}]{Wu:2016vtq}%
  \BibitemOpen
  \bibfield  {author} {\bibinfo {author} {\bibfnamefont {J.}~\bibnamefont
  {Wu}}, \bibinfo {author} {\bibfnamefont {Y.-R.}\ \bibnamefont {Liu}},
  \bibinfo {author} {\bibfnamefont {K.}~\bibnamefont {Chen}}, \bibinfo {author}
  {\bibfnamefont {X.}~\bibnamefont {Liu}},\ and\ \bibinfo {author}
  {\bibfnamefont {S.-L.}\ \bibnamefont {Zhu}},\ }\href
  {https://doi.org/10.1103/PhysRevD.97.094015} {\bibfield  {journal} {\bibinfo
  {journal} {Phys. Rev. D}\ }\textbf {\bibinfo {volume} {97}},\ \bibinfo
  {pages} {094015} (\bibinfo {year} {2018})}\BibitemShut {NoStop}%
\bibitem [{\citenamefont {Richard}\ \emph {et~al.}(2017)\citenamefont
  {Richard}, \citenamefont {Valcarce},\ and\ \citenamefont
  {Vijande}}]{Richard:2017vry}%
  \BibitemOpen
  \bibfield  {author} {\bibinfo {author} {\bibfnamefont {J.-M.}\ \bibnamefont
  {Richard}}, \bibinfo {author} {\bibfnamefont {A.}~\bibnamefont {Valcarce}},\
  and\ \bibinfo {author} {\bibfnamefont {J.}~\bibnamefont {Vijande}},\ }\href
  {https://doi.org/10.1103/PhysRevD.95.054019} {\bibfield  {journal} {\bibinfo
  {journal} {Phys. Rev. D}\ }\textbf {\bibinfo {volume} {95}},\ \bibinfo
  {pages} {054019} (\bibinfo {year} {2017})}\BibitemShut {NoStop}%
\bibitem [{\citenamefont {Bai}\ \emph {et~al.}(2019)\citenamefont {Bai},
  \citenamefont {Lu},\ and\ \citenamefont {Osborne}}]{Bai:2016int}%
  \BibitemOpen
  \bibfield  {author} {\bibinfo {author} {\bibfnamefont {Y.}~\bibnamefont
  {Bai}}, \bibinfo {author} {\bibfnamefont {S.}~\bibnamefont {Lu}},\ and\
  \bibinfo {author} {\bibfnamefont {J.}~\bibnamefont {Osborne}},\ }\href
  {https://doi.org/10.1016/j.physletb.2019.134930} {\bibfield  {journal}
  {\bibinfo  {journal} {Phys. Lett. B}\ }\textbf {\bibinfo {volume} {798}},\
  \bibinfo {pages} {134930} (\bibinfo {year} {2019})}\BibitemShut {NoStop}%
\bibitem [{\citenamefont {Chen}(2019)}]{Chen:2019dvd}%
  \BibitemOpen
  \bibfield  {author} {\bibinfo {author} {\bibfnamefont {X.-Y.}\ \bibnamefont
  {Chen}},\ }\href {https://doi.org/10.1140/epja/i2019-12807-2} {\bibfield
  {journal} {\bibinfo  {journal} {Eur. Phys. J. A}\ }\textbf {\bibinfo {volume}
  {55}},\ \bibinfo {pages} {106} (\bibinfo {year} {2019})}\BibitemShut
  {NoStop}%
\bibitem [{\citenamefont {Debastiani}\ and\ \citenamefont
  {Navarra}(2019)}]{Debastiani:2017msn}%
  \BibitemOpen
  \bibfield  {author} {\bibinfo {author} {\bibfnamefont {V.~R.}\ \bibnamefont
  {Debastiani}}\ and\ \bibinfo {author} {\bibfnamefont {F.~S.}\ \bibnamefont
  {Navarra}},\ }\href {https://doi.org/10.1088/1674-1137/43/1/013105}
  {\bibfield  {journal} {\bibinfo  {journal} {Chin. Phys. C}\ }\textbf
  {\bibinfo {volume} {43}},\ \bibinfo {pages} {013105} (\bibinfo {year}
  {2019})}\BibitemShut {NoStop}%
\bibitem [{\citenamefont {Chao}(1981)}]{Chao:1980dv}%
  \BibitemOpen
  \bibfield  {author} {\bibinfo {author} {\bibfnamefont {K.-T.}\ \bibnamefont
  {Chao}},\ }\href {https://doi.org/10.1007/BF01431564} {\bibfield  {journal}
  {\bibinfo  {journal} {Z. Phys. C}\ }\textbf {\bibinfo {volume} {7}},\
  \bibinfo {pages} {317} (\bibinfo {year} {1981})}\BibitemShut {NoStop}%
\bibitem [{\citenamefont {Ader}\ \emph {et~al.}(1982)\citenamefont {Ader},
  \citenamefont {Richard},\ and\ \citenamefont {Taxil}}]{Ader:1981db}%
  \BibitemOpen
  \bibfield  {author} {\bibinfo {author} {\bibfnamefont {J.~P.}\ \bibnamefont
  {Ader}}, \bibinfo {author} {\bibfnamefont {J.~M.}\ \bibnamefont {Richard}},\
  and\ \bibinfo {author} {\bibfnamefont {P.}~\bibnamefont {Taxil}},\ }\href
  {https://doi.org/10.1103/PhysRevD.25.2370} {\bibfield  {journal} {\bibinfo
  {journal} {Phys. Rev. D}\ }\textbf {\bibinfo {volume} {25}},\ \bibinfo
  {pages} {2370} (\bibinfo {year} {1982})}\BibitemShut {NoStop}%
\bibitem [{\citenamefont {{Silvestre-Brac}}(1992)}]{Silvestre-Brac:1992kaa}%
  \BibitemOpen
  \bibfield  {author} {\bibinfo {author} {\bibfnamefont {B.}~\bibnamefont
  {{Silvestre-Brac}}},\ }\href {https://doi.org/10.1103/PhysRevD.46.2179}
  {\bibfield  {journal} {\bibinfo  {journal} {Phys. Rev. D}\ }\textbf {\bibinfo
  {volume} {46}},\ \bibinfo {pages} {2179} (\bibinfo {year}
  {1992})}\BibitemShut {NoStop}%
\bibitem [{\citenamefont {Lloyd}\ and\ \citenamefont
  {Vary}(2004)}]{Lloyd:2003yc}%
  \BibitemOpen
  \bibfield  {author} {\bibinfo {author} {\bibfnamefont {R.~J.}\ \bibnamefont
  {Lloyd}}\ and\ \bibinfo {author} {\bibfnamefont {J.~P.}\ \bibnamefont
  {Vary}},\ }\href {https://doi.org/10.1103/PhysRevD.70.014009} {\bibfield
  {journal} {\bibinfo  {journal} {Phys. Rev. D}\ }\textbf {\bibinfo {volume}
  {70}},\ \bibinfo {pages} {014009} (\bibinfo {year} {2004})},\ \Eprint
  {https://arxiv.org/abs/hep-ph/0311179} {arXiv:hep-ph/0311179} \BibitemShut
  {NoStop}%
\bibitem [{\citenamefont {Barnea}\ \emph {et~al.}(2006)\citenamefont {Barnea},
  \citenamefont {Vijande},\ and\ \citenamefont {Valcarce}}]{Barnea:2006sd}%
  \BibitemOpen
  \bibfield  {author} {\bibinfo {author} {\bibfnamefont {N.}~\bibnamefont
  {Barnea}}, \bibinfo {author} {\bibfnamefont {J.}~\bibnamefont {Vijande}},\
  and\ \bibinfo {author} {\bibfnamefont {A.}~\bibnamefont {Valcarce}},\ }\href
  {https://doi.org/10.1103/PhysRevD.73.054004} {\bibfield  {journal} {\bibinfo
  {journal} {Phys. Rev. D}\ }\textbf {\bibinfo {volume} {73}},\ \bibinfo
  {pages} {054004} (\bibinfo {year} {2006})}\BibitemShut {NoStop}%
\bibitem [{\citenamefont {Berezhnoy}\ \emph {et~al.}(2012)\citenamefont
  {Berezhnoy}, \citenamefont {Luchinsky},\ and\ \citenamefont
  {Novoselov}}]{Berezhnoy:2011xn}%
  \BibitemOpen
  \bibfield  {author} {\bibinfo {author} {\bibfnamefont {A.~V.}\ \bibnamefont
  {Berezhnoy}}, \bibinfo {author} {\bibfnamefont {A.~V.}\ \bibnamefont
  {Luchinsky}},\ and\ \bibinfo {author} {\bibfnamefont {A.~A.}\ \bibnamefont
  {Novoselov}},\ }\href {https://doi.org/10.1103/PhysRevD.86.034004} {\bibfield
   {journal} {\bibinfo  {journal} {Phys. Rev. D}\ }\textbf {\bibinfo {volume}
  {86}},\ \bibinfo {pages} {034004} (\bibinfo {year} {2012})},\ \Eprint
  {https://arxiv.org/abs/1111.1867} {arXiv:1111.1867 [hep-ph]} \BibitemShut
  {NoStop}%
\bibitem [{\citenamefont {Aaij}\ \emph {et~al.}(2020)\citenamefont {Aaij} \emph
  {et~al.}}]{LHCb:2020bwg}%
  \BibitemOpen
  \bibfield  {author} {\bibinfo {author} {\bibfnamefont {R.}~\bibnamefont
  {Aaij}} \emph {et~al.} (\bibinfo {collaboration} {LHCb}),\ }\href
  {https://doi.org/10.1016/j.scib.2020.08.032} {\bibfield  {journal} {\bibinfo
  {journal} {Sci. Bull.}\ }\textbf {\bibinfo {volume} {65}},\ \bibinfo {pages}
  {1983} (\bibinfo {year} {2020})},\ \Eprint {https://arxiv.org/abs/2006.16957}
  {arXiv:2006.16957 [hep-ex]} \BibitemShut {NoStop}%
\bibitem [{\citenamefont {Aad}\ \emph {et~al.}(2023)\citenamefont {Aad} \emph
  {et~al.}}]{ATLAS:2023bft}%
  \BibitemOpen
  \bibfield  {author} {\bibinfo {author} {\bibfnamefont {G.}~\bibnamefont
  {Aad}} \emph {et~al.} (\bibinfo {collaboration} {ATLAS}),\ }\href
  {https://doi.org/10.1103/PhysRevLett.131.151902} {\bibfield  {journal}
  {\bibinfo  {journal} {Phys. Rev. Lett.}\ }\textbf {\bibinfo {volume} {131}},\
  \bibinfo {pages} {151902} (\bibinfo {year} {2023})}\BibitemShut {NoStop}%
\bibitem [{\citenamefont {Hayrapetyan}\ \emph {et~al.}(2023)\citenamefont
  {Hayrapetyan} \emph {et~al.}}]{CMS:2023owd}%
  \BibitemOpen
  \bibfield  {author} {\bibinfo {author} {\bibfnamefont {A.}~\bibnamefont
  {Hayrapetyan}} \emph {et~al.} (\bibinfo {collaboration} {CMS}),\ }\href@noop
  {} \ \Eprint {https://arxiv.org/abs/2306.07164}
  {arXiv:2306.07164 [hep-ex]} \BibitemShut {NoStop}%
\bibitem [{\citenamefont {Albuquerque}\ \emph {et~al.}(2020)\citenamefont
  {Albuquerque}, \citenamefont {Narison}, \citenamefont {Rabemananjara},
  \citenamefont {Rabetiarivony},\ and\ \citenamefont
  {Randriamanatrika}}]{Albuquerque:2020hio}%
  \BibitemOpen
  \bibfield  {author} {\bibinfo {author} {\bibfnamefont {R.}~\bibnamefont
  {Albuquerque}}, \bibinfo {author} {\bibfnamefont {S.}~\bibnamefont
  {Narison}}, \bibinfo {author} {\bibfnamefont {A.}~\bibnamefont
  {Rabemananjara}}, \bibinfo {author} {\bibfnamefont {D.}~\bibnamefont
  {Rabetiarivony}},\ and\ \bibinfo {author} {\bibfnamefont {G.}~\bibnamefont
  {Randriamanatrika}},\ }\href {https://doi.org/10.1103/PhysRevD.102.094001}
  {\bibfield  {journal} {\bibinfo  {journal} {Phys. Rev. D}\ }\textbf {\bibinfo
  {volume} {102}},\ \bibinfo {pages} {094001} (\bibinfo {year}
  {2020})}\BibitemShut {NoStop}%
\bibitem [{\citenamefont {Giron}\ and\ \citenamefont
  {Lebed}(2020)}]{Giron:2020wpx}%
  \BibitemOpen
  \bibfield  {author} {\bibinfo {author} {\bibfnamefont {J.~F.}\ \bibnamefont
  {Giron}}\ and\ \bibinfo {author} {\bibfnamefont {R.~F.}\ \bibnamefont
  {Lebed}},\ }\href {https://doi.org/10.1103/PhysRevD.102.074003} {\bibfield
  {journal} {\bibinfo  {journal} {Phys. Rev. D}\ }\textbf {\bibinfo {volume}
  {102}},\ \bibinfo {pages} {074003} (\bibinfo {year} {2020})},\ \Eprint
  {https://arxiv.org/abs/2008.01631} {arXiv:2008.01631} \BibitemShut {NoStop}%
\bibitem [{\citenamefont {Gordillo}\ \emph {et~al.}(2020)\citenamefont
  {Gordillo}, \citenamefont {De~Soto},\ and\ \citenamefont
  {Segovia}}]{Gordillo:2020sgc}%
  \BibitemOpen
  \bibfield  {author} {\bibinfo {author} {\bibfnamefont {M.~C.}\ \bibnamefont
  {Gordillo}}, \bibinfo {author} {\bibfnamefont {F.}~\bibnamefont {De~Soto}},\
  and\ \bibinfo {author} {\bibfnamefont {J.}~\bibnamefont {Segovia}},\ }\href
  {https://doi.org/10.1103/PhysRevD.102.114007} {\bibfield  {journal} {\bibinfo
   {journal} {Phys. Rev. D}\ }\textbf {\bibinfo {volume} {102}},\ \bibinfo
  {pages} {114007} (\bibinfo {year} {2020})},\ \Eprint
  {https://arxiv.org/abs/2009.11889} {arXiv:2009.11889} \BibitemShut {NoStop}%
\bibitem [{\citenamefont {Guo}\ and\ \citenamefont
  {Oller}(2021)}]{Guo:2020pvt}%
  \BibitemOpen
  \bibfield  {author} {\bibinfo {author} {\bibfnamefont {Z.-H.}\ \bibnamefont
  {Guo}}\ and\ \bibinfo {author} {\bibfnamefont {J.~A.}\ \bibnamefont
  {Oller}},\ }\href {https://doi.org/10.1103/PhysRevD.103.034024} {\bibfield
  {journal} {\bibinfo  {journal} {Phys. Rev. D}\ }\textbf {\bibinfo {volume}
  {103}},\ \bibinfo {pages} {034024} (\bibinfo {year} {2021})},\ \Eprint
  {https://arxiv.org/abs/2011.00978} {arXiv:2011.00978} \BibitemShut {NoStop}%
\bibitem [{\citenamefont {Jin}\ \emph {et~al.}(2020)\citenamefont {Jin},
  \citenamefont {Xue}, \citenamefont {Huang},\ and\ \citenamefont
  {Ping}}]{Jin:2020jfc}%
  \BibitemOpen
  \bibfield  {author} {\bibinfo {author} {\bibfnamefont {X.}~\bibnamefont
  {Jin}}, \bibinfo {author} {\bibfnamefont {Y.}~\bibnamefont {Xue}}, \bibinfo
  {author} {\bibfnamefont {H.}~\bibnamefont {Huang}},\ and\ \bibinfo {author}
  {\bibfnamefont {J.}~\bibnamefont {Ping}},\ }\href
  {https://doi.org/10.1140/epjc/s10052-020-08650-z} {\bibfield  {journal}
  {\bibinfo  {journal} {Eur. Phys. J. C}\ }\textbf {\bibinfo {volume} {80}},\
  \bibinfo {pages} {1083} (\bibinfo {year} {2020})},\ \Eprint
  {https://arxiv.org/abs/2006.13745} {arXiv:2006.13745} \BibitemShut {NoStop}%
\bibitem [{\citenamefont {Karliner}\ and\ \citenamefont
  {Rosner}(2020)}]{Karliner:2020dta}%
  \BibitemOpen
  \bibfield  {author} {\bibinfo {author} {\bibfnamefont {M.}~\bibnamefont
  {Karliner}}\ and\ \bibinfo {author} {\bibfnamefont {J.~L.}\ \bibnamefont
  {Rosner}},\ }\href {https://doi.org/10.1103/PhysRevD.102.114039} {\bibfield
  {journal} {\bibinfo  {journal} {Phys. Rev. D}\ }\textbf {\bibinfo {volume}
  {102}},\ \bibinfo {pages} {114039} (\bibinfo {year} {2020})}\BibitemShut
  {NoStop}%
\bibitem [{\citenamefont {Faustov}\ \emph {et~al.}(2020)\citenamefont
  {Faustov}, \citenamefont {Galkin},\ and\ \citenamefont
  {Savchenko}}]{Faustov:2020qfm}%
  \BibitemOpen
  \bibfield  {author} {\bibinfo {author} {\bibfnamefont {R.~N.}\ \bibnamefont
  {Faustov}}, \bibinfo {author} {\bibfnamefont {V.~O.}\ \bibnamefont
  {Galkin}},\ and\ \bibinfo {author} {\bibfnamefont {E.~M.}\ \bibnamefont
  {Savchenko}},\ }\href {https://doi.org/10.1103/PhysRevD.102.114030}
  {\bibfield  {journal} {\bibinfo  {journal} {Phys. Rev. D}\ }\textbf {\bibinfo
  {volume} {102}},\ \bibinfo {pages} {114030} (\bibinfo {year}
  {2020})}\BibitemShut {NoStop}%
\bibitem [{\citenamefont {Ke}\ \emph {et~al.}(2021)\citenamefont {Ke},
  \citenamefont {Han}, \citenamefont {Liu},\ and\ \citenamefont
  {Shi}}]{Ke:2021iyh}%
  \BibitemOpen
  \bibfield  {author} {\bibinfo {author} {\bibfnamefont {H.-W.}\ \bibnamefont
  {Ke}}, \bibinfo {author} {\bibfnamefont {X.}~\bibnamefont {Han}}, \bibinfo
  {author} {\bibfnamefont {X.-H.}\ \bibnamefont {Liu}},\ and\ \bibinfo {author}
  {\bibfnamefont {Y.-L.}\ \bibnamefont {Shi}},\ }\href
  {https://doi.org/10.1140/epjc/s10052-021-09229-y} {\bibfield  {journal}
  {\bibinfo  {journal} {Eur. Phys. J. C}\ }\textbf {\bibinfo {volume} {81}},\
  \bibinfo {pages} {427} (\bibinfo {year} {2021})},\ \Eprint
  {https://arxiv.org/abs/2103.13140} {arXiv:2103.13140} \BibitemShut {NoStop}%
\bibitem [{\citenamefont {Li}\ \emph {et~al.}(2021)\citenamefont {Li},
  \citenamefont {Chang}, \citenamefont {Wang},\ and\ \citenamefont
  {Wang}}]{Li:2021ygk}%
  \BibitemOpen
  \bibfield  {author} {\bibinfo {author} {\bibfnamefont {Q.}~\bibnamefont
  {Li}}, \bibinfo {author} {\bibfnamefont {C.-H.}\ \bibnamefont {Chang}},
  \bibinfo {author} {\bibfnamefont {G.-L.}\ \bibnamefont {Wang}},\ and\
  \bibinfo {author} {\bibfnamefont {T.}~\bibnamefont {Wang}},\ }\href
  {https://doi.org/10.1103/PhysRevD.104.014018} {\bibfield  {journal} {\bibinfo
   {journal} {Phys. Rev. D}\ }\textbf {\bibinfo {volume} {104}},\ \bibinfo
  {pages} {014018} (\bibinfo {year} {2021})}\BibitemShut {NoStop}%
\bibitem [{\citenamefont {Liang}\ \emph {et~al.}(2021)\citenamefont {Liang},
  \citenamefont {Wu},\ and\ \citenamefont {Yao}}]{Liang:2021fzr}%
  \BibitemOpen
  \bibfield  {author} {\bibinfo {author} {\bibfnamefont {Z.-R.}\ \bibnamefont
  {Liang}}, \bibinfo {author} {\bibfnamefont {X.-Y.}\ \bibnamefont {Wu}},\ and\
  \bibinfo {author} {\bibfnamefont {D.-L.}\ \bibnamefont {Yao}},\ }\href
  {https://doi.org/10.1103/PhysRevD.104.034034} {\bibfield  {journal} {\bibinfo
   {journal} {Phys. Rev. D}\ }\textbf {\bibinfo {volume} {104}},\ \bibinfo
  {pages} {034034} (\bibinfo {year} {2021})}\BibitemShut {NoStop}%
\bibitem [{\citenamefont {Liu}\ \emph {et~al.}(2019{\natexlab{b}})\citenamefont
  {Liu}, \citenamefont {L{\"u}}, \citenamefont {Zhong},\ and\ \citenamefont
  {Zhao}}]{Liu:2019zuc}%
  \BibitemOpen
  \bibfield  {author} {\bibinfo {author} {\bibfnamefont {M.-S.}\ \bibnamefont
  {Liu}}, \bibinfo {author} {\bibfnamefont {Q.-F.}\ \bibnamefont {L{\"u}}},
  \bibinfo {author} {\bibfnamefont {X.-H.}\ \bibnamefont {Zhong}},\ and\
  \bibinfo {author} {\bibfnamefont {Q.}~\bibnamefont {Zhao}},\ }\href
  {https://doi.org/10.1103/PhysRevD.100.016006} {\bibfield  {journal} {\bibinfo
   {journal} {Phys. Rev. D}\ }\textbf {\bibinfo {volume} {100}},\ \bibinfo
  {pages} {016006} (\bibinfo {year} {2019}{\natexlab{b}})}\BibitemShut
  {NoStop}%
\bibitem [{\citenamefont {Liu}\ \emph {et~al.}(2020)\citenamefont {Liu},
  \citenamefont {Liu}, \citenamefont {Zhong},\ and\ \citenamefont
  {Zhao}}]{liu:2020eha}%
  \BibitemOpen
  \bibfield  {author} {\bibinfo {author} {\bibfnamefont {M.-S.}\ \bibnamefont
  {Liu}}, \bibinfo {author} {\bibfnamefont {F.-X.}\ \bibnamefont {Liu}},
  \bibinfo {author} {\bibfnamefont {X.-H.}\ \bibnamefont {Zhong}},\ and\
  \bibinfo {author} {\bibfnamefont {Q.}~\bibnamefont {Zhao}},\ }\href@noop {}
  \ \Eprint {https://arxiv.org/abs/2006.11952}
  {arXiv:2006.11952} \BibitemShut {NoStop}%
\bibitem [{\citenamefont {Pal}\ \emph {et~al.}(2021)\citenamefont {Pal},
  \citenamefont {Ghosh}, \citenamefont {Chakrabarti},\ and\ \citenamefont
  {Bhattacharya}}]{Pal:2021gkr}%
  \BibitemOpen
  \bibfield  {author} {\bibinfo {author} {\bibfnamefont {S.}~\bibnamefont
  {Pal}}, \bibinfo {author} {\bibfnamefont {R.}~\bibnamefont {Ghosh}}, \bibinfo
  {author} {\bibfnamefont {B.}~\bibnamefont {Chakrabarti}},\ and\ \bibinfo
  {author} {\bibfnamefont {A.}~\bibnamefont {Bhattacharya}},\ }\href
  {https://doi.org/10.1140/epjp/s13360-021-01302-5} {\bibfield  {journal}
  {\bibinfo  {journal} {Eur. Phys. J. Plus}\ }\textbf {\bibinfo {volume}
  {136}},\ \bibinfo {pages} {625} (\bibinfo {year} {2021})}\BibitemShut
  {NoStop}%
\bibitem [{\citenamefont {Sonnenschein}\ and\ \citenamefont
  {Weissman}(2021)}]{Sonnenschein:2020nwn}%
  \BibitemOpen
  \bibfield  {author} {\bibinfo {author} {\bibfnamefont {J.}~\bibnamefont
  {Sonnenschein}}\ and\ \bibinfo {author} {\bibfnamefont {D.}~\bibnamefont
  {Weissman}},\ }\href {https://doi.org/10.1140/epjc/s10052-020-08818-7}
  {\bibfield  {journal} {\bibinfo  {journal} {Eur. Phys. J. C}\ }\textbf
  {\bibinfo {volume} {81}},\ \bibinfo {pages} {25} (\bibinfo {year} {2021})},\
  \Eprint {https://arxiv.org/abs/2008.01095} {arXiv:2008.01095} \BibitemShut
  {NoStop}%
\bibitem [{\citenamefont {Wan}\ and\ \citenamefont {Qiao}(2021)}]{Wan:2020fsk}%
  \BibitemOpen
  \bibfield  {author} {\bibinfo {author} {\bibfnamefont {B.-D.}\ \bibnamefont
  {Wan}}\ and\ \bibinfo {author} {\bibfnamefont {C.-F.}\ \bibnamefont {Qiao}},\
  }\href {https://doi.org/10.1016/j.physletb.2021.136339} {\bibfield  {journal}
  {\bibinfo  {journal} {Phys. Lett. B}\ }\textbf {\bibinfo {volume} {817}},\
  \bibinfo {pages} {136339} (\bibinfo {year} {2021})},\ \Eprint
  {https://arxiv.org/abs/2012.00454} {arXiv:2012.00454} \BibitemShut {NoStop}%
\bibitem [{\citenamefont {Wang}\ and\ \citenamefont {Di}(2019)}]{Wang:2018poa}%
  \BibitemOpen
  \bibfield  {author} {\bibinfo {author} {\bibfnamefont {Z.-G.}\ \bibnamefont
  {Wang}}\ and\ \bibinfo {author} {\bibfnamefont {Z.-Y.}\ \bibnamefont {Di}},\
  }\href {https://doi.org/10.5506/APhysPolB.50.1335} {\bibfield  {journal}
  {\bibinfo  {journal} {Acta Phys. Polon. B}\ }\textbf {\bibinfo {volume}
  {50}},\ \bibinfo {pages} {1335} (\bibinfo {year} {2019})},\ \Eprint
  {https://arxiv.org/abs/1807.08520} {arXiv:1807.08520} \BibitemShut {NoStop}%
\bibitem [{\citenamefont {Wang}\ \emph {et~al.}(2019)\citenamefont {Wang},
  \citenamefont {Meng},\ and\ \citenamefont {Zhu}}]{Wang:2019rdo}%
  \BibitemOpen
  \bibfield  {author} {\bibinfo {author} {\bibfnamefont {G.-J.}\ \bibnamefont
  {Wang}}, \bibinfo {author} {\bibfnamefont {L.}~\bibnamefont {Meng}},\ and\
  \bibinfo {author} {\bibfnamefont {S.-L.}\ \bibnamefont {Zhu}},\ }\href
  {https://doi.org/10.1103/PhysRevD.100.096013} {\bibfield  {journal} {\bibinfo
   {journal} {Phys. Rev. D}\ }\textbf {\bibinfo {volume} {100}},\ \bibinfo
  {pages} {096013} (\bibinfo {year} {2019})}\BibitemShut {NoStop}%
\bibitem [{\citenamefont {Wang}(2020)}]{Wang:2020ols}%
  \BibitemOpen
  \bibfield  {author} {\bibinfo {author} {\bibfnamefont {Z.-G.}\ \bibnamefont
  {Wang}},\ }\href {https://doi.org/10.1088/1674-1137/abb080} {\bibfield
  {journal} {\bibinfo  {journal} {Chin. Phys. C}\ }\textbf {\bibinfo {volume}
  {44}},\ \bibinfo {pages} {113106} (\bibinfo {year} {2020})},\ \Eprint
  {https://arxiv.org/abs/2006.13028} {arXiv:2006.13028} \BibitemShut {NoStop}%
\bibitem [{\citenamefont {Wang}\ \emph
  {et~al.}(2021{\natexlab{a}})\citenamefont {Wang}, \citenamefont {Meng},
  \citenamefont {Oka},\ and\ \citenamefont {Zhu}}]{Wang:2021kfv}%
  \BibitemOpen
  \bibfield  {author} {\bibinfo {author} {\bibfnamefont {G.-J.}\ \bibnamefont
  {Wang}}, \bibinfo {author} {\bibfnamefont {L.}~\bibnamefont {Meng}}, \bibinfo
  {author} {\bibfnamefont {M.}~\bibnamefont {Oka}},\ and\ \bibinfo {author}
  {\bibfnamefont {S.-L.}\ \bibnamefont {Zhu}},\ }\href
  {https://doi.org/10.1103/PhysRevD.104.036016} {\bibfield  {journal} {\bibinfo
   {journal} {Phys. Rev. D}\ }\textbf {\bibinfo {volume} {104}},\ \bibinfo
  {pages} {036016} (\bibinfo {year} {2021}{\natexlab{a}})},\ \Eprint
  {https://arxiv.org/abs/2105.13109} {arXiv:2105.13109} \BibitemShut {NoStop}%
\bibitem [{\citenamefont {Weng}\ \emph {et~al.}(2021)\citenamefont {Weng},
  \citenamefont {Chen}, \citenamefont {Deng},\ and\ \citenamefont
  {Zhu}}]{Weng:2020jao}%
  \BibitemOpen
  \bibfield  {author} {\bibinfo {author} {\bibfnamefont {X.-Z.}\ \bibnamefont
  {Weng}}, \bibinfo {author} {\bibfnamefont {X.-L.}\ \bibnamefont {Chen}},
  \bibinfo {author} {\bibfnamefont {W.-Z.}\ \bibnamefont {Deng}},\ and\
  \bibinfo {author} {\bibfnamefont {S.-L.}\ \bibnamefont {Zhu}},\ }\href
  {https://doi.org/10.1103/PhysRevD.103.034001} {\bibfield  {journal} {\bibinfo
   {journal} {Phys. Rev. D}\ }\textbf {\bibinfo {volume} {103}},\ \bibinfo
  {pages} {034001} (\bibinfo {year} {2021})},\ \Eprint
  {https://arxiv.org/abs/2010.05163} {arXiv:2010.05163} \BibitemShut {NoStop}%
\bibitem [{\citenamefont {Yang}\ \emph {et~al.}(2020)\citenamefont {Yang},
  \citenamefont {Ping}, \citenamefont {He},\ and\ \citenamefont
  {Wang}}]{Yang:2020rih}%
  \BibitemOpen
  \bibfield  {author} {\bibinfo {author} {\bibfnamefont {G.}~\bibnamefont
  {Yang}}, \bibinfo {author} {\bibfnamefont {J.-L.}\ \bibnamefont {Ping}},
  \bibinfo {author} {\bibfnamefont {L.-Y.}\ \bibnamefont {He}},\ and\ \bibinfo
  {author} {\bibfnamefont {Q.}~\bibnamefont {Wang}},\ }\href@noop {} \ \Eprint {https://arxiv.org/abs/2006.13756} {arXiv:2006.13756}
  \BibitemShut {NoStop}%
\bibitem [{\citenamefont {Yang}\ \emph
  {et~al.}(2021{\natexlab{a}})\citenamefont {Yang}, \citenamefont {Tang},\ and\
  \citenamefont {Qiao}}]{Yang:2020wkh}%
  \BibitemOpen
  \bibfield  {author} {\bibinfo {author} {\bibfnamefont {B.-C.}\ \bibnamefont
  {Yang}}, \bibinfo {author} {\bibfnamefont {L.}~\bibnamefont {Tang}},\ and\
  \bibinfo {author} {\bibfnamefont {C.-F.}\ \bibnamefont {Qiao}},\ }\href
  {https://doi.org/10.1140/epjc/s10052-021-09096-7} {\bibfield  {journal}
  {\bibinfo  {journal} {Eur. Phys. J. C}\ }\textbf {\bibinfo {volume} {81}},\
  \bibinfo {pages} {324} (\bibinfo {year} {2021}{\natexlab{a}})},\ \Eprint
  {https://arxiv.org/abs/2012.04463} {arXiv:2012.04463} \BibitemShut {NoStop}%
\bibitem [{\citenamefont {Zhang}(2021)}]{Zhang:2020xtb}%
  \BibitemOpen
  \bibfield  {author} {\bibinfo {author} {\bibfnamefont {J.-R.}\ \bibnamefont
  {Zhang}},\ }\href {https://doi.org/10.1103/PhysRevD.103.014018} {\bibfield
  {journal} {\bibinfo  {journal} {Phys. Rev. D}\ }\textbf {\bibinfo {volume}
  {103}},\ \bibinfo {pages} {014018} (\bibinfo {year} {2021})},\ \Eprint
  {https://arxiv.org/abs/2010.07719} {arXiv:2010.07719} \BibitemShut {NoStop}%
\bibitem [{\citenamefont {Zhao}\ \emph
  {et~al.}(2021{\natexlab{a}})\citenamefont {Zhao}, \citenamefont {Xu},
  \citenamefont {Kaewsnod}, \citenamefont {Liu}, \citenamefont {Limphirat},\
  and\ \citenamefont {Yan}}]{Zhao:2020zjh}%
  \BibitemOpen
  \bibfield  {author} {\bibinfo {author} {\bibfnamefont {Z.}~\bibnamefont
  {Zhao}}, \bibinfo {author} {\bibfnamefont {K.}~\bibnamefont {Xu}}, \bibinfo
  {author} {\bibfnamefont {A.}~\bibnamefont {Kaewsnod}}, \bibinfo {author}
  {\bibfnamefont {X.}~\bibnamefont {Liu}}, \bibinfo {author} {\bibfnamefont
  {A.}~\bibnamefont {Limphirat}},\ and\ \bibinfo {author} {\bibfnamefont
  {Y.}~\bibnamefont {Yan}},\ }\href
  {https://doi.org/10.1103/PhysRevD.103.116027} {\bibfield  {journal} {\bibinfo
   {journal} {Phys. Rev. D}\ }\textbf {\bibinfo {volume} {103}},\ \bibinfo
  {pages} {116027} (\bibinfo {year} {2021}{\natexlab{a}})}\BibitemShut
  {NoStop}%
\bibitem [{\citenamefont {Zhao}\ \emph {et~al.}(2020)\citenamefont {Zhao},
  \citenamefont {Shi},\ and\ \citenamefont {Zhuang}}]{Zhao:2020nwy}%
  \BibitemOpen
  \bibfield  {author} {\bibinfo {author} {\bibfnamefont {J.}~\bibnamefont
  {Zhao}}, \bibinfo {author} {\bibfnamefont {S.}~\bibnamefont {Shi}},\ and\
  \bibinfo {author} {\bibfnamefont {P.}~\bibnamefont {Zhuang}},\ }\href
  {https://doi.org/10.1103/PhysRevD.102.114001} {\bibfield  {journal} {\bibinfo
   {journal} {Phys. Rev. D}\ }\textbf {\bibinfo {volume} {102}},\ \bibinfo
  {pages} {114001} (\bibinfo {year} {2020})}\BibitemShut {NoStop}%
\bibitem [{\citenamefont {Zhu}(2021)}]{Zhu:2020xni}%
  \BibitemOpen
  \bibfield  {author} {\bibinfo {author} {\bibfnamefont {R.}~\bibnamefont
  {Zhu}},\ }\href {https://doi.org/10.1016/j.nuclphysb.2021.115393} {\bibfield
  {journal} {\bibinfo  {journal} {Nucl. Phys. B}\ }\textbf {\bibinfo {volume}
  {966}},\ \bibinfo {pages} {115393} (\bibinfo {year} {2021})},\ \Eprint
  {https://arxiv.org/abs/2010.09082} {arXiv:2010.09082} \BibitemShut {NoStop}%
\bibitem [{\citenamefont {Cao}\ \emph {et~al.}(2021)\citenamefont {Cao},
  \citenamefont {Chen}, \citenamefont {Qi},\ and\ \citenamefont
  {Zheng}}]{Cao:2020gul}%
  \BibitemOpen
  \bibfield  {author} {\bibinfo {author} {\bibfnamefont {Q.-F.}\ \bibnamefont
  {Cao}}, \bibinfo {author} {\bibfnamefont {H.}~\bibnamefont {Chen}}, \bibinfo
  {author} {\bibfnamefont {H.-R.}\ \bibnamefont {Qi}},\ and\ \bibinfo {author}
  {\bibfnamefont {H.-Q.}\ \bibnamefont {Zheng}},\ }\href
  {https://doi.org/10.1088/1674-1137/ac0ee5} {\bibfield  {journal} {\bibinfo
  {journal} {Chin. Phys. C}\ }\textbf {\bibinfo {volume} {45}},\ \bibinfo
  {pages} {103102} (\bibinfo {year} {2021})},\ \Eprint
  {https://arxiv.org/abs/2011.04347} {arXiv:2011.04347 [hep-ph]} \BibitemShut
  {NoStop}%
\bibitem [{\citenamefont {Lesteiro-Tejeda}\ \emph {et~al.}(2021)\citenamefont
  {Lesteiro-Tejeda} \emph {et~al.}}]{Lesteiro-Tejeda:2021wne}%
  \BibitemOpen
  \bibfield  {author} {\bibinfo {author} {\bibfnamefont {J.~A.}\ \bibnamefont
  {Lesteiro-Tejeda}} \emph {et~al.},\ }\href@noop {} \ \Eprint {https://arxiv.org/abs/2101.03192} {arXiv:2101.03192
  [hep-ph]} \BibitemShut {NoStop}%
\bibitem [{\citenamefont {Albuquerque}\ \emph {et~al.}(2021)\citenamefont
  {Albuquerque}, \citenamefont {Narison}, \citenamefont {Rabemananjara},
  \citenamefont {Rabetiarivony},\ and\ \citenamefont
  {Randriamanatrika}}]{Albuquerque:2021erv}%
  \BibitemOpen
  \bibfield  {author} {\bibinfo {author} {\bibfnamefont {R.~M.}\ \bibnamefont
  {Albuquerque}}, \bibinfo {author} {\bibfnamefont {S.}~\bibnamefont
  {Narison}}, \bibinfo {author} {\bibfnamefont {A.}~\bibnamefont
  {Rabemananjara}}, \bibinfo {author} {\bibfnamefont {D.}~\bibnamefont
  {Rabetiarivony}},\ and\ \bibinfo {author} {\bibfnamefont {G.}~\bibnamefont
  {Randriamanatrika}},\ }\href
  {https://doi.org/10.1016/j.nuclphysbps.2021.05.031} {\bibfield  {journal}
  {\bibinfo  {journal} {Nucl. Part. Phys. Proc.}\ }\textbf {\bibinfo {volume}
  {312-317}},\ \bibinfo {pages} {120} (\bibinfo {year} {2021})},\ \Eprint
  {https://arxiv.org/abs/2102.08776} {arXiv:2102.08776 [hep-ph]} \BibitemShut
  {NoStop}%
\bibitem [{\citenamefont {Wang}\ \emph
  {et~al.}(2021{\natexlab{b}})\citenamefont {Wang}, \citenamefont {Yang},\ and\
  \citenamefont {Chen}}]{Wang:2021mma}%
  \BibitemOpen
  \bibfield  {author} {\bibinfo {author} {\bibfnamefont {Q.-N.}\ \bibnamefont
  {Wang}}, \bibinfo {author} {\bibfnamefont {Z.-Y.}\ \bibnamefont {Yang}},\
  and\ \bibinfo {author} {\bibfnamefont {W.}~\bibnamefont {Chen}},\ }\href
  {https://doi.org/10.1103/PhysRevD.104.114037} {\bibfield  {journal} {\bibinfo
   {journal} {Phys. Rev. D}\ }\textbf {\bibinfo {volume} {104}},\ \bibinfo
  {pages} {114037} (\bibinfo {year} {2021}{\natexlab{b}})}\BibitemShut
  {NoStop}%
\bibitem [{\citenamefont {Wang}\ \emph
  {et~al.}(2021{\natexlab{c}})\citenamefont {Wang}, \citenamefont {Yang},
  \citenamefont {Chen},\ and\ \citenamefont {Chen}}]{Wang:2021taf}%
  \BibitemOpen
  \bibfield  {author} {\bibinfo {author} {\bibfnamefont {Q.-N.}\ \bibnamefont
  {Wang}}, \bibinfo {author} {\bibfnamefont {Z.-Y.}\ \bibnamefont {Yang}},
  \bibinfo {author} {\bibfnamefont {W.}~\bibnamefont {Chen}},\ and\ \bibinfo
  {author} {\bibfnamefont {H.-X.}\ \bibnamefont {Chen}},\ }\href
  {https://doi.org/10.1103/PhysRevD.104.014020} {\bibfield  {journal} {\bibinfo
   {journal} {Phys. Rev. D}\ }\textbf {\bibinfo {volume} {104}},\ \bibinfo
  {pages} {014020} (\bibinfo {year} {2021}{\natexlab{c}})},\ \Eprint
  {https://arxiv.org/abs/2106.05550} {arXiv:2106.05550 [hep-ph]} \BibitemShut
  {NoStop}%
\bibitem [{\citenamefont {Yang}\ \emph
  {et~al.}(2021{\natexlab{b}})\citenamefont {Yang}, \citenamefont {Wang},
  \citenamefont {Chen},\ and\ \citenamefont {Chen}}]{Yang:2021zrc}%
  \BibitemOpen
  \bibfield  {author} {\bibinfo {author} {\bibfnamefont {Z.-Y.}\ \bibnamefont
  {Yang}}, \bibinfo {author} {\bibfnamefont {Q.-N.}\ \bibnamefont {Wang}},
  \bibinfo {author} {\bibfnamefont {W.}~\bibnamefont {Chen}},\ and\ \bibinfo
  {author} {\bibfnamefont {H.-X.}\ \bibnamefont {Chen}},\ }\href
  {https://doi.org/10.1103/PhysRevD.104.014003} {\bibfield  {journal} {\bibinfo
   {journal} {Phys. Rev. D}\ }\textbf {\bibinfo {volume} {104}},\ \bibinfo
  {pages} {014003} (\bibinfo {year} {2021}{\natexlab{b}})},\ \Eprint
  {https://arxiv.org/abs/2102.10605} {arXiv:2102.10605 [hep-ph]} \BibitemShut
  {NoStop}%
\bibitem [{\citenamefont {An}\ \emph {et~al.}(2023)\citenamefont {An},
  \citenamefont {Luo}, \citenamefont {Liu},\ and\ \citenamefont
  {Liu}}]{An:2022qpt}%
  \BibitemOpen
  \bibfield  {author} {\bibinfo {author} {\bibfnamefont {H.-T.}\ \bibnamefont
  {An}}, \bibinfo {author} {\bibfnamefont {S.-Q.}\ \bibnamefont {Luo}},
  \bibinfo {author} {\bibfnamefont {Z.-W.}\ \bibnamefont {Liu}},\ and\ \bibinfo
  {author} {\bibfnamefont {X.}~\bibnamefont {Liu}},\ }\href
  {http://dx.doi.org/10.1140/epjc/s10052-023-11847-7} {\bibfield  {journal}
  {\bibinfo  {journal} {Eur. Phys. J. C}\ }\textbf {\bibinfo {volume} {83}}
  (\bibinfo {year} {2023})}\BibitemShut {NoStop}%
\bibitem [{\citenamefont {Wu}\ \emph {et~al.}(2022)\citenamefont {Wu},
  \citenamefont {Zuo}, \citenamefont {Wang}, \citenamefont {Meng},
  \citenamefont {Ma},\ and\ \citenamefont {Chao}}]{Wu:2022qwd}%
  \BibitemOpen
  \bibfield  {author} {\bibinfo {author} {\bibfnamefont {R.-H.}\ \bibnamefont
  {Wu}}, \bibinfo {author} {\bibfnamefont {Y.-S.}\ \bibnamefont {Zuo}},
  \bibinfo {author} {\bibfnamefont {C.-Y.}\ \bibnamefont {Wang}}, \bibinfo
  {author} {\bibfnamefont {C.}~\bibnamefont {Meng}}, \bibinfo {author}
  {\bibfnamefont {Y.-Q.}\ \bibnamefont {Ma}},\ and\ \bibinfo {author}
  {\bibfnamefont {K.-T.}\ \bibnamefont {Chao}},\ }\href
  {https://doi.org/10.1007/jhep11(2022)023} {\bibfield  {journal} {\bibinfo
  {journal} {JHEP}\ }\textbf {\bibinfo {volume} {2022}}\bibinfo  {number} {
  (11)}}\BibitemShut {NoStop}%
\bibitem [{\citenamefont {Kuang}\ \emph {et~al.}(2022)\citenamefont {Kuang},
  \citenamefont {Serafin}, \citenamefont {Zhao},\ and\ \citenamefont
  {Vary}}]{Kuang:2022vdy}%
  \BibitemOpen
\bibfield  {number} {  }\bibfield  {author} {\bibinfo {author} {\bibfnamefont
  {Z.}~\bibnamefont {Kuang}}, \bibinfo {author} {\bibfnamefont
  {K.}~\bibnamefont {Serafin}}, \bibinfo {author} {\bibfnamefont
  {X.}~\bibnamefont {Zhao}},\ and\ \bibinfo {author} {\bibfnamefont {J.~P.}\
  \bibnamefont {Vary}} (\bibinfo {collaboration} {BLFQ Collaboration}),\ }\href
  {https://doi.org/10.1103/PhysRevD.105.094028} {\bibfield  {journal} {\bibinfo
   {journal} {Phys. Rev. D}\ }\textbf {\bibinfo {volume} {105}},\ \bibinfo
  {pages} {094028} (\bibinfo {year} {2022})}\BibitemShut {NoStop}%
\bibitem{Faustov:2022mvs}
R.~N.~Faustov, V.~O.~Galkin and E.~M.~Savchenko,
Symmetry \textbf{14}, no.12, 2504 (2022)
doi:10.3390/sym14122504
\bibitem{Dong:2022sef}
W.~C.~Dong and Z.~G.~Wang,
Phys. Rev. D \textbf{107}, no.7, 074010 (2023)
\bibitem [{\citenamefont {Lu}\ \emph {et~al.}(2023)\citenamefont {Lu},
  \citenamefont {Chen}, \citenamefont {Kang}, \citenamefont {Qin},\ and\
  \citenamefont {Zheng}}]{Lu:2023ccs}%
  \BibitemOpen
  \bibfield  {author} {\bibinfo {author} {\bibfnamefont {Y.}~\bibnamefont
  {Lu}}, \bibinfo {author} {\bibfnamefont {C.}~\bibnamefont {Chen}}, \bibinfo
  {author} {\bibfnamefont {K.-G.}\ \bibnamefont {Kang}}, \bibinfo {author}
  {\bibfnamefont {G.-Y.}\ \bibnamefont {Qin}},\ and\ \bibinfo {author}
  {\bibfnamefont {H.-Q.}\ \bibnamefont {Zheng}},\ }\href
  {https://doi.org/10.1103/PhysRevD.107.094006} {\bibfield  {journal} {\bibinfo
   {journal} {Phys. Rev. D}\ }\textbf {\bibinfo {volume} {107}},\ \bibinfo
  {pages} {094006} (\bibinfo {year} {2023})}\BibitemShut {NoStop}%
\bibitem [{\citenamefont {{Feng, Feng and Huang, Yingsheng and Jia, Yu and
  Sang, Wen-Long and Yang, De-Shan and Zhang, Jia-Yue}}(2023)}]{Feng:2023agq}%
  \BibitemOpen
  \bibfield  {author} {\bibinfo {author} {\bibnamefont {{Feng, Feng and Huang,
  Yingsheng and Jia, Yu and Sang, Wen-Long and Yang, De-Shan and Zhang,
  Jia-Yue}}},\ }\href {https://doi.org/10.1103/PhysRevD.108.L051501} {\bibfield
   {journal} {\bibinfo  {journal} {Phys. Rev. D}\ }\textbf {\bibinfo {volume}
  {108}},\ \bibinfo {pages} {L051501} (\bibinfo {year} {2023})},\ \Eprint
  {https://arxiv.org/abs/2304.11142} {arXiv:2304.11142 [hep-ph]} \BibitemShut
  {NoStop}%
\bibitem [{\citenamefont {Wang}\ \emph {et~al.}(2023)\citenamefont {Wang},
  \citenamefont {Oka},\ and\ \citenamefont {Jido}}]{Wang:2023igz}%
  \BibitemOpen
  \bibfield  {author} {\bibinfo {author} {\bibfnamefont {G.-J.}\ \bibnamefont
  {Wang}}, \bibinfo {author} {\bibfnamefont {M.}~\bibnamefont {Oka}},\ and\
  \bibinfo {author} {\bibfnamefont {D.}~\bibnamefont {Jido}},\ }\href@noop {}
  \ \Eprint {https://arxiv.org/abs/2307.04310}
  {arXiv:2307.04310 [hep-ph]} \BibitemShut {NoStop}%
\bibitem [{\citenamefont {Mutuk}(2021)}]{Mutuk:2021hmi}%
  \BibitemOpen
  \bibfield  {author} {\bibinfo {author} {\bibfnamefont {H.}~\bibnamefont
  {Mutuk}},\ }\href {https://doi.org/10.1140/epjc/s10052-021-09176-8}
  {\bibfield  {journal} {\bibinfo  {journal} {Eur. Phys. J. C}\ }\textbf
  {\bibinfo {volume} {81}},\ \bibinfo {pages} {367} (\bibinfo {year} {2021})},\
  \Eprint {https://arxiv.org/abs/2104.11823} {arXiv:2104.11823 [hep-ph]}
  \BibitemShut {NoStop}%
\bibitem [{\citenamefont {Yang}\ \emph
  {et~al.}(2021{\natexlab{c}})\citenamefont {Yang}, \citenamefont {Ping},\ and\
  \citenamefont {Segovia}}]{Yang:2021hrb}%
  \BibitemOpen
  \bibfield  {author} {\bibinfo {author} {\bibfnamefont {G.}~\bibnamefont
  {Yang}}, \bibinfo {author} {\bibfnamefont {J.}~\bibnamefont {Ping}},\ and\
  \bibinfo {author} {\bibfnamefont {J.}~\bibnamefont {Segovia}},\ }\href
  {https://doi.org/10.1103/PhysRevD.104.014006} {\bibfield  {journal} {\bibinfo
   {journal} {Phys. Rev. D}\ }\textbf {\bibinfo {volume} {104}},\ \bibinfo
  {pages} {014006} (\bibinfo {year} {2021}{\natexlab{c}})},\ \Eprint
  {https://arxiv.org/abs/2104.08814} {arXiv:2104.08814 [hep-ph]} \BibitemShut
  {NoStop}%
\bibitem [{\citenamefont {Galkin}\ and\ \citenamefont
  {Savchenko}(2023)}]{Galkin:2023wox}%
  \BibitemOpen
  \bibfield  {author} {\bibinfo {author} {\bibfnamefont {V.~O.}\ \bibnamefont
  {Galkin}}\ and\ \bibinfo {author} {\bibfnamefont {E.~M.}\ \bibnamefont
  {Savchenko}},\ }\href@noop {} \ \Eprint
  {https://arxiv.org/abs/2310.20247} {arXiv:2310.20247 [hep-ph]} \BibitemShut
  {NoStop}%
\bibitem [{\citenamefont {Anwar}\ and\ \citenamefont
  {Burns}(2023)}]{Anwar:2023fbp}%
  \BibitemOpen
  \bibfield  {author} {\bibinfo {author} {\bibfnamefont {M.~N.}\ \bibnamefont
  {Anwar}}\ and\ \bibinfo {author} {\bibfnamefont {T.~J.}\ \bibnamefont
  {Burns}},\ }\href@noop {} \ \Eprint
  {https://arxiv.org/abs/2311.15853} {arXiv:2311.15853 [hep-ph]} \BibitemShut
  {NoStop}%
\bibitem [{\citenamefont {Wu}\ \emph {et~al.}(2024)\citenamefont {Wu},
  \citenamefont {Chen}, \citenamefont {Meng},\ and\ \citenamefont
  {Zhu}}]{Wu:2024euj}%
  \BibitemOpen
  \bibfield  {author} {\bibinfo {author} {\bibfnamefont {W.-L.}\ \bibnamefont
  {Wu}}, \bibinfo {author} {\bibfnamefont {Y.-K.}\ \bibnamefont {Chen}},
  \bibinfo {author} {\bibfnamefont {L.}~\bibnamefont {Meng}},\ and\ \bibinfo
  {author} {\bibfnamefont {S.-L.}\ \bibnamefont {Zhu}},\ }\href@noop {}\ \Eprint {https://arxiv.org/abs/2401.14899}
  {arXiv:2401.14899 [hep-ph]} \BibitemShut {NoStop}%
\bibitem [{\citenamefont {Agaev}\ \emph
  {et~al.}(2023{\natexlab{a}})\citenamefont {Agaev}, \citenamefont {Azizi},
  \citenamefont {Barsbay},\ and\ \citenamefont {Sundu}}]{Agaev:2023tzi}%
  \BibitemOpen
  \bibfield  {author} {\bibinfo {author} {\bibfnamefont {S.~S.}\ \bibnamefont
  {Agaev}}, \bibinfo {author} {\bibfnamefont {K.}~\bibnamefont {Azizi}},
  \bibinfo {author} {\bibfnamefont {B.}~\bibnamefont {Barsbay}},\ and\ \bibinfo
  {author} {\bibfnamefont {H.}~\bibnamefont {Sundu}},\ }\href@noop {} \ \Eprint
  {https://arxiv.org/abs/2311.10534} {arXiv:2311.10534 [hep-ph]} \BibitemShut
  {NoStop}%
\bibitem [{\citenamefont {Agaev}\ \emph {et~al.}(2024)\citenamefont {Agaev},
  \citenamefont {Azizi},\ and\ \citenamefont {Sundu}}]{Agaev:2024pej}%
  \BibitemOpen
  \bibfield  {author} {\bibinfo {author} {\bibfnamefont {S.~S.}\ \bibnamefont
  {Agaev}}, \bibinfo {author} {\bibfnamefont {K.}~\bibnamefont {Azizi}},\ and\
  \bibinfo {author} {\bibfnamefont {H.}~\bibnamefont {Sundu}},\ }\href@noop {}
  \ \Eprint {https://arxiv.org/abs/2401.05162}
  {arXiv:2401.05162 [hep-ph]} \BibitemShut {NoStop}%
\bibitem [{\citenamefont {L{\"u}}\ \emph {et~al.}(2020)\citenamefont {L{\"u}},
  \citenamefont {Chen},\ and\ \citenamefont {Dong}}]{Lu:2020cns}%
  \BibitemOpen
  \bibfield  {author} {\bibinfo {author} {\bibfnamefont {Q.-F.}\ \bibnamefont
  {L{\"u}}}, \bibinfo {author} {\bibfnamefont {D.-Y.}\ \bibnamefont {Chen}},\
  and\ \bibinfo {author} {\bibfnamefont {Y.-B.}\ \bibnamefont {Dong}},\ }\href
  {https://doi.org/10.1140/epjc/s10052-020-08454-1} {\bibfield  {journal}
  {\bibinfo  {journal} {Eur. Phys. J. C}\ }\textbf {\bibinfo {volume} {80}},\
  \bibinfo {pages} {871} (\bibinfo {year} {2020})}\BibitemShut {NoStop}%
\bibitem [{\citenamefont {Li}\ \emph {et~al.}(2019)\citenamefont {Li},
  \citenamefont {Wang},\ and\ \citenamefont {Xing}}]{Li:2019uch}%
  \BibitemOpen
  \bibfield  {author} {\bibinfo {author} {\bibfnamefont {G.}~\bibnamefont
  {Li}}, \bibinfo {author} {\bibfnamefont {X.-F.}\ \bibnamefont {Wang}},\ and\
  \bibinfo {author} {\bibfnamefont {Y.}~\bibnamefont {Xing}},\ }\href
  {https://doi.org/10.1140/epjc/s10052-019-7150-4} {\bibfield  {journal}
  {\bibinfo  {journal} {Eur. Phys. J. C}\ }\textbf {\bibinfo {volume} {79}},\
  \bibinfo {pages} {645} (\bibinfo {year} {2019})}\BibitemShut {NoStop}%
\bibitem [{\citenamefont {Chen}\ \emph {et~al.}(2020)\citenamefont {Chen},
  \citenamefont {Chen}, \citenamefont {Liu},\ and\ \citenamefont
  {Zhu}}]{Chen:2020xwe}%
  \BibitemOpen
  \bibfield  {author} {\bibinfo {author} {\bibfnamefont {H.-X.}\ \bibnamefont
  {Chen}}, \bibinfo {author} {\bibfnamefont {W.}~\bibnamefont {Chen}}, \bibinfo
  {author} {\bibfnamefont {X.}~\bibnamefont {Liu}},\ and\ \bibinfo {author}
  {\bibfnamefont {S.-L.}\ \bibnamefont {Zhu}},\ }\href
  {https://doi.org/https://doi.org/10.1016/j.scib.2020.08.038} {\bibfield
  {journal} {\bibinfo  {journal} {Sci. Bull.}\ }\textbf {\bibinfo {volume}
  {65}},\ \bibinfo {pages} {1994} (\bibinfo {year} {2020})}\BibitemShut
  {NoStop}%
\bibitem [{\citenamefont {Becchi}\ \emph {et~al.}(2020)\citenamefont {Becchi},
  \citenamefont {Ferretti}, \citenamefont {Giachino}, \citenamefont {Maiani},\
  and\ \citenamefont {Santopinto}}]{Becchi:2020uvq}%
  \BibitemOpen
  \bibfield  {author} {\bibinfo {author} {\bibfnamefont {C.}~\bibnamefont
  {Becchi}}, \bibinfo {author} {\bibfnamefont {J.}~\bibnamefont {Ferretti}},
  \bibinfo {author} {\bibfnamefont {A.}~\bibnamefont {Giachino}}, \bibinfo
  {author} {\bibfnamefont {L.}~\bibnamefont {Maiani}},\ and\ \bibinfo {author}
  {\bibfnamefont {E.}~\bibnamefont {Santopinto}},\ }\href
  {https://doi.org/10.1016/j.physletb.2020.135952} {\bibfield  {journal}
  {\bibinfo  {journal} {Phys. Lett. B}\ }\textbf {\bibinfo {volume} {811}},\
  \bibinfo {pages} {135952} (\bibinfo {year} {2020})}\BibitemShut
  {NoStop}%
\bibitem [{\citenamefont {Chen}\ \emph {et~al.}(2022)\citenamefont {Chen},
  \citenamefont {Yan},\ and\ \citenamefont {Chen}}]{Chen:2022sbf}%
  \BibitemOpen
  \bibfield  {author} {\bibinfo {author} {\bibfnamefont {H.-X.}\ \bibnamefont
  {Chen}}, \bibinfo {author} {\bibfnamefont {Y.-X.}\ \bibnamefont {Yan}},\ and\
  \bibinfo {author} {\bibfnamefont {W.}~\bibnamefont {Chen}},\ }\href
  {https://doi.org/10.1103/PhysRevD.106.094019} {\bibfield  {journal} {\bibinfo
   {journal} {Phys. Rev. D}\ }\textbf {\bibinfo {volume} {106}},\ \bibinfo
  {pages} {094019} (\bibinfo {year} {2022})}\BibitemShut {NoStop}%
\bibitem [{\citenamefont {Biloshytskyi}\ \emph {et~al.}(2022)\citenamefont
  {Biloshytskyi}, \citenamefont {Harland-Lang}, \citenamefont {Malaescu},
  \citenamefont {Pascalutsa}, \citenamefont {Schmieden},\ and\ \citenamefont
  {Schott}}]{Biloshytskyi:2022pdl}%
  \BibitemOpen
  \bibfield  {author} {\bibinfo {author} {\bibfnamefont {V.}~\bibnamefont
  {Biloshytskyi}}, \bibinfo {author} {\bibfnamefont {L.}~\bibnamefont
  {Harland-Lang}}, \bibinfo {author} {\bibfnamefont {B.}~\bibnamefont
  {Malaescu}}, \bibinfo {author} {\bibfnamefont {V.}~\bibnamefont
  {Pascalutsa}}, \bibinfo {author} {\bibfnamefont {K.}~\bibnamefont
  {Schmieden}},\ and\ \bibinfo {author} {\bibfnamefont {M.}~\bibnamefont
  {Schott}},\ }\href {https://doi.org/10.1051/epjconf/202227406007} {\bibfield
  {journal} {\bibinfo  {journal} {EPJ Web Conf.}\ }\textbf {\bibinfo {volume}
  {274}},\ \bibinfo {pages} {06007} (\bibinfo {year} {2022})}\BibitemShut
  {NoStop}%
\bibitem [{\citenamefont {Sang}\ \emph {et~al.}(2023)\citenamefont {Sang},
  \citenamefont {Wang}, \citenamefont {Zhang},\ and\ \citenamefont
  {Feng}}]{Sang:2023ncm}%
  \BibitemOpen
  \bibfield  {author} {\bibinfo {author} {\bibfnamefont {W.-L.}\ \bibnamefont
  {Sang}}, \bibinfo {author} {\bibfnamefont {T.}~\bibnamefont {Wang}}, \bibinfo
  {author} {\bibfnamefont {Y.-D.}\ \bibnamefont {Zhang}},\ and\ \bibinfo
  {author} {\bibfnamefont {F.}~\bibnamefont {Feng}},\ }\href@noop {} \ \Eprint {https://arxiv.org/abs/2307.16150}
  {arXiv:2307.16150 [hep-ph]} \BibitemShut {NoStop}%
\bibitem [{\citenamefont {Agaev}\ \emph
  {et~al.}(2023{\natexlab{b}})\citenamefont {Agaev}, \citenamefont {Azizi},
  \citenamefont {Barsbay},\ and\ \citenamefont {Sundu}}]{Agaev:2023rpj}%
  \BibitemOpen
  \bibfield  {author} {\bibinfo {author} {\bibfnamefont {S.~S.}\ \bibnamefont
  {Agaev}}, \bibinfo {author} {\bibfnamefont {K.}~\bibnamefont {Azizi}},
  \bibinfo {author} {\bibfnamefont {B.}~\bibnamefont {Barsbay}},\ and\ \bibinfo
  {author} {\bibfnamefont {H.}~\bibnamefont {Sundu}},\ }\href@noop {}
  {\bibfield  {journal} {\bibinfo  {journal} {Eur. Phys. J. C}\ }\textbf
  {\bibinfo {volume} {83}},\ \bibinfo {pages} {994} (\bibinfo {year}
  {2023}{\natexlab{b}})},\ \Eprint {https://arxiv.org/abs/2307.01857}
  {arXiv:2307.01857 [hep-ph]} \BibitemShut {NoStop}%
\bibitem [{\citenamefont {Huang}\ \emph {et~al.}(2021)\citenamefont {Huang},
  \citenamefont {Feng}, \citenamefont {Jia}, \citenamefont {Sang},
  \citenamefont {Yang},\ and\ \citenamefont {Zhang}}]{Huang:2021vtb}%
  \BibitemOpen
  \bibfield  {author} {\bibinfo {author} {\bibfnamefont {Y.}~\bibnamefont
  {Huang}}, \bibinfo {author} {\bibfnamefont {F.}~\bibnamefont {Feng}},
  \bibinfo {author} {\bibfnamefont {Y.}~\bibnamefont {Jia}}, \bibinfo {author}
  {\bibfnamefont {W.-L.}\ \bibnamefont {Sang}}, \bibinfo {author}
  {\bibfnamefont {D.-S.}\ \bibnamefont {Yang}},\ and\ \bibinfo {author}
  {\bibfnamefont {J.-Y.}\ \bibnamefont {Zhang}},\ }\href
  {https://doi.org/10.1088/1674-1137/ac0b38} {\bibfield  {journal} {\bibinfo
  {journal} {Chin. Phys. C}\ }\textbf {\bibinfo {volume} {45}},\ \bibinfo
  {pages} {093101} (\bibinfo {year} {2021})}\BibitemShut
  {NoStop}%
\bibitem [{\citenamefont {Feng}\ \emph {et~al.}(2020)\citenamefont {Feng},
  \citenamefont {Huang}, \citenamefont {Jia}, \citenamefont {Sang},
  \citenamefont {Xiong},\ and\ \citenamefont {Zhang}}]{Feng:2020riv}%
  \BibitemOpen
  \bibfield  {author} {\bibinfo {author} {\bibfnamefont {F.}~\bibnamefont
  {Feng}}, \bibinfo {author} {\bibfnamefont {Y.}~\bibnamefont {Huang}},
  \bibinfo {author} {\bibfnamefont {Y.}~\bibnamefont {Jia}}, \bibinfo {author}
  {\bibfnamefont {W.-L.}\ \bibnamefont {Sang}}, \bibinfo {author}
  {\bibfnamefont {X.}~\bibnamefont {Xiong}},\ and\ \bibinfo {author}
  {\bibfnamefont {J.-Y.}\ \bibnamefont {Zhang}},\ }\href@noop {} \ \Eprint {https://arxiv.org/abs/2009.08450} {arXiv:2009.08450}
  \BibitemShut {NoStop}%
\bibitem [{\citenamefont {Wang}\ \emph {et~al.}(2020)\citenamefont {Wang},
  \citenamefont {Lin}, \citenamefont {Xu}, \citenamefont {Xie}, \citenamefont
  {Huang},\ and\ \citenamefont {Chen}}]{Wang:2020gmd}%
  \BibitemOpen
  \bibfield  {author} {\bibinfo {author} {\bibfnamefont {X.-Y.}\ \bibnamefont
  {Wang}}, \bibinfo {author} {\bibfnamefont {Q.-Y.}\ \bibnamefont {Lin}},
  \bibinfo {author} {\bibfnamefont {H.}~\bibnamefont {Xu}}, \bibinfo {author}
  {\bibfnamefont {Y.-P.}\ \bibnamefont {Xie}}, \bibinfo {author} {\bibfnamefont
  {Y.}~\bibnamefont {Huang}},\ and\ \bibinfo {author} {\bibfnamefont
  {X.}~\bibnamefont {Chen}},\ }\href
  {https://doi.org/10.1103/PhysRevD.102.116014} {\bibfield  {journal} {\bibinfo
   {journal} {Phys. Rev. D}\ }\textbf {\bibinfo {volume} {102}},\ \bibinfo
  {pages} {116014} (\bibinfo {year} {2020})},\ \Eprint
  {https://arxiv.org/abs/2007.09697} {arXiv:2007.09697} \BibitemShut {NoStop}%
\bibitem [{\citenamefont {Feng}\ \emph {et~al.}(2021)\citenamefont {Feng},
  \citenamefont {Huang}, \citenamefont {Jia}, \citenamefont {Sang},\ and\
  \citenamefont {Zhang}}]{Feng:2020qee}%
  \BibitemOpen
  \bibfield  {author} {\bibinfo {author} {\bibfnamefont {F.}~\bibnamefont
  {Feng}}, \bibinfo {author} {\bibfnamefont {Y.}~\bibnamefont {Huang}},
  \bibinfo {author} {\bibfnamefont {Y.}~\bibnamefont {Jia}}, \bibinfo {author}
  {\bibfnamefont {W.-L.}\ \bibnamefont {Sang}},\ and\ \bibinfo {author}
  {\bibfnamefont {J.-Y.}\ \bibnamefont {Zhang}},\ }\href
  {https://doi.org/10.1016/j.physletb.2021.136368} {\bibfield  {journal}
  {\bibinfo  {journal} {Phys. Lett. B}\ }\textbf {\bibinfo {volume} {818}},\
  \bibinfo {pages} {136368} (\bibinfo {year} {2021})},\ \Eprint
  {https://arxiv.org/abs/2011.03039} {arXiv:2011.03039 [hep-ph]} \BibitemShut
  {NoStop}%
\bibitem [{\citenamefont {Maciu\l{}a}\ \emph {et~al.}(2021)\citenamefont
  {Maciu\l{}a}, \citenamefont {Sch\"afer},\ and\ \citenamefont
  {Szczurek}}]{Maciula:2020wri}%
  \BibitemOpen
  \bibfield  {author} {\bibinfo {author} {\bibfnamefont {R.}~\bibnamefont
  {Maciu\l{}a}}, \bibinfo {author} {\bibfnamefont {W.}~\bibnamefont
  {Sch\"afer}},\ and\ \bibinfo {author} {\bibfnamefont {A.}~\bibnamefont
  {Szczurek}},\ }\href {https://doi.org/10.1016/j.physletb.2020.136010}
  {\bibfield  {journal} {\bibinfo  {journal} {Phys. Lett. B}\ }\textbf
  {\bibinfo {volume} {812}},\ \bibinfo {pages} {136010} (\bibinfo {year}
  {2021})},\ \Eprint {https://arxiv.org/abs/2009.02100} {arXiv:2009.02100
  [hep-ph]} \BibitemShut {NoStop}%
\bibitem [{\citenamefont {Gon\c{c}alves}\ and\ \citenamefont
  {Moreira}(2021)}]{Goncalves:2021ytq}%
  \BibitemOpen
  \bibfield  {author} {\bibinfo {author} {\bibfnamefont {V.~P.}\ \bibnamefont
  {Gon\c{c}alves}}\ and\ \bibinfo {author} {\bibfnamefont {B.~D.}\ \bibnamefont
  {Moreira}},\ }\href {https://doi.org/10.1016/j.physletb.2021.136249}
  {\bibfield  {journal} {\bibinfo  {journal} {Phys. Lett. B}\ }\textbf
  {\bibinfo {volume} {816}},\ \bibinfo {pages} {136249} (\bibinfo {year}
  {2021})},\ \Eprint {https://arxiv.org/abs/2101.03798} {arXiv:2101.03798
  [hep-ph]} \BibitemShut {NoStop}%
\bibitem [{\citenamefont {Ma}\ and\ \citenamefont {Zhang}(2020)}]{Ma:2020kwb}%
  \BibitemOpen
  \bibfield  {author} {\bibinfo {author} {\bibfnamefont {Y.-Q.}\ \bibnamefont
  {Ma}}\ and\ \bibinfo {author} {\bibfnamefont {H.-F.}\ \bibnamefont {Zhang}},\
  }\href@noop {} \ \Eprint {https://arxiv.org/abs/2009.08376}
  {arXiv:2009.08376} \BibitemShut {NoStop}%
\bibitem [{\citenamefont {Wang}\ \emph
  {et~al.}(2021{\natexlab{d}})\citenamefont {Wang}, \citenamefont {Liu},\ and\
  \citenamefont {Matsuki}}]{Wang:2020tpt}%
  \BibitemOpen
  \bibfield  {author} {\bibinfo {author} {\bibfnamefont {J.-Z.}\ \bibnamefont
  {Wang}}, \bibinfo {author} {\bibfnamefont {X.}~\bibnamefont {Liu}},\ and\
  \bibinfo {author} {\bibfnamefont {T.}~\bibnamefont {Matsuki}},\ }\href
  {https://doi.org/10.1016/j.physletb.2021.136209} {\bibfield  {journal}
  {\bibinfo  {journal} {Phys. Lett. B}\ }\textbf {\bibinfo {volume} {816}},\
  \bibinfo {pages} {136209} (\bibinfo {year} {2021}{\natexlab{d}})},\ \Eprint
  {https://arxiv.org/abs/2012.03281} {arXiv:2012.03281} \BibitemShut {NoStop}%
\bibitem [{\citenamefont {Zhu}\ \emph {et~al.}(2020{\natexlab{a}})\citenamefont
  {Zhu}, \citenamefont {Guo}, \citenamefont {Zhang}, \citenamefont {Ma},\ and\
  \citenamefont {Li}}]{Zhu:2020sn}%
  \BibitemOpen
  \bibfield  {author} {\bibinfo {author} {\bibfnamefont {J.-W.}\ \bibnamefont
  {Zhu}}, \bibinfo {author} {\bibfnamefont {X.-D.}\ \bibnamefont {Guo}},
  \bibinfo {author} {\bibfnamefont {R.-Y.}\ \bibnamefont {Zhang}}, \bibinfo
  {author} {\bibfnamefont {W.-G.}\ \bibnamefont {Ma}},\ and\ \bibinfo {author}
  {\bibfnamefont {X.-Q.}\ \bibnamefont {Li}},\ }\href@noop {} \ \Eprint {https://arxiv.org/abs/2011.07799}
  {arXiv:2011.07799} \BibitemShut {NoStop}%
\bibitem [{\citenamefont {Gong}\ \emph {et~al.}(2020)\citenamefont {Gong},
  \citenamefont {Du}, \citenamefont {Zhou}, \citenamefont {Zhao},\ and\
  \citenamefont {Zhong}}]{Gong:2020bmg}%
  \BibitemOpen
  \bibfield  {author} {\bibinfo {author} {\bibfnamefont {C.}~\bibnamefont
  {Gong}}, \bibinfo {author} {\bibfnamefont {M.-C.}\ \bibnamefont {Du}},
  \bibinfo {author} {\bibfnamefont {B.}~\bibnamefont {Zhou}}, \bibinfo {author}
  {\bibfnamefont {Q.}~\bibnamefont {Zhao}},\ and\ \bibinfo {author}
  {\bibfnamefont {X.-H.}\ \bibnamefont {Zhong}},\ }\href@noop {} \ \Eprint {https://arxiv.org/abs/2011.11374} {arXiv:2011.11374}
  \BibitemShut {NoStop}%
\bibitem [{\citenamefont {Feng}\ \emph {et~al.}(2023)\citenamefont {Feng},
  \citenamefont {Huang}, \citenamefont {Jia}, \citenamefont {Sang},
  \citenamefont {Yang},\ and\ \citenamefont {Zhang}}]{Feng:2023ghc}%
  \BibitemOpen
  \bibfield  {author} {\bibinfo {author} {\bibfnamefont {F.}~\bibnamefont
  {Feng}}, \bibinfo {author} {\bibfnamefont {Y.}~\bibnamefont {Huang}},
  \bibinfo {author} {\bibfnamefont {Y.}~\bibnamefont {Jia}}, \bibinfo {author}
  {\bibfnamefont {W.-L.}\ \bibnamefont {Sang}}, \bibinfo {author}
  {\bibfnamefont {D.-S.}\ \bibnamefont {Yang}},\ and\ \bibinfo {author}
  {\bibfnamefont {J.-Y.}\ \bibnamefont {Zhang}},\ }\href@noop {} \ \Eprint {https://arxiv.org/abs/2311.08292}
  {arXiv:2311.08292 [hep-ph]} \BibitemShut {NoStop}%
\bibitem [{\citenamefont {Maiani}(2020)}]{Maiani:2020pur}%
  \BibitemOpen
  \bibfield  {author} {\bibinfo {author} {\bibfnamefont {L.}~\bibnamefont
  {Maiani}},\ }\href {https://doi.org/10.1016/j.scib.2020.08.019} {\bibfield
  {journal} {\bibinfo  {journal} {Sci. Bull.}\ }\textbf {\bibinfo {volume}
  {65}},\ \bibinfo {pages} {1949} (\bibinfo {year} {2020})},\ \Eprint
  {https://arxiv.org/abs/2008.01637} {arXiv:2008.01637 [hep-ph]} \BibitemShut
  {NoStop}%
\bibitem [{\citenamefont {Chao}\ and\ \citenamefont
  {Zhu}(2020)}]{Chao:2020dml}%
  \BibitemOpen
  \bibfield  {author} {\bibinfo {author} {\bibfnamefont {K.-T.}\ \bibnamefont
  {Chao}}\ and\ \bibinfo {author} {\bibfnamefont {S.-L.}\ \bibnamefont {Zhu}},\
  }\href {https://doi.org/10.1016/j.scib.2020.08.031} {\bibfield  {journal}
  {\bibinfo  {journal} {Sci. Bull.}\ }\textbf {\bibinfo {volume} {65}},\
  \bibinfo {pages} {1952} (\bibinfo {year} {2020})},\ \Eprint
  {https://arxiv.org/abs/2008.07670} {arXiv:2008.07670 [hep-ph]} \BibitemShut
  {NoStop}%
\bibitem [{\citenamefont {Richard}(2020)}]{Richard:2020hdw}%
  \BibitemOpen
  \bibfield  {author} {\bibinfo {author} {\bibfnamefont {J.-M.}\ \bibnamefont
  {Richard}},\ }\href {https://doi.org/10.1016/j.scib.2020.08.020} {\bibfield
  {journal} {\bibinfo  {journal} {Sci. Bull.}\ }\textbf {\bibinfo {volume}
  {65}},\ \bibinfo {pages} {1954} (\bibinfo {year} {2020})},\ \Eprint
  {https://arxiv.org/abs/2008.01962} {arXiv:2008.01962 [hep-ph]} \BibitemShut
  {NoStop}%
\bibitem [{\citenamefont {Zhuang}\ \emph {et~al.}(2022)\citenamefont {Zhuang},
  \citenamefont {Zhang}, \citenamefont {Ma},\ and\ \citenamefont
  {Wang}}]{Zhuang:2021pci}%
  \BibitemOpen
  \bibfield  {author} {\bibinfo {author} {\bibfnamefont {Z.}~\bibnamefont
  {Zhuang}}, \bibinfo {author} {\bibfnamefont {Y.}~\bibnamefont {Zhang}},
  \bibinfo {author} {\bibfnamefont {Y.}~\bibnamefont {Ma}},\ and\ \bibinfo
  {author} {\bibfnamefont {Q.}~\bibnamefont {Wang}},\ }\href
  {https://doi.org/10.1103/PhysRevD.105.054026} {\bibfield  {journal} {\bibinfo
   {journal} {Phys. Rev. D}\ }\textbf {\bibinfo {volume} {105}},\ \bibinfo
  {pages} {054026} (\bibinfo {year} {2022})},\ \Eprint
  {https://arxiv.org/abs/2111.14028} {arXiv:2111.14028 [hep-ph]} \BibitemShut
  {NoStop}%
\bibitem [{\citenamefont {Zhu}\ \emph {et~al.}(2020{\natexlab{b}})\citenamefont
  {Zhu}, \citenamefont {Guo}, \citenamefont {Zhang}, \citenamefont {Ma},\ and\
  \citenamefont {Li}}]{Zhu:2020snb}%
  \BibitemOpen
  \bibfield  {author} {\bibinfo {author} {\bibfnamefont {J.-W.}\ \bibnamefont
  {Zhu}}, \bibinfo {author} {\bibfnamefont {X.-D.}\ \bibnamefont {Guo}},
  \bibinfo {author} {\bibfnamefont {R.-Y.}\ \bibnamefont {Zhang}}, \bibinfo
  {author} {\bibfnamefont {W.-G.}\ \bibnamefont {Ma}},\ and\ \bibinfo {author}
  {\bibfnamefont {X.-Q.}\ \bibnamefont {Li}},\ }\href@noop {}\ \Eprint {https://arxiv.org/abs/2011.07799}
  {arXiv:2011.07799 [hep-ph]} \BibitemShut {NoStop}%
\bibitem [{\citenamefont {Zhang}\ \emph {et~al.}(2022)\citenamefont {Zhang}
  \emph {et~al.}}]{Zhang:2022qtp}%
  \BibitemOpen
  \bibfield  {author} {\bibinfo {author} {\bibfnamefont {J.}~\bibnamefont
  {Zhang}} \emph {et~al.},\ }\href
  {http://dx.doi.org/10.1140/epjc/s10052-022-11111-4} {\bibfield  {journal}
  {\bibinfo  {journal} {Eur. Phys. J. C}\ }\textbf {\bibinfo {volume} {82}}
  (\bibinfo {year} {2022})}\BibitemShut {NoStop}%
\bibitem [{\citenamefont {Yan}\ \emph {et~al.}(2023)\citenamefont {Yan},
  \citenamefont {Zhang},\ and\ \citenamefont {Jia}}]{Yan:2023lvm}%
  \BibitemOpen
  \bibfield  {author} {\bibinfo {author} {\bibfnamefont {T.-Q.}\ \bibnamefont
  {Yan}}, \bibinfo {author} {\bibfnamefont {W.-X.}\ \bibnamefont {Zhang}},\
  and\ \bibinfo {author} {\bibfnamefont {D.}~\bibnamefont {Jia}},\ }\href@noop
  {} {\bibfield  {journal} {\bibinfo  {journal} {Eur. Phys. J. C}\ }\textbf
  {\bibinfo {volume} {83}},\ \bibinfo {pages} {810} (\bibinfo {year} {2023})},\
  \Eprint {https://arxiv.org/abs/2304.01684} {arXiv:2304.01684 [hep-ph]}
  \BibitemShut {NoStop}%
\bibitem [{\citenamefont {Yu}\ \emph {et~al.}(2023)\citenamefont {Yu},
  \citenamefont {Li}, \citenamefont {Wang}, \citenamefont {Jie},\ and\
  \citenamefont {Meng}}]{Yu:2022lak}%
  \BibitemOpen
  \bibfield  {author} {\bibinfo {author} {\bibfnamefont {G.-L.}\ \bibnamefont
  {Yu}}, \bibinfo {author} {\bibfnamefont {Z.-Y.}\ \bibnamefont {Li}}, \bibinfo
  {author} {\bibfnamefont {Z.-G.}\ \bibnamefont {Wang}}, \bibinfo {author}
  {\bibfnamefont {L.}~\bibnamefont {Jie}},\ and\ \bibinfo {author}
  {\bibfnamefont {Y.}~\bibnamefont {Meng}},\ }\href
  {http://dx.doi.org/10.1140/epjc/s10052-023-11445-7} {\bibfield  {journal}
  {\bibinfo  {journal} {Eur. Phys. J. C}\ }\textbf {\bibinfo {volume} {83}}
  (\bibinfo {year} {2023})}\BibitemShut {NoStop}%
\bibitem [{\citenamefont {Bedolla}\ \emph {et~al.}(2020)\citenamefont
  {Bedolla}, \citenamefont {Ferretti}, \citenamefont {Roberts},\ and\
  \citenamefont {Santopinto}}]{Bedolla:2019zwg}%
  \BibitemOpen
  \bibfield  {author} {\bibinfo {author} {\bibfnamefont {M.~A.}\ \bibnamefont
  {Bedolla}}, \bibinfo {author} {\bibfnamefont {J.}~\bibnamefont {Ferretti}},
  \bibinfo {author} {\bibfnamefont {C.~D.}\ \bibnamefont {Roberts}},\ and\
  \bibinfo {author} {\bibfnamefont {E.}~\bibnamefont {Santopinto}},\ }\href
  {http://dx.doi.org/10.1140/epjc/s10052-020-08579-3} {\bibfield  {journal}
  {\bibinfo  {journal} {Eur. Phys. J. C}\ }\textbf {\bibinfo {volume} {80}}
  (\bibinfo {year} {2020})}\BibitemShut {NoStop}%
\bibitem [{\citenamefont {Tiwari}\ \emph {et~al.}(2023)\citenamefont {Tiwari},
  \citenamefont {Rathaud},\ and\ \citenamefont {Rai}}]{Tiwari:2021tmz}%
  \BibitemOpen
  \bibfield  {author} {\bibinfo {author} {\bibfnamefont {R.}~\bibnamefont
  {Tiwari}}, \bibinfo {author} {\bibfnamefont {D.~P.}\ \bibnamefont
  {Rathaud}},\ and\ \bibinfo {author} {\bibfnamefont {A.~K.}\ \bibnamefont
  {Rai}},\ }\href {https://doi.org/10.1007/s12648-022-02427-8} {\bibfield
  {journal} {\bibinfo  {journal} {Indian J. Phys.}\ }\textbf {\bibinfo {volume}
  {97}},\ \bibinfo {pages} {943} (\bibinfo {year} {2023})},\ \Eprint
  {https://arxiv.org/abs/2108.04017} {arXiv:2108.04017 [hep-ph]} \BibitemShut
  {NoStop}%
\bibitem [{\citenamefont {Kuchta}(2023)}]{Kuchta:2023anj}%
  \BibitemOpen
  \bibfield  {author} {\bibinfo {author} {\bibfnamefont {M.}~\bibnamefont
  {Kuchta}},\ }\href@noop {} \ \Eprint
  {https://arxiv.org/abs/2309.04794} {arXiv:2309.04794 [hep-ph]} \BibitemShut
  {NoStop}%
\bibitem [{\citenamefont {Niu}\ \emph {et~al.}(2022)\citenamefont {Niu},
  \citenamefont {Wang}, \citenamefont {Wang},\ and\ \citenamefont
  {Yang}}]{Niu:2022cug}%
  \BibitemOpen
  \bibfield  {author} {\bibinfo {author} {\bibfnamefont {P.-Y.}\ \bibnamefont
  {Niu}}, \bibinfo {author} {\bibfnamefont {E.}~\bibnamefont {Wang}}, \bibinfo
  {author} {\bibfnamefont {Q.}~\bibnamefont {Wang}},\ and\ \bibinfo {author}
  {\bibfnamefont {S.}~\bibnamefont {Yang}},\ }\href@noop {} \ \Eprint {https://arxiv.org/abs/2209.01924}
  {arXiv:2209.01924 [hep-ph]} \BibitemShut {NoStop}%
\bibitem [{\citenamefont {Zhou}\ \emph {et~al.}(2022)\citenamefont {Zhou},
  \citenamefont {Guo}, \citenamefont {Kuang}, \citenamefont {Yang},\ and\
  \citenamefont {Dai}}]{Zhou:2022xpd}%
  \BibitemOpen
  \bibfield  {author} {\bibinfo {author} {\bibfnamefont {Q.}~\bibnamefont
  {Zhou}}, \bibinfo {author} {\bibfnamefont {D.}~\bibnamefont {Guo}}, \bibinfo
  {author} {\bibfnamefont {S.-Q.}\ \bibnamefont {Kuang}}, \bibinfo {author}
  {\bibfnamefont {Q.-H.}\ \bibnamefont {Yang}},\ and\ \bibinfo {author}
  {\bibfnamefont {L.-Y.}\ \bibnamefont {Dai}},\ }\href
  {https://doi.org/10.1103/PhysRevD.106.L111502} {\bibfield  {journal}
  {\bibinfo  {journal} {Phys. Rev. D}\ }\textbf {\bibinfo {volume} {106}},\
  \bibinfo {pages} {L111502} (\bibinfo {year} {2022})}\BibitemShut {NoStop}%
\bibitem [{\citenamefont {Kuang}\ \emph {et~al.}(2023)\citenamefont {Kuang},
  \citenamefont {Zhou}, \citenamefont {Guo}, \citenamefont {Yang},\ and\
  \citenamefont {Dai}}]{Kuang:2023vac}%
  \BibitemOpen
  \bibfield  {author} {\bibinfo {author} {\bibfnamefont {S.-Q.}\ \bibnamefont
  {Kuang}}, \bibinfo {author} {\bibfnamefont {Q.}~\bibnamefont {Zhou}},
  \bibinfo {author} {\bibfnamefont {D.}~\bibnamefont {Guo}}, \bibinfo {author}
  {\bibfnamefont {Q.-H.}\ \bibnamefont {Yang}},\ and\ \bibinfo {author}
  {\bibfnamefont {L.-Y.}\ \bibnamefont {Dai}},\ }\href
  {https://doi.org/10.1140/epjc/s10052-023-11473-3} {\bibfield  {journal}
  {\bibinfo  {journal} {Eur. Phys. J. C}\ }\textbf {\bibinfo {volume} {83}},\
  \bibinfo {pages} {383} (\bibinfo {year} {2023})},\ \Eprint
  {https://arxiv.org/abs/2302.03968} {arXiv:2302.03968 [hep-ph]} \BibitemShut
  {NoStop}%
\bibitem [{\citenamefont {Zhao}\ \emph
  {et~al.}(2021{\natexlab{b}})\citenamefont {Zhao}, \citenamefont {Xu},
  \citenamefont {Kaewsnod}, \citenamefont {Liu}, \citenamefont {Limphirat},\
  and\ \citenamefont {Yan}}]{Zhao:2020jvl}%
  \BibitemOpen
  \bibfield  {author} {\bibinfo {author} {\bibfnamefont {Z.}~\bibnamefont
  {Zhao}}, \bibinfo {author} {\bibfnamefont {K.}~\bibnamefont {Xu}}, \bibinfo
  {author} {\bibfnamefont {A.}~\bibnamefont {Kaewsnod}}, \bibinfo {author}
  {\bibfnamefont {X.}~\bibnamefont {Liu}}, \bibinfo {author} {\bibfnamefont
  {A.}~\bibnamefont {Limphirat}},\ and\ \bibinfo {author} {\bibfnamefont
  {Y.}~\bibnamefont {Yan}},\ }\href
  {https://doi.org/10.1103/PhysRevD.103.116027} {\bibfield  {journal} {\bibinfo
   {journal} {Phys. Rev. D}\ }\textbf {\bibinfo {volume} {103}},\ \bibinfo
  {pages} {116027} (\bibinfo {year} {2021}{\natexlab{b}})}\BibitemShut
  {NoStop}%
\bibitem [{\citenamefont {Liu}\ \emph {et~al.}(2021)\citenamefont {Liu},
  \citenamefont {Liu}, \citenamefont {Zhong},\ and\ \citenamefont
  {Zhao}}]{Liu:2021rtn}%
  \BibitemOpen
  \bibfield  {author} {\bibinfo {author} {\bibfnamefont {F.-X.}\ \bibnamefont
  {Liu}}, \bibinfo {author} {\bibfnamefont {M.-S.}\ \bibnamefont {Liu}},
  \bibinfo {author} {\bibfnamefont {X.-H.}\ \bibnamefont {Zhong}},\ and\
  \bibinfo {author} {\bibfnamefont {Q.}~\bibnamefont {Zhao}},\ }\href
  {https://doi.org/10.1103/PhysRevD.104.116029} {\bibfield  {journal} {\bibinfo
   {journal} {Phys. Rev. D}\ }\textbf {\bibinfo {volume} {104}},\ \bibinfo
  {pages} {116029} (\bibinfo {year} {2021})},\ \Eprint
  {https://arxiv.org/abs/2110.09052} {arXiv:2110.09052 [hep-ph]} \BibitemShut
  {NoStop}%
\bibitem [{\citenamefont {Jaffe}(2005)}]{Jaffe:2004ph}%
  \BibitemOpen
  \bibfield  {author} {\bibinfo {author} {\bibfnamefont {R.~L.}\ \bibnamefont
  {Jaffe}},\ }\href {https://doi.org/10.1016/j.physrep.2004.11.005} {\bibfield
  {journal} {\bibinfo  {journal} {Phys. Rept.}\ }\textbf {\bibinfo {volume}
  {409}},\ \bibinfo {pages} {1} (\bibinfo {year} {2005})},\ \Eprint
  {https://arxiv.org/abs/hep-ph/0409065} {arXiv:hep-ph/0409065} \BibitemShut
  {NoStop}%
\bibitem [{\citenamefont {Du}\ \emph {et~al.}(2013)\citenamefont {Du},
  \citenamefont {Chen}, \citenamefont {Chen},\ and\ \citenamefont
  {Zhu}}]{Du:2012wp}%
  \BibitemOpen
  \bibfield  {author} {\bibinfo {author} {\bibfnamefont {M.-L.}\ \bibnamefont
  {Du}}, \bibinfo {author} {\bibfnamefont {W.}~\bibnamefont {Chen}}, \bibinfo
  {author} {\bibfnamefont {X.-L.}\ \bibnamefont {Chen}},\ and\ \bibinfo
  {author} {\bibfnamefont {S.-L.}\ \bibnamefont {Zhu}},\ }\href
  {https://doi.org/10.1103/PhysRevD.87.014003} {\bibfield  {journal} {\bibinfo
  {journal} {Phys. Rev. D}\ }\textbf {\bibinfo {volume} {87}},\ \bibinfo
  {pages} {014003} (\bibinfo {year} {2013})},\ \Eprint
  {https://arxiv.org/abs/1209.5134} {arXiv:1209.5134} \BibitemShut {NoStop}%
\bibitem{Wang:2018ejf}
  Z.~G.~Wang,
  Eur. Phys. J. C \textbf{79} (2019) no.1, 29
\bibitem [{\citenamefont {Wang}\ \emph
  {et~al.}(2021{\natexlab{e}})\citenamefont {Wang}, \citenamefont {Meng},
  \citenamefont {Oka},\ and\ \citenamefont {Zhu}}]{PhysRevD.104.036016}%
  \BibitemOpen
  \bibfield  {author} {\bibinfo {author} {\bibfnamefont {G.-J.}\ \bibnamefont
  {Wang}}, \bibinfo {author} {\bibfnamefont {L.}~\bibnamefont {Meng}}, \bibinfo
  {author} {\bibfnamefont {M.}~\bibnamefont {Oka}},\ and\ \bibinfo {author}
  {\bibfnamefont {S.-L.}\ \bibnamefont {Zhu}},\ }\href
  {https://doi.org/10.1103/PhysRevD.104.036016} {\bibfield  {journal} {\bibinfo
   {journal} {Phys. Rev. D}\ }\textbf {\bibinfo {volume} {104}},\ \bibinfo
  {pages} {036016} (\bibinfo {year} {2021}{\natexlab{e}})}\BibitemShut
  {NoStop}%
\bibitem [{\citenamefont {Shifman}\ \emph {et~al.}(1979)\citenamefont
  {Shifman}, \citenamefont {Vainshtein},\ and\ \citenamefont
  {Zakharov}}]{Shifman:1978bx}%
  \BibitemOpen
  \bibfield  {author} {\bibinfo {author} {\bibfnamefont {M.~A.}\ \bibnamefont
  {Shifman}}, \bibinfo {author} {\bibfnamefont {A.~I.}\ \bibnamefont
  {Vainshtein}},\ and\ \bibinfo {author} {\bibfnamefont {V.~I.}\ \bibnamefont
  {Zakharov}},\ }\href {https://doi.org/10.1016/0550-3213(79)90022-1}
  {\bibfield  {journal} {\bibinfo  {journal} {Nucl. Phys. B}\ }\textbf
  {\bibinfo {volume} {147}},\ \bibinfo {pages} {385} (\bibinfo {year}
  {1979})}\BibitemShut {NoStop}%
\bibitem [{\citenamefont {Reinders}\ \emph {et~al.}(1985)\citenamefont
  {Reinders}, \citenamefont {Rubinstein},\ and\ \citenamefont
  {Yazaki}}]{Reinders:1984sr}%
  \BibitemOpen
  \bibfield  {author} {\bibinfo {author} {\bibfnamefont {L.~J.}\ \bibnamefont
  {Reinders}}, \bibinfo {author} {\bibfnamefont {H.}~\bibnamefont
  {Rubinstein}},\ and\ \bibinfo {author} {\bibfnamefont {S.}~\bibnamefont
  {Yazaki}},\ }\href {https://doi.org/10.1016/0370-1573(85)90065-1} {\bibfield
  {journal} {\bibinfo  {journal} {Phys. Rept.}\ }\textbf {\bibinfo {volume}
  {127}},\ \bibinfo {pages} {1} (\bibinfo {year} {1985})}\BibitemShut {NoStop}%
\bibitem [{\citenamefont {Wilson}(1969)}]{Wilson:1969zs}%
  \BibitemOpen
  \bibfield  {author} {\bibinfo {author} {\bibfnamefont {K.~G.}\ \bibnamefont
  {Wilson}},\ }\href {https://doi.org/10.1103/PhysRev.179.1499} {\bibfield
  {journal} {\bibinfo  {journal} {Phys. Rev.}\ }\textbf {\bibinfo {volume}
  {179}},\ \bibinfo {pages} {1499} (\bibinfo {year} {1969})}\BibitemShut
  {NoStop}%
\bibitem [{\citenamefont {Ioffe}(1981)}]{Ioffe:1981kw}%
  \BibitemOpen
  \bibfield  {author} {\bibinfo {author} {\bibfnamefont {B.}~\bibnamefont
  {Ioffe}},\ }\href {https://doi.org/10.1016/0550-3213(81)90259-5} {\bibfield
  {journal} {\bibinfo  {journal} {Nucl. Phys. B}\ }\textbf {\bibinfo {volume}
  {188}},\ \bibinfo {pages} {317} (\bibinfo {year} {1981})}\BibitemShut
  {NoStop}%
\bibitem [{\citenamefont {Ovchinnikov}\ and\ \citenamefont
  {Pivovarov}(1988)}]{Ovchinnikov:1988gk}%
  \BibitemOpen
  \bibfield  {author} {\bibinfo {author} {\bibfnamefont {A.~A.}\ \bibnamefont
  {Ovchinnikov}}\ and\ \bibinfo {author} {\bibfnamefont {A.~A.}\ \bibnamefont
  {Pivovarov}},\ }\href@noop {} {\bibfield  {journal} {\bibinfo  {journal}
  {Sov. J. Nucl. Phys.}\ }\textbf {\bibinfo {volume} {48}},\ \bibinfo {pages}
  {721} (\bibinfo {year} {1988})}\BibitemShut {NoStop}%
\bibitem [{\citenamefont {Narison}(2018)}]{Narison:2018dcr}%
  \BibitemOpen
  \bibfield  {author} {\bibinfo {author} {\bibfnamefont {S.}~\bibnamefont
  {Narison}},\ }\href {https://doi.org/10.1142/S0217751X18500458} {\bibfield
  {journal} {\bibinfo  {journal} {Int. J. Mod. Phys. A}\ }\textbf {\bibinfo
  {volume} {33}},\ \bibinfo {pages} {1850045} (\bibinfo {year} {2018})},\
  \Eprint {https://arxiv.org/abs/1801.00592} {arXiv:1801.00592 [hep-ph]}
  \BibitemShut {NoStop}%
\bibitem [{\citenamefont {{Particle Data Group}}\ \emph
  {et~al.}(2022)\citenamefont {{Particle Data Group}} \emph
  {et~al.}}]{10.1093/ptep/ptac097}%
  \BibitemOpen
  \bibfield  {author} {\bibinfo {author} {\bibfnamefont{R.~L.} \bibnamefont{Workman}}
  \emph {et~al.} (Particle Data Group),\ }\href {https://doi.org/10.1093/ptep/ptac097} {\bibfield
  {journal} {\bibinfo  {journal} {PTEP}\ }\textbf {\bibinfo {volume} {2022}},\
  \bibinfo {pages} {083C01} (\bibinfo {year} {2022})}\BibitemShut {NoStop}%
\bibitem [{\citenamefont {Chen}\ \emph
  {et~al.}(2017{\natexlab{b}})\citenamefont {Chen}, \citenamefont {Chen},
  \citenamefont {Liu}, \citenamefont {Steele},\ and\ \citenamefont
  {Zhu}}]{Chen:2017rhl}%
  \BibitemOpen
  \bibfield  {author} {\bibinfo {author} {\bibfnamefont {W.}~\bibnamefont
  {Chen}}, \bibinfo {author} {\bibfnamefont {H.-X.}\ \bibnamefont {Chen}},
  \bibinfo {author} {\bibfnamefont {X.}~\bibnamefont {Liu}}, \bibinfo {author}
  {\bibfnamefont {T.~G.}\ \bibnamefont {Steele}},\ and\ \bibinfo {author}
  {\bibfnamefont {S.-L.}\ \bibnamefont {Zhu}},\ }\href
  {https://doi.org/10.1103/PhysRevD.95.114005} {\bibfield  {journal} {\bibinfo
  {journal} {Phys. Rev. D}\ }\textbf {\bibinfo {volume} {95}},\ \bibinfo
  {pages} {114005} (\bibinfo {year} {2017}{\natexlab{b}})}\BibitemShut
  {NoStop}%
\bibitem [{\citenamefont {Duan}\ \emph {et~al.}(2024)\citenamefont {Duan},
  \citenamefont {Wang}, \citenamefont {Yang}, \citenamefont {Chen},\ and\
  \citenamefont {Chen}}]{Duan:2024uuf}%
  \BibitemOpen
  \bibfield  {author} {\bibinfo {author} {\bibfnamefont {F.-B.}\ \bibnamefont
  {Duan}}, \bibinfo {author} {\bibfnamefont {Q.-N.}\ \bibnamefont {Wang}},
  \bibinfo {author} {\bibfnamefont {Z.-Y.}\ \bibnamefont {Yang}}, \bibinfo
  {author} {\bibfnamefont {X.-L.}\ \bibnamefont {Chen}},\ and\ \bibinfo
  {author} {\bibfnamefont {W.}~\bibnamefont {Chen}},\ }\href@noop {} \ \Eprint {https://arxiv.org/abs/2401.10078}
  {arXiv:2401.10078 [hep-ph]} \BibitemShut {NoStop}%
\end{thebibliography}

\end{document}